# Electrical-Field Distributions in Waveguide Arrays Exact and Approximate

*Uri Levy and Yaron Silberberg*
*Department of Physics of Complex Systems, Weizmann Institute of Science, Rehovot 76100, Israel*
*(Dated: November 26, 2013)*
*uri.levy@weizmann.ac.il*

**Abstract**

Five methods of calculating electrical field distributions in one dimensional wave-guide arrays are reviewed. We analytically solve the scalar Helmholtz Equation and, based on the computed Bloch functions and associated bands of propagation constants, generate the exact field distribution maps.

For the approximated slowly varying envelope equation we show that the base Bloch functions are identical to those in the exact case, and study the differences in the bands of propagation constants. We demonstrate that by selecting the reference refractive index value, it is possible to minimize the error in propagation constants of any desired band.

For the distributions calculated by the coupled mode theory, we reveal the similarity and differences of the band made of eigenvalues of the coupled mode equations matrix when compared to the first band of propagation constants found by the exact solution.

Analysis of two numeric beam propagation methods shows that the relative accuracy of the calculated field distributions of each of these methods depends on excitation conditions. The presented analysis of the slowly varying envelope equation provides guide lines for selecting the value of the reference refractive index to be incorporated in these numeric methods where an analytic solution is difficult to work out or in the frequently occurring cases where an analytic solution does not exist at all.

## 1. Overview

Propagation of an electrical field across an array of closely spaced waveguides (WG's) attracts a "visible" volume of research [1]. Not only the associated optical phenomena are of interest, but the dynamics of several other physical systems are mapped, one-to-one, onto the spatial distributions of the electrical field amplitudes in these WG arrays.

We review and compare five known methods for calculating electrical-field distributions in waveguide arrays:

A. Full (scalar) Helmholtz Equation ("EXAC")
B. Approximated Helmholtz Equation (slowly varying envelope) ("SVEA")
C. Coupled-mode theory ("CMT")
D. Beam propagation method – finite difference ("BPM-fd")
E. Beam propagation method – split step ("BPM-ss")

And while all five methods are discussed in detail in the following sections, we start by briefly reviewing each method here in the overview section.

### A. Full (scalar) Helmholtz Equation

Maxwell Equation for the electrical field [$\boldsymbol{E}(\boldsymbol{r})$], in the absence of currents and in the absence of free charges, can be arranged to read:

$$\nabla^2 \boldsymbol{E}(\boldsymbol{r}) + k_0^2 \cdot \epsilon(\boldsymbol{r}) \cdot \boldsymbol{E}(\boldsymbol{r}) = -\boldsymbol{\nabla}[\boldsymbol{\nabla} \circ \boldsymbol{E}(\boldsymbol{r})]$$

*(1)*

where $k_0$ is the vacuum wave-vector ($= \omega/C$) and $\epsilon(x, y, z)$ is the medium permittivity.

The scalar approximation $\{-\boldsymbol{\nabla}[\boldsymbol{\nabla} \circ \boldsymbol{E}(\boldsymbol{r})] \sim 0\}$ to the so arranged Maxwell Equation [equation *(1)*] leads to (homogeneous) Helmholtz Equation [for each of the three electric field components -$E(x, y, z)$] **[2],[3],[4]**:

$$\nabla^2 E(x, y, z) + k_0^2 \cdot \epsilon(x, y, z) \cdot E(x, y, z) = 0$$

*(2)*

If the permittivity function [$\epsilon(x, y, z)$] can be written as a product of three single-variable functions - $\epsilon(x, y, z) = f(x) \cdot g(y) \cdot h(z)$, equation *(2)* can be separated into three single-variable equations (and possibly be solved analytically).

For $y$-independent permittivity [$\epsilon(x, y, z) \Rightarrow \epsilon(x, z)$] and $y$-independent initial conditions, we can assume $\partial E/\partial y = 0$ such that equation *(2)* is reduced to -

$$\frac{\partial^2 E(x,z)}{\partial x^2} + \frac{\partial^2 E(x,z)}{\partial z^2} + k_0^2 \cdot \epsilon(x, z) \cdot E(x, z) = 0$$

*(3)*

In the case of z-independent permittivity [$\epsilon(x,z) \Rightarrow \epsilon(x)$], the separation of variables condition is trivially satisfied, and equation *(3)* gets its analytically separable form:

$$\frac{\partial^2 E_e(x,z)}{\partial x^2} + \frac{\partial^2 E_e(x,z)}{\partial z^2} + k_0^2 \cdot \epsilon(x) \cdot E_e(x,z) = 0$$

*(4)*

We solve equation *(4)*, the "Full (scalar) Helmholtz Equation" (in the context of this paper), for periodic step-index waveguides, by separation of variables. We regard the obtained solution as the "exact" solution (hence the subscript "*e*"), and use it as a reference for evaluating approximate solutions obtained by the other methods. (Below, for simplification, we use "EXAC").

## B. Approximated Helmholtz Equation (slowly varying envelope)

The known approach of using the Slowly Varying Envelope Approximation (SVEA) in order to simplify equation *(2)* is to try a solution of the form -

$$E_s(x,y,z) \equiv U_s(x,y,z) \cdot e^{i \cdot k_0 \cdot n_{ref} \cdot z}.$$

*(5)*

Inserting the trial sulution [equation *(5)*] into equation *(2)* and adopting the slowly varying envelope approximation:

$$\left| \frac{\partial^2 U_s(x,y,z)}{\partial z^2} \right| \ll \left| 2 \cdot i \cdot k_0 \cdot n_{ref} \cdot \frac{\partial U_s(x,y,z)}{\partial z} \right|$$

*(6)*

eliminates the second order z-derivative from equation *(2)* and thus leads to the frequently quoted SVEA Helmholtz Equation –

$$i \cdot \frac{\partial U_s(x,y,z)}{\partial z}$$
$$= -\frac{1}{2 \cdot k_0 \cdot n_{ref}} \cdot \left[ \frac{\partial^2 U_s(x,y,z)}{\partial x^2} + \frac{\partial^2 U_s(x,y,z)}{\partial y^2} \right]$$
$$- \frac{k_0^2}{2 \cdot k_0 \cdot n_{ref}} \cdot \left[ \epsilon(x,y,z) - n_{ref}^2 \right] \cdot U_s(x,y,z)$$

*(7)*

The wave-vector ($k_0 \cdot n_{ref}$) in the trial solution [equation *(5)*] is commonly referred to as the "reference wave-vector" **[5]** and the associated refractive index ($n_{ref}$) is referred to as the "reference refractive index" **[6]**. Since the value of the reference refractive index is not a priori fixed and is somewhat arbitrary, its proper choice is a subject of discussion in these papers (particularly **[5]**) as well as down below in this paper.

With the proper choice of variables, equation **[7]** can be transformed to an "Optical Schrödinger Equation" - exactly matching the "physical" Schrödinger Equation **[7]**:

$$i \cdot \hbar \cdot \frac{\partial \phi_s(x,y,\tau)}{\partial \tau} = -\frac{\hbar^2}{2 \cdot m_{opt}} \cdot \left[ \frac{\partial^2 \phi_s(x,y,\tau)}{\partial x^2} + \frac{\partial^2 \phi_s(x,y,\tau)}{\partial y^2} \right] + V_{opt}(x,y,\tau) \cdot \phi_s(x,y,\tau)$$

*(8)*

The optical length coordinate ($z$) is related to the Schrödinger "time" coordinate ($\tau$) as $z \equiv \tau/(\hbar \cdot k_0 \cdot n_{ref})$, the "optical mass" is defined as $m_{opt} \equiv (\hbar \cdot k_0 \cdot n_{ref})^2$, and the "optical potential" ($V_{opt}$) in equation *(8)* is defined as **[7]**:

$$V_{opt}(x,y,\tau) \equiv -\frac{1}{2} \cdot \left[ \frac{n^2(x,y,\tau) - n_{ref}^2}{n_{ref}^2} \right]$$

*(9)*

Looking at equation *(8)* and at equation *(9)*, we already see the importance of a proper choice of a value to the reference refractive index. We show below the relations of the chosen value to the reference index and the accuracy of the calculated SVEA propagation constants. In more detail – we show below that with careful selection of a reference index, any band can be targeted for minimum errors in its calculated set of propagation constants.

In the detailed discussion below we again assume x-only dependent permittivity [$\epsilon(x,y,z) \Rightarrow \epsilon(x)$] and analytically solve the SVEA Helmholtz Equation [equation *(7)*] by separation of variables.

## C. Coupled-mode theory

For calculating electrical field distributions in periodic one dimensional single-mode-supporting waveguide arrays by Coupled Mode Theory (CMT), we solve the nearest neighbors and next-nearest neighbors CMT equation -

$$i \cdot \frac{dA_j(z)}{dz} + C_{babc} \cdot [A_{j+1}(z) + A_{j-1}(z)] + C_{acb} \cdot [A_{j+2}(z) + A_{j-2}(z)] = 0$$

*(10)*

for a finite number of WGs, and for various excitation scenarios.

To determine the value of each of the coupling constants [$C_{babc}$ and $C_{acb}$] in equation *(10)*, we insert the single waveguide (WG) parameters and the WG-to-WG distance into analytic expressions that take into account the presence of neighboring WGs. The analytic expressions (to calculate $C_{babc}$ and $C_{acb}$) are briefly discussed and explicitly stated below.

For the initial conditions, to allow fair comparison, we suggest integration of the product of the (complex) input field and the WG mode across the full WG width.

Once the CMT amplitudes are calculated, we impose the single-mode electric field onto each WG-amplitude and normalize the overall resulted field.

## D. Beam propagation method – finite difference

The versatile and very frequently used Beam Propagation Method (BPM) is a numerical method applied to the SVEA Helmholtz Equation [equation *(7)*] **[8]**,**[9]**. Several BPM "methods" are common, one of which is a "finite difference" (fd) method.
Here we use a trapezoidal integration rule suggested by Youngchul Chung and Nadir Nagli **[10]** (resulting in a Crank Nicolson set of equations), "leading to a tridiagonal system of linear equations, which can be solved very efficiently". Listing of the key numerical code lines for the BPM-fd method is "pasted" in *Appendix 5*.

### E. Beam propagation method – split step

The "split step" (ss) version of the BPM is also very frequently used. Here we apply the more accurate version of "PQP" per step as discussed, for example, by Youngchul Chung and Nadir Nagli **[10]** and by Debjani Bhattacharya and Anurag Sharma **[11]**. And for the split-step too, listing of the key numerical code lines is "pasted" in an appendix (*Appendix 6*).

In the following sections we solve the three equations, one for each method [equations *(4)*, *(7)*, and *(10)*], after which we arrive at the method-comparison sections.

We discuss and compare the bands of propagation constants produced by each of the three analytic methods, compare the "base functions", and of course compare the resulting maps of electric field distributions produced by the above-mentioned five methods.

Essentially, and as expected, we find differences in calculated electrical field distributions (EXAC vs. the other methods). Relatively small differences in intensity distributions (depending on excitation conditions, and on the *z*-extent of the WG array), and rather large differences in amplitude distributions. Obviously, intensity inaccuracies and mainly phase inaccuracies are expected (for the predictions by the numeric methods), also in cases where analytic methods can no longer be applied.

## 2. Analytic Solutions

Let's now solve the three analytic equations, starting with the full (scalar) Helmholtz Equation.

### 2.1. Full (scalar) Helmholtz Equation

We wish to solve now equation *(4)* (repeated here for convenience):

$$\frac{\partial^2 E_e(x,z)}{\partial x^2} + \frac{\partial^2 E_e(x,z)}{\partial z^2} + k_0^2 \cdot \epsilon(x) \cdot E_e(x,z) = 0$$

*(11)*

#### 2.1.1. Homogeneous Permittivity

On passing, let's see the basic case of homogeneous permittivity (or "homogeneous space" or "uniform space") $\epsilon(x) \Rightarrow \epsilon_H$:

The solution to equation *(11)*, in the case of homogeneous permittivity is

$$E_{e;j}(x,z) = E_{e;0j} \cdot e^{i \cdot (k_{xj} \cdot x + k_{zj} \cdot z)} \; ; \; k_{x;j}^2 + k_{z;j}^2 = k_0^2 \cdot \epsilon_H$$

*(12)*

which represents an infinite set of plane-waves. Obviously, any linear combination of the solutions *(12)* will also solve the homogeneous equation [*(11)*]. And here the values of the propagation constants $[k_{z;j}]$ are continuous (no quantization and no bands).

#### 2.1.2. Non-homogeneous Permittivity

In the case of non-homogeneous permittivity, equation *(11)* is solved by separation of variables. Representing the electrical field $E_e(x,z)$ as a product of two functions -

$$E_e(x,z) = \psi_e(x) \cdot g_e(z)$$

*(13)*

equation *(11)* can be separated into two single-variable equations:

$$\frac{1}{g_e(z)} \frac{d^2 g_e(z)}{dz^2} = -K^2$$

$$\frac{1}{\psi_e(x)} \cdot \frac{d^2 \psi_e(x)}{dx^2} + k_0^2 \cdot \epsilon(x) = K^2$$

*(14)*

And the overall solution now reads:

$$E_e(x,z) = \psi_e(x) \cdot e^{i \cdot K \cdot z}$$

*(15)*

With $\psi_e(x)$ satisfying –

$$\frac{\partial^2 \psi_e(x)}{\partial x^2} + (k_0^2 \cdot \epsilon(x) - K^2) \cdot \psi_e(x) = 0$$

*(16)*

Throughout this paper we focus solely on the case of one dimensional lossless $[\epsilon(x) = n(x)^2]$ finite periodic structure:

$$n(x \geq 0) = \begin{cases} n_2 & (j-1) \cdot d + \frac{b}{2} < x \leq j \cdot d - \frac{b}{2} \\ n_1 & (j-1) \cdot d \leq x \leq (j-1) \cdot d + \frac{b}{2} \text{ and } (j-1) \cdot d + a + b/2 < x \leq j \cdot d \\ & j = 1,2, \ldots (N_p - 1), N_p \; ; \; d = a + b \end{cases}$$

$$n(-x) = n(x)$$

*(17)*

Defined this way, the array is perfectly symmetric (with respect to x=0), and spreads across $2 \cdot N_p$ periods (cf. *Figure 1*). However, to solve equation *(16)* (without loss of generality), it is convenient to shift the array (cf. *Figure 2*):

$$n(x) = \begin{cases} n_2 & j \cdot d < x \leq j \cdot d + a \\ n_1 & j \cdot d - b < x \leq j \cdot d \\ & j = -N_p, -(N_p - 1), \ldots 1,0,1, \ldots (N_p - 1) \; ; \; d = a + b \end{cases}$$

*(18)*

After equation *(16)* is solved for the shifted array [equation *(18)*], the solving functions are shifted back to fit the symmetric array. This way, the same symmetric structure appears in all five methods, allowing easy comparison [for the CMT WG array and for the two BPM WG arrays, we added two half-WGs at the edges, (cf. *Figure 1*), ending up with $2 \cdot N_p + 1$ WGs].

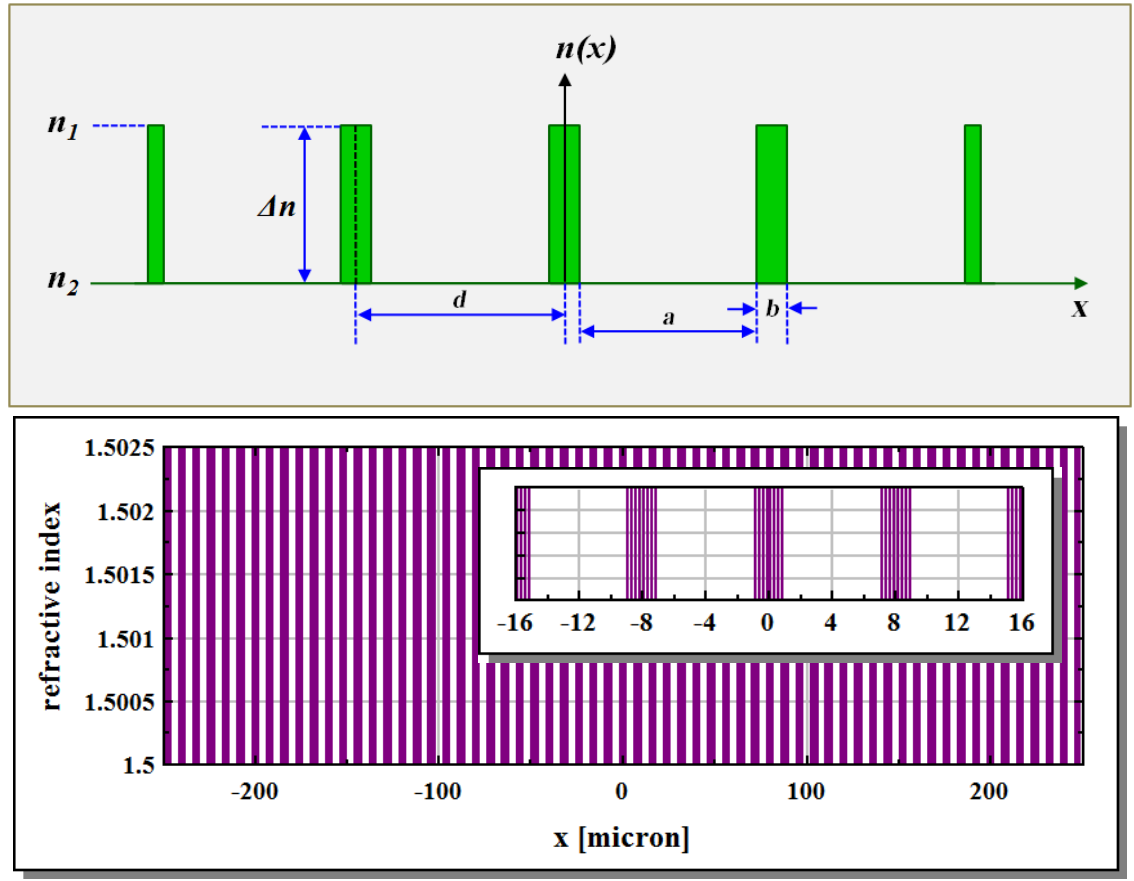

*Figure 1:* Top - parameters of the waveguide array assumed for all simulations reported in this paper (see also the list numbered *(22)*]. Bottom – the entire field (62 periods). Inset – zoom showing four periods.

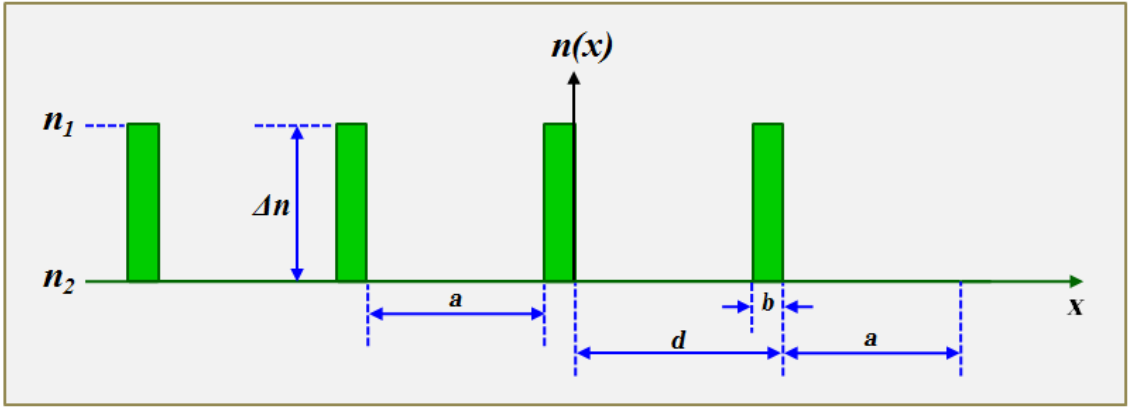

*Figure 2: Shifted array. Analytic solutions, found first for this shifted array, are shifted back to fit the symmetric array (cf. **Figure 1** and cf. **Appendix 1** and **Appendix 2**).*

The solution to equation *(16)*, with 1D periodic permittivity [equation *(18)* followed by equation *(17)*] is detailed in *Appendix 1*. In the appendix, we follow the Kronig–Penney model **[13]** and show that a complete set of orthogonal Bloch functions [$\psi_{e;n,j}(x)$] solve equation *(16)*. Inserting these [$\psi_{e;n,j}(x)$] solutions into equation *(15)*, we find the general solution to the Full (scalar) Helmholtz Equation [equation *(4)* or equation *(11)*]:

$$E_e(x, z) = \sum_{n,j} C_{e;n,j} \cdot \psi_{e;n,j}(x) \cdot e^{i \cdot K_{n,j} \cdot z}$$

*(19)*

The coefficients $C_{e;n,j}$ in equation *(19)* are determined by the initial conditions (the external electrical field at z = 0). Given the external electrical field [$E_{ext}(x)$] at $z = 0$, and as the set of Bloch functions forms a complete orthonormal basis **[14]**, the coefficients in *(19)* are calculated by an overlap integral:

$$C_{e;n,j} = \int_{x_{min}}^{x_{max}} E_{ext}(x) \cdot \psi^*_{e;n,j}(x) \cdot dx$$

*(20)*

Equations *(19)* and *(20)* constitute the complete solution for the (scalar) electrical field distribution in the EXAC case.

In the illustrating examples that follow we represent the WG array as a "potential" array (in most cases arbitrarily scaled), defined for the EXAC case as -

$$V_{exac}(x) \equiv V_e(x) \equiv -\frac{1}{2} \cdot \left[\frac{n^2(x) - n_1^2}{n_1^2}\right]$$

*(21)*

Throughout this paper, the wavelength and the WG array parameters are fixed at:

- $\lambda_0 = 0.8 \mu m$
- $k_0 = \frac{2 \cdot \pi}{\lambda_0} \mu m^{-1}$
- $a = 6 \mu m$
- $b = 2 \mu m$
- $d = a + b = 8 \mu m$
- $n_2 = 1.5$
- $n_1 = 1.5025$

*(22)*

The following four figures (*Figure 3*, *Figure 4*, *Figure 5*, and *Figure 6*) show examples of "cell functions" [$u_e(x)$], Bloch functions, and intensity functions. *Figure 3* also shows the basic mode supported by an isolated WG and, for illustration (similarity with quantum mechanics), shows "atoms".

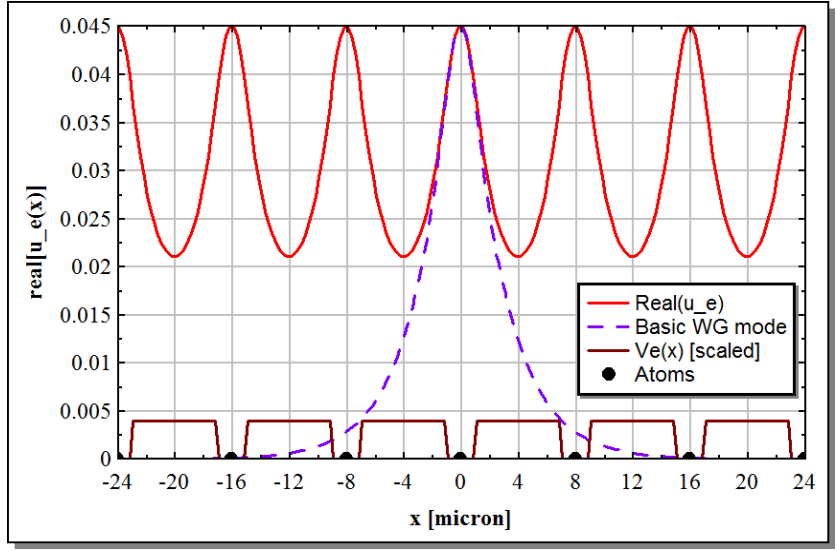

**Figure 3:** *First-band Bloch function at the center of the Brillouin Zone [$\psi_{e;1,N_p+1}$ ; $k_g = 0$]. The function solves equation (4) with the 1D WG array given by equation (17) [$a = 6$ ; $b = 2$]. Dashed purple – basic mode supported by a (stand-alone) waveguide of the array. The brown curve represents a scaled "potential" [equation (21)]. Note that the way the potential is defined [equation (21)], potential wells represent permittivity (or refractive index)* **ridges**. *Black circles – "atoms".*

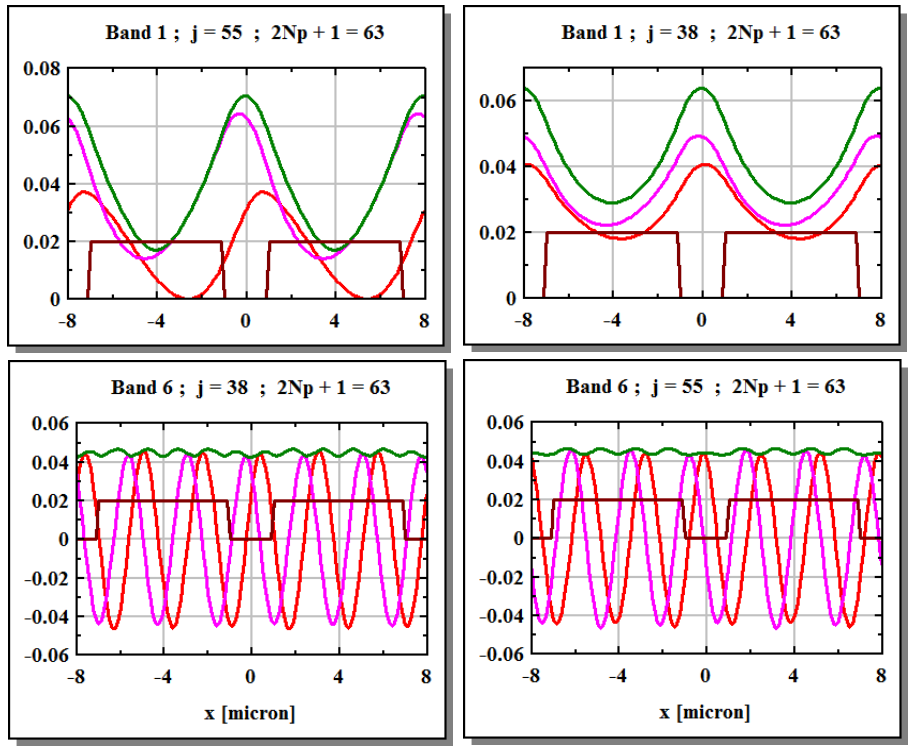

**Figure 4:** *Examples of four (out of 6*63 in this example) Bloch "cell functions" – $u_e(x)$ [cf. equation (67) of Appendix 1]. Red – real. Magenta – imaginary. Green - $|u_e(x)|^2$. Brown – scaled "potential".*

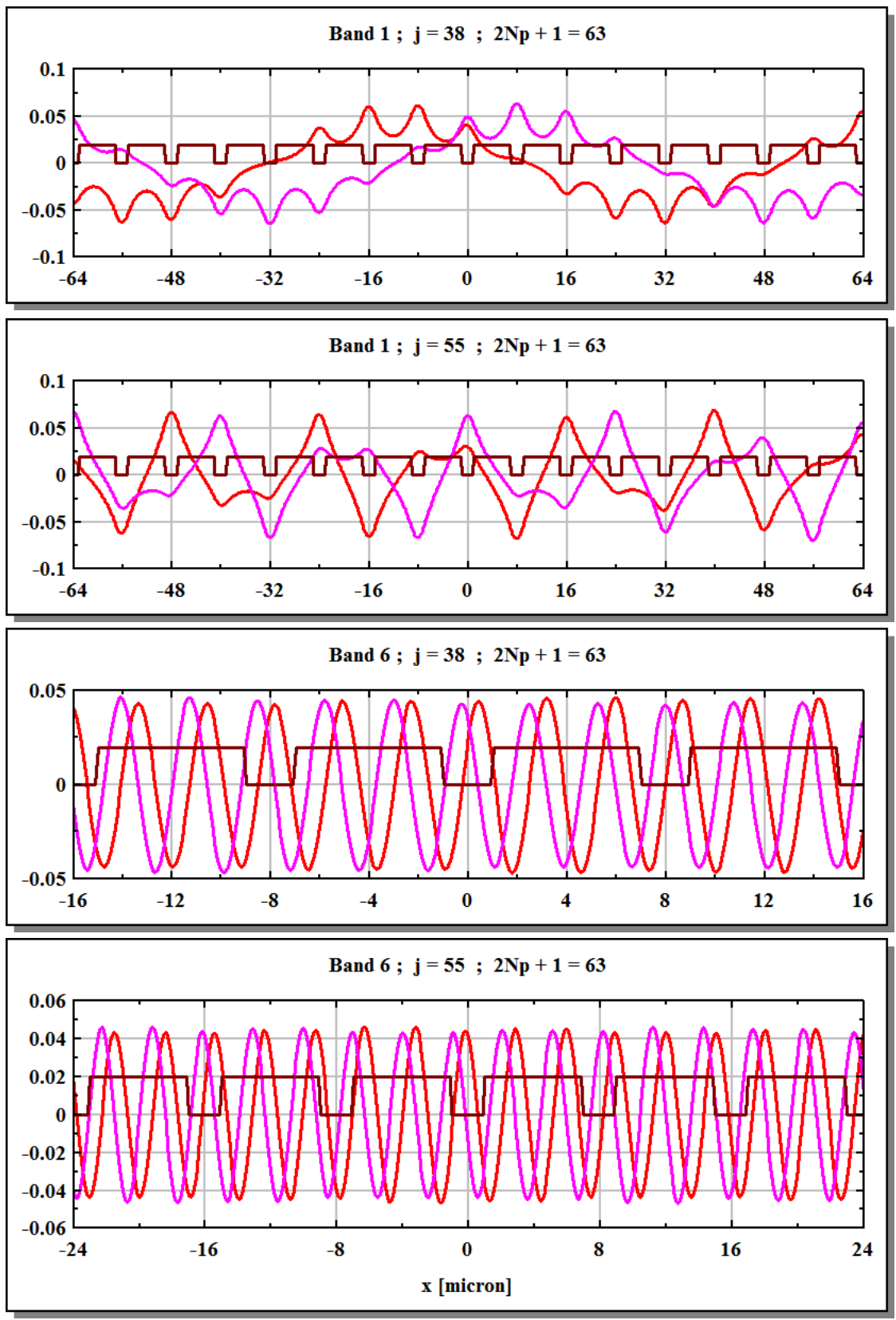

*Figure 5:* Examples of four Bloch functions [cf. equation **(67)** of *Appendix 1*]. Red – real part. Magenta – imaginary part. Brown – scaled "potential".

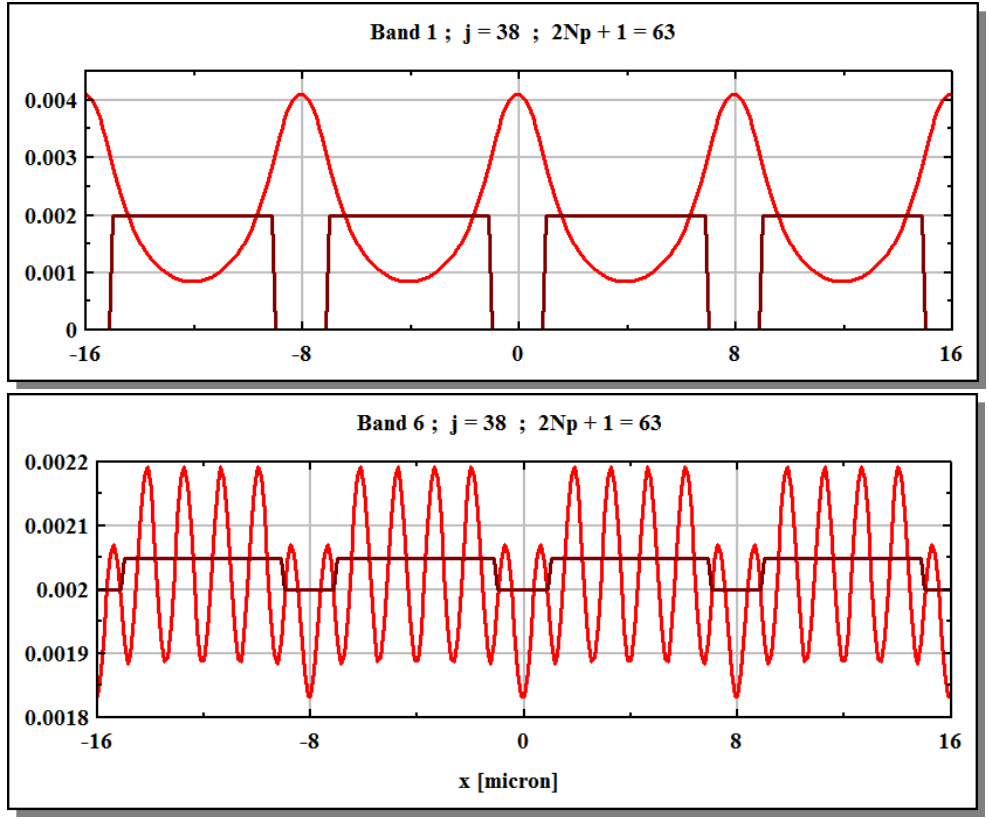

*Figure 6: Examples of two "intensity" (or "probability") distributions. Bloch-function intensities exhibit cell periodicity.*

The following two figures (*Figure 7* and *Figure 8*) involve the input field. The input field, throughout this paper, is a Gaussian characterized by three parameters $[\sigma\,;k_{tilt}\,;x_{shift}]$:

$$E_{ext}(x) = e^{i \cdot k_{tilt} \cdot x} \cdot e^{-\frac{(x-x_{shift})^2}{2 \cdot \sigma^2}}$$

*(23)*

*Figure 7* shows various input Gaussians (green) and their "reconstruction" by the calculated Bloch functions (red). For the shown Gaussians, Bloch functions "contributed" by six bands seem to adequately reconstruct the input field (right of *Figure 7*). However, as *Figure 7* shows, a set of Bloch functions contributed by the first band only do NOT faithfully reconstruct the input Gaussian, particularly the shifted one (center left of *Figure 7*).

*Figure 8* shows bands of expansion coefficients [equation *(20)*] for a narrow Gaussian $[\sigma = 2\mu m]$. The top two sets, related to a centered Gaussian, show strong excitation of the first band and very weak excitation of all other bands. In contrast, the bottom two sets, related to a shifted Gaussian excitation (in between two WGs), show strong excitation of the second band and even the third band.

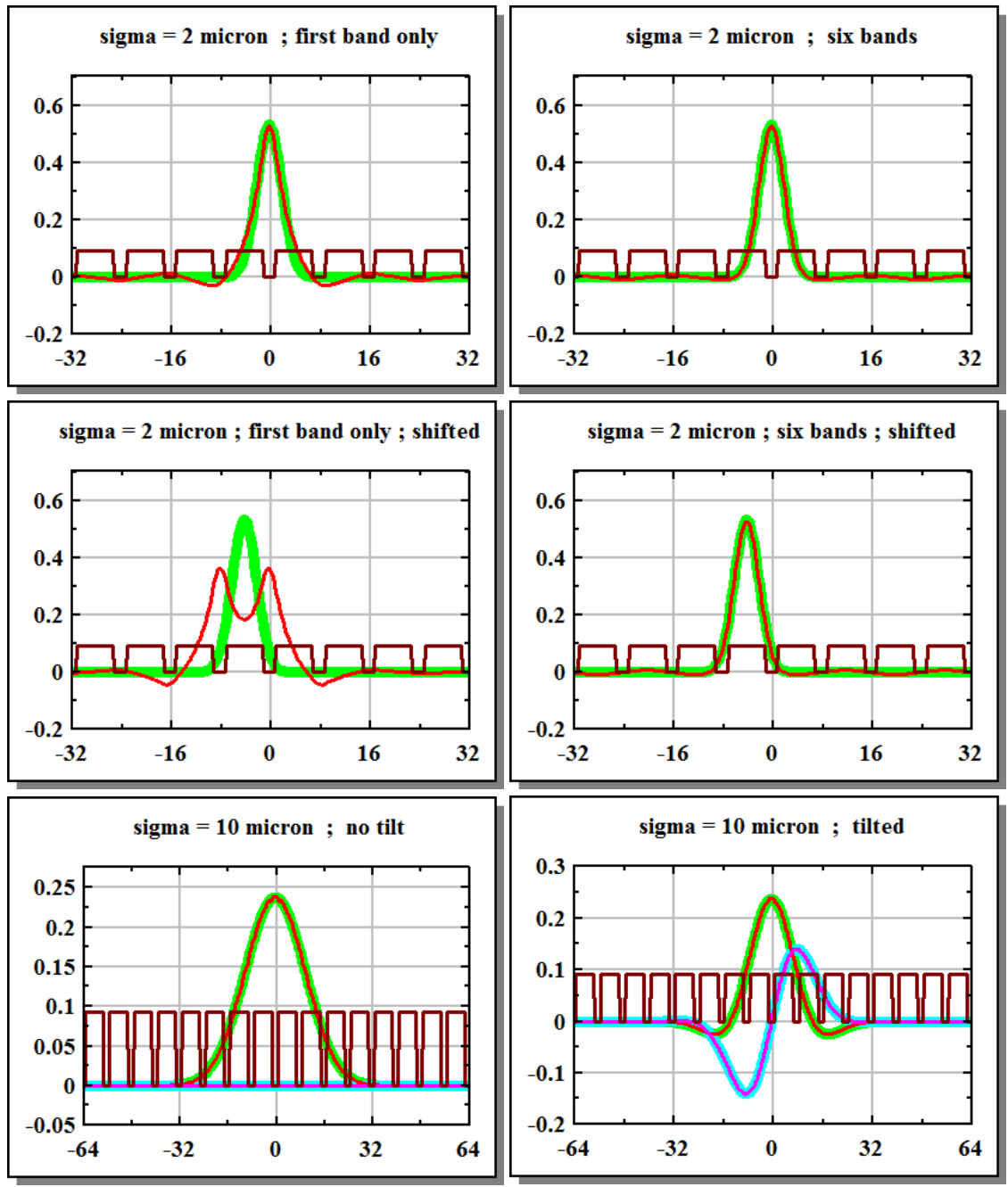

**Figure 7:** *"Reconstruction" of the input electrical field by the set of Bloch functions. {Green - Re[$E_{ext}(x)$] ; blue – Im[$E_{ext}(x)$] ; red – Re[$E_e(x,0)$] ; magenta – Im[$E_e(x,0)$]}. Top-left and center-left – first band only.*

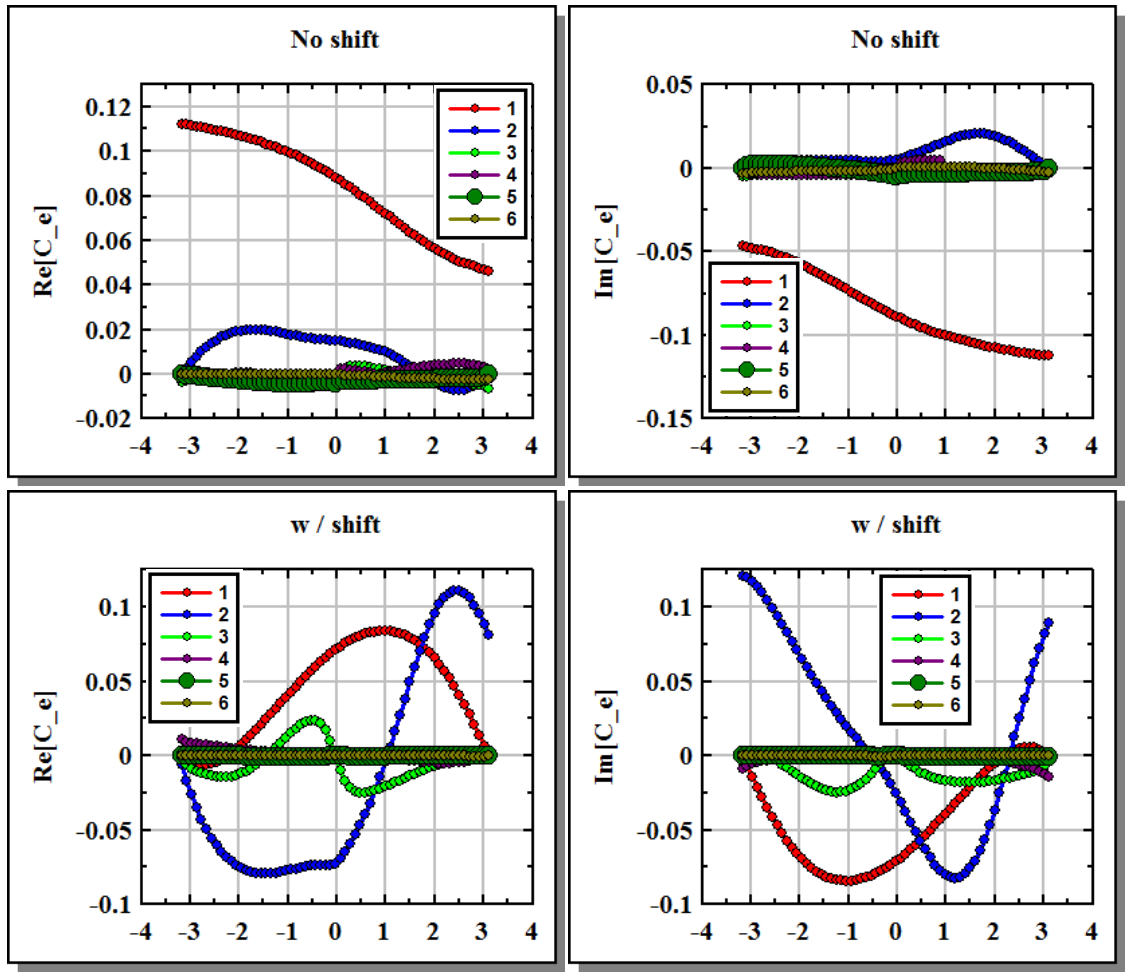

***Figure 8:*** *Bands of expansion coefficients for a narrow Gaussian [σ = 2μm]. Top – centered Gaussian. Essentially Bloch functions of only the first band are excited (second band excitation strength ~ 1/4 of the first band excitation strength). Bottom – same Gaussian shifted half period to the left (cf. center graphs of **Figure 7**). Here, second band Bloch functions dominate with "visible" contribution of Bloch functions from the third band too. Electrical field distribution maps associated with such shifted-Gaussian excitation are shown in **Figure 29** of section **4.1.4.***

Our discussion of the EXAC solution ends here. The electrical field distributions generated by the EXAC-solving functions are presented in section *4* below and are used as references against which approximated distributions (calculated by SVEA, CMT, BPM-fd, and BPM-ss) are compared.

We proceed now to discussing the SVEA case.

## 2.2. Approximated Helmholtz Equation (slowly varying envelope)

We wish to solve equation *(7)* for the case of y-independent and z-independent permittivity $[\epsilon(x,y,z) \Rightarrow \epsilon(x)]$ (and for y-independent initial conditions):

$$i \cdot \frac{\partial U_s(x,z)}{\partial z} = -\frac{1}{2 \cdot k_0 \cdot n_{ref}} \cdot \frac{\partial^2 U_s(x,z)}{\partial x^2} - \frac{k_0^2}{2 \cdot k_0 \cdot n_{ref}} \cdot [\epsilon(x) - n_{ref}^2] \cdot U_s(x,z)$$

*(24)*

### 2.2.1. Homogeneous Permittivity

On passing, let's see the basic case of homogeneous permittivity $\epsilon(x) \Rightarrow \epsilon_H$. For $\epsilon(x) = \epsilon_H$ the natural choice for the reference index is $n_{ref}^2 = n_H^2 = \epsilon_H$. With this choice of the reference index, the last term on the RHS of equation *(24)* drops, and we are left with -

$$i \cdot \frac{\partial U_s(x,z)}{\partial z} = -\frac{1}{2 \cdot k_H} \cdot \frac{\partial^2 U_s(x,z)}{\partial x^2} \quad ; \quad k_H \equiv k_0 \cdot n_H$$

*(25)*

#### 2.2.1.1. Plane-waves solution

If we now solve equation *(25)* by separation of variables and insert the solution into a suitably modified equation *(5)*:

$$E_s(x,z) \equiv U_s(x,z) \cdot e^{i \cdot k_H \cdot z}.$$

*(26)*

we get:

$$E_{s;j}(x,z) = E_{s;0j} \cdot e^{i \cdot (k_{x;j} \cdot x + k_{z;j} \cdot z)}$$

$$k_{z;j} \equiv k_H - \frac{k_{x;j}^2}{2 \cdot k_H}$$

and for $|k_{x;j}| \ll |k_H|$

$$k_{x;j}^2 + k_{z;j}^2 = k_H^2 + \frac{k_{x;j}^4}{4 \cdot k_H^2} \approx k_H^2$$

*(27)*

The solution *(27)* is similar to the solutions *(12)* under the paraxial approximation **[15],[16]**. Here too the values of the propagation constants $[k_{z;j}]$ are continuous (but must, as required by the paraxial approximation, stay "close" to $k_H$), and no bands are formed.

### 2.2.1.2. Gaussian beams

Equation *(25)* is commonly solved by Gaussian beams **[17],[18]**:

$$U_{s;j}(x,z) = A_{H-G}^{j} \cdot \left[\frac{\sigma(0)}{\sigma(z)}\right]^{\frac{1}{2}} \cdot H_j\left(\frac{x}{\sigma(z)}\right)$$
$$\cdot exp\left[-\frac{x^2}{2 \cdot \sigma(z)^2} + \frac{i \cdot k_H \cdot x^2}{2 \cdot R(z)} - i \cdot (j + 0.5) \cdot \text{atan}\left(\frac{z}{L_F}\right)\right]$$

*(28)*

where $A_{H-G}^{j}$ are normalizing amplitudes (for Hermite-Gaussian polynomials), $H_j\left(\frac{x}{\sigma(z)}\right)$ is Hermite polynomial of order $j$ and with the parameters as commonly defined **[16][17]**: $L_F \equiv \frac{\pi \cdot w_0^2}{\lambda} \equiv \frac{1}{2} \cdot k_H \cdot w_0^2$ ; $\frac{w(z)}{\sqrt{2}} = \sigma(z) \equiv \sigma_0 \cdot \left[1 + \left(\frac{z}{L_F}\right)^2\right]^{\frac{1}{2}}$ ; $R(z) \equiv z \cdot \left[1 + \left(\frac{L_F}{z}\right)^2\right]$.

Thus, the full solution to the SVEA (or paraxial) Helmholtz Equation, given homogeneous permittivity [equation *(25)*], can (**also**) be written as:

$$E_{s;j}(x,z) = A_{H-G}^{j} \cdot \left[\frac{\sigma(0)}{\sigma(z)}\right]^{\frac{1}{2}} \cdot H_j\left(\frac{x}{\sigma(z)}\right) \cdot \exp^{i \cdot k_H \cdot z}$$
$$\cdot exp\left[-\frac{x^2}{2 \cdot \sigma(z)^2} + \frac{i \cdot k_H \cdot x^2}{2 \cdot R(z)} - i \cdot (j + 0.5) \cdot \text{atan}\left(\frac{z}{L_F}\right)\right]$$

*(29)*

The set of Hermite-Gaussian solutions [equation *(29)*] is obviously very different from the set of plane-wave solutions [equation *(27)*], yet both solve the SVEA Helmholtz Equation [equation *(25)* with the preceding assumption *(5)*→*(26)*].

The maps in *Figure 9* show the propagation of high-order Hermite-Gaussian beams (order 4 with five lobes and order 7 with eight lobes) in a homogeneous medium [equation *(29)*].

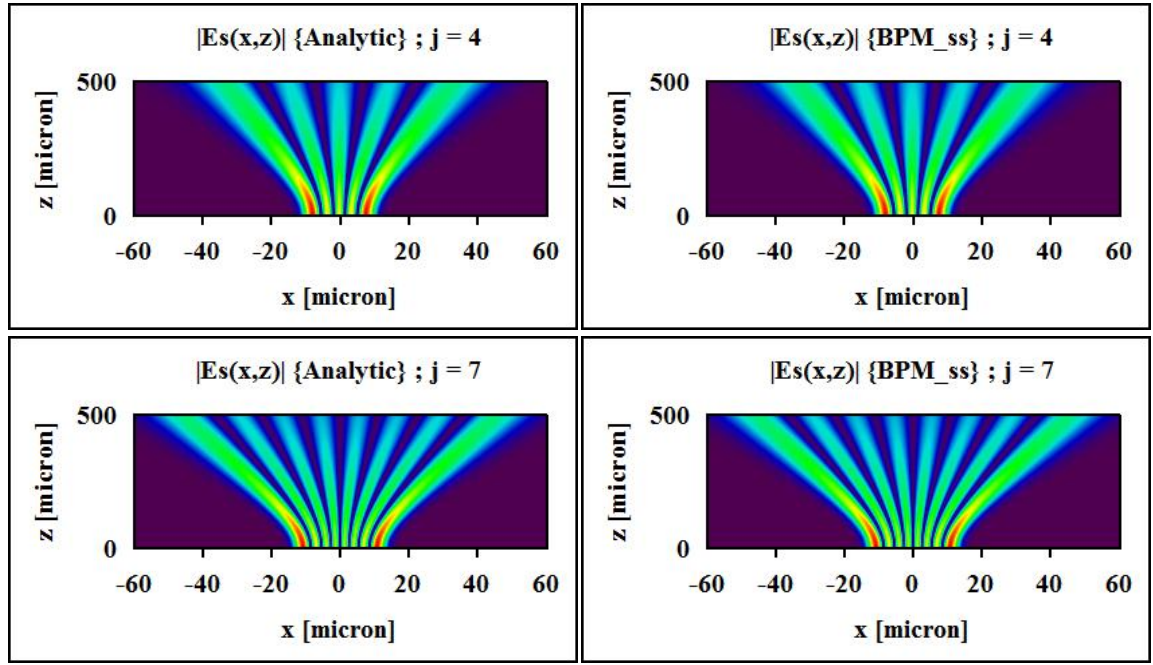

*Figure 9:* Propagation of Hermite-Gaussian fields in a homogeneous medium [equation *(29)*].

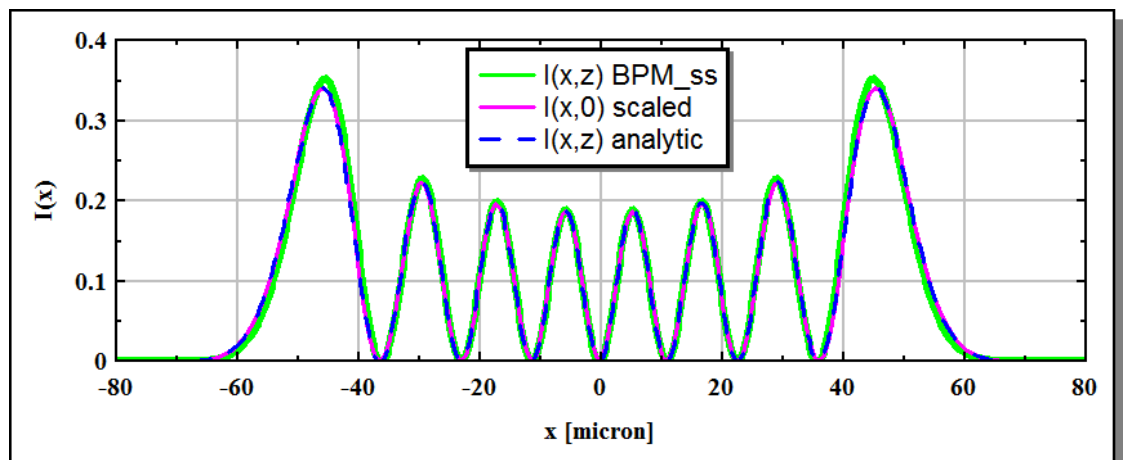

*Figure 10:* Cross-sectional intensity for Hermite-Gaussian of order 7 (cf. lower row of *Figure 9*) at distance z. As equation *(29)* predicts, for a pure Hermite-Gaussian excitation, the functional intensity distribution is maintained at any distance z, apart from an *x*-scaling factor.

*Figure 10* shows the cross-sectional intensity at distance $z$. As the theoretical solution [equation *(29)*] states, for any distance $z$ the initial Hermite-Gaussian functional distribution is preserved, apart from an $x$-scaling factor.

### 2.2.2. Non-homogeneous Permittivity

But we wish to concentrate in this paper on WG arrays. So we want to solve the SVEA Helmholtz Equation [equation *(24)*], repeated here for convenience:

$$i \cdot \frac{\partial U_s(x,z)}{\partial z} = -\frac{1}{2 \cdot k_0 \cdot n_{ref}} \cdot \frac{\partial^2 U_s(x,z)}{\partial x^2} - \frac{k_0^2}{2 \cdot k_0 \cdot n_{ref}} \cdot \left[\epsilon(x) - n_{ref}^2\right] \cdot U_s(x,z)$$

*(30)*

And here again we assume (similar to the assumption for the Full Helmholtz case), a finite periodic 1D WG array as given by equation *(18)*. Once again we solve for the non-symmetric WG array [equation *(18)*] and shift the solving functions to fit the symmetric array [equation *(17)*].

The solution to equation *(30)*, with 1D periodic permittivity [equation *(18)*] is detailed in *Appendix 2*.

The Kronig–Penney model **[13]**, following separation of variables, applies also to the SVEA Helmholtz Equation. So that a complete set of orthogonal Bloch functions [$\psi_{s;n,j}(x)$] solve equation *(30)* with permittivity as defined by equation *(18)*. Inserting these [$\psi_{s;n,j}(x)$] solutions into equation *(5)*, we find the general solution to the SVEA Helmholtz Equation [equation *(7)* or equation *(30)*]:

$$E_s(x,z) = \sum_{n,j} C_{s;n,j} \cdot \psi_{s;n,j}(x) \cdot e^{i \cdot k_0 \cdot n_{ref} \cdot (1-q_{n,j}) \cdot z}$$

*(31)*

The parameters $q_{n,j}$ in the exponent of the solution *(31)* determine the propagation bands and are derived and discussed in *Appendix 2*.
The coefficients $C_{s;n,j}$ in equation *(31)* are determined by the initial conditions (the external electrical field at z = 0). Given the external electrical field [$E_{ext}(x)$] at $z = 0$, and as the set of Bloch functions forms a complete orthonormal basis **[14]**, the coefficients in *(31)* are calculated by an overlap integral:

$$C_{s;n,j} = \int_{x_{min}}^{x_{max}} E_{ext}(x) \cdot \psi_{s;n,j}^*(x) \cdot dx$$

*(32)*

Equations *(31)* and *(32)* constitute the complete solution for the (scalar) electrical field distribution in the SVEA case.

In the illustrating example that follow (cf. *Figure 11*) we represent the WG array as a "potential" array - $V_s(x)$ (in most cases arbitrarily scaled), defined for the SVEA case as -

$$V_s(x) \equiv -\frac{1}{2} \cdot \left[ \frac{n^2(x) - n_{ref}^2}{n_{ref}^2} \right]$$

*(33)*

While the solution for the electrical field distribution in the EXAC case is determined solely by the properties of the WG array, the solution for the electrical field distribution in the SVEA case is determined also by the choice if the reference index. However, as shown below (*Figure 11*) and as verified by comparing equation *(59)* to equation *(76)*, the two sets of Bloch-functions are identical:

$$\psi_{e;n,j}(x) = \psi_{s;n,j}(x)$$

*(34)*

From equation *(34)*, equality of the expansion coefficients (for a given external electrical field) follows [cf. equation *(20)* and equation *(32)*]:

$$C_{e;n,j}(x) = C_{s;n,j}(x)$$

*(35)*

The difference in the electrical field distributions (EXAC vs. SVEA) stems solely from the differences in the propagation constants ($k_{ze;n,j} \neq k_{zs;n,j}$). These differences are discussed in detail in section *3* below.

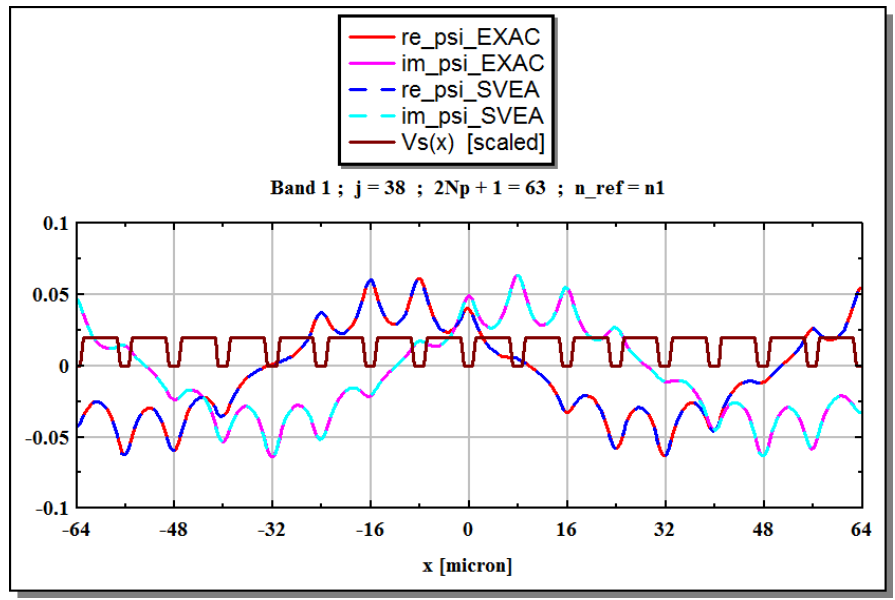

*Figure 11:* *Bloch functions for the EXAC case (red and magenta) vs. Bloch functions for the SVEA case (blue and cyan). Regardless of the choice of the reference index, the two sets of Bloch functions match [equation (34)].*

Note the difference between the "optical situation" discussed here and the quantum mechanical situation {described by exactly the same SVEA equation [equation *(7)*]}. In the optical case, the permittivity array is fixed, and the SVEA potential is an approximation, depending on the choice of the reference index. In the quantum mechanical case, the potential is given as well as the mass of the particle. The reference index (for a fixed wavelength) is determined by the optical mass [see text following equation *(8)*]. The freedom to choose the reference index (for numerical simulations) is therefore removed. The two situations are numerically identical if and only if the two potentials are identical given the **mass-determined** reference index.

So much for the analytic SVEA solution. Let's turn now to discussing the third method of analytically calculating electrical field distributions in waveguide arrays – the Coupled Mode Theory (CMT).

## 2.3. Coupled-Mode Theory

The coupled mode theory treats $2 \cdot N_p + 1$ discrete functions that obey equation *(10)* (repeated here for reader's convenience):

$$i \cdot \frac{dA_j(z)}{dz} + C_{babc} \cdot [A_{j+1}(z) + A_{j-1}(z)] + C_{acb} \cdot [A_{j+2}(z) + A_{j-2}(z)] = 0$$

*(36)*

Written in a matrix form, equation *(36)* reads -

$$\frac{d[\bar{A}(z)]}{dz} = (B) \cdot \bar{A}(z)$$

$$\bar{A}(0) \equiv \bar{A}_0 \equiv [A_{10}, A_{20}, \ldots A_{N0}] \ ; \ N = 2 \cdot N_p + 1$$

*(37)*

If the propagation matrix $[(B)]$ is diagonalizable, the solution to equation *(36)* [or *(37)*] is given as:

$$\bar{A}(z) = (\xi) \cdot e^{(I_K) \cdot z} \cdot (\xi)^{-1} \cdot \bar{A}_0$$

*(38)*

where $(\xi)$ is the diagonalizing matrix [of matrix $(B)$], made-up of the $N$ eigenvectors of $(B)$:

$$(\xi)^{-1} \cdot (B) \cdot (\xi) = (I_K)$$

*(39)*

$(I_K)$ is the unity matrix with 1's replaced by $K_j$'s – the eigenvalues of $(B)$, and

$$e^{(I_K) \cdot z} \equiv \begin{bmatrix} e^{K_1 \cdot z}, 0, 0, 0, \ldots, 0 \\ 0, e^{K_2 \cdot z}, 0, 0, \ldots, 0 \\ 0, 0, e^{K_3 \cdot z}, 0, \ldots, 0 \\ \ldots \\ 0, 0, 0, 0, \ldots, e^{K_N \cdot z} \end{bmatrix} \cdot$$

*(40)*

In the case of equation *(36)*, the propagation matrix $(B)$ is a hollow penta-diagonal symmetric purely complex (and hence energy-conserving anti-Hermitian) matrix with $i \cdot C_{babc}$ in the $\pm 2$ diagonals and $i \cdot C_{acb}$ in the $\pm 3$ diagonals:

$$(B) = \begin{bmatrix} 0, i \cdot C_{babc}, i \cdot C_{acb}, 0,0,0,0,0,0,0,0, \ldots, 0 \\ i \cdot C_{babc}, 0, i \cdot C_{babc}, i \cdot C_{acb}, 0, 0,0,0, \ldots, 0 \\ i \cdot C_{acb}, i \cdot C_{babc}, 0, i \cdot C_{babc}, i \cdot C_{acb}, \ldots, 0 \\ \ldots \\ 0,0,0,0, 0,0,0,0, 0,0, \ldots i \cdot C_{acb}, i \cdot C_{babc}, 0 \end{bmatrix}.$$

*(41)*

The coupling coefficients $[C_{babc} ; C_{acb}]$ are calculated by analytic expressions, which are extensions of the basic expression [equation *(42)*] developed in the seventies by A. Yariv **[19]**:

$$C_{ab}(\Delta n, b, a) = \frac{2 \cdot \delta_x^2 \cdot \gamma_x \cdot e^{-\gamma \cdot a}}{k_{z,WG} \cdot (b + 2/\gamma_x) \cdot k_0^2 \cdot (n_1^2 - n_2^2)}$$

*(42)*

$$C_{babc}(\Delta n, b, a) = C_{ab} + 2 \cdot \frac{k_0^2}{2 \cdot k_{z,WG} \cdot I_a} \cdot \cos^2(\delta_x \cdot b/2) \cdot (n_1^2 - n_2^2) \cdot \frac{1}{2 \cdot \gamma_x} \cdot [e^{\gamma_x \cdot b} - e^{-\gamma_x \cdot b}] \cdot e^{[-\gamma_x \cdot (2 \cdot b + 3 \cdot a)]}$$

*(43)*

$$C_{ac}(\Delta n, b, a) = \frac{2 \cdot \delta_x^2 \cdot \gamma_x \cdot e^{-\gamma_x \cdot (b+2 \cdot a)}}{k_{z,WG} \cdot (b + 2/\gamma_x) \cdot k_0^2 \cdot (n_1^2 - n_2^2)}$$

*(44)*

$$C_{acb}(\Delta n, b, a) = C_{ac} + \frac{k_0^2}{2 \cdot k_{z,WG} \cdot I_a} \cdot \cos^2(\delta_x \cdot b/2) \cdot w \cdot (n_1^2 - n_2^2) \cdot e^{[-\gamma \cdot (b+2 \cdot a)]}$$

*(45)*

The parameters in equations *(42)*,*(43)*,*(44)*,*(45)* are the WG parameters (cf. *Figure 1*), and the parameters of the basic mode $[A_{wg}(x)]$ guided by one such (isolated) WG [equation *(46)*]:

$$A_{wg}(x) = \begin{cases} \cos(\delta_x \cdot x) & |x| \leq b/2 \\ e^{-\gamma_x \cdot (x-b/2)} & x > b/2 \\ e^{\gamma_x \cdot (x+b/2)} & x < b/2 \end{cases}$$

$$k_{z,WG}^2 = k_0^2 \cdot n_1^2 - \delta_x^2$$

*(46)*

The parameter $I_a$ in equations *(43)*, and *(45)* is the integral:

$$I_a \equiv \int_{-\infty}^{\infty} A_{wg}^2(x) \cdot dx = \frac{b}{2} \cdot \left[1 + \frac{\sin(\delta_x \cdot b)}{\delta_x \cdot b}\right] + \frac{\cos^2(\delta_x \cdot b/2)}{\gamma_x}$$

*(47)*

In the examples to follow, the values for the initial "vector" [$\bar{A}(0)$, cf. equation *(37)*] are calculated by an overlap integral of the external electrical field [$E_{ext}(x)$] at $z = 0$ and the WG mode across the respective WG:

$$A_{j0} \equiv A_j(0) = \int_{-b/2}^{b/2} E_{ext}(x) \cdot A_{wg;j}^*(x) \cdot dx$$

*(48)*

The CMT field at $z = 0$ [$E_{CMT}(x, 0)$] is then composed of the discrete amplitudes – $A_j(0)$, each "carrying" the basic (shifted) mode – $A_{wg;j}(x)$:

$$E_{CMT}(x, 0) = \sum_j A_j(0) \cdot A_{wg;j}(x)$$

*(49)*

The coefficients $A_j(0)$ are now normalized (to yield $A_{norm;j}(0)$] such that

$$\int_{x_{min}}^{x_{max}} \left(\sum_j A_{norm;j}(0) \cdot A_{wg,j}(x)\right) \cdot \left(\sum_j A_{norm;j}(0) \cdot A_{wg,j}(x)\right)^* \cdot dx$$
$$= 1$$

*(50)*

And so now the CMT electrical field $[E_{CMT}(x,z)]$ is normalized (at any distance $z$):

$$E_{CMT}(x,z) = \sum_j A_{norm;j}(z) \cdot A_{wg;j}(x)$$

*(51)*

*Figure 12* shows intensity of the "constructed" CMT field at $z = 0$ for an off-axis wide $[\sigma = 10\mu m]$ Gaussian excitation. The inset shows the calculated amplitudes (green – real, blue – imaginary) for the excited waveguides [equation *(48)*]. The full CMT field at $z = 0$ is constructed by multiplying the amplitude and the basic (shifted) WG mode for every waveguide [equation *(49)*].

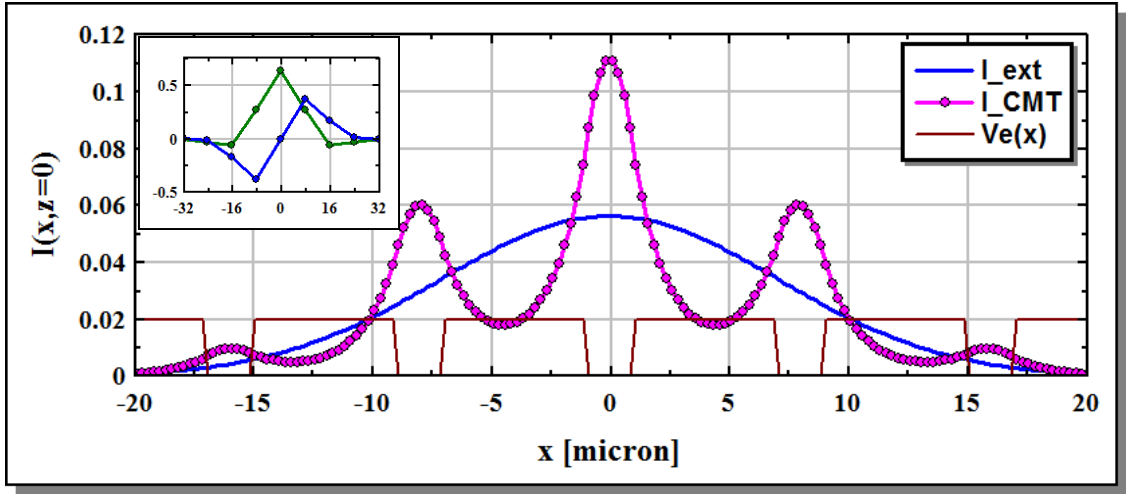

*Figure 12:* *Calculated CMT intensity cross-section at $z = 0$ for an off-axis wide $[\sigma = 10\mu m]$ Gaussian excitation. The inset shows the calculated amplitudes (green – real, blue – imaginary) for the excited waveguides [equation (48)].*

At this point, our review of the three analytic methods, for calculating the electrical field distributions in 1D waveguide arrays (EXAC, SVEA, CMT) ends. The mathematical aspects of the two numerical methods (BPM-fd, BPM-ss) will not be reviewed in this paper. However, listing of the key code lines for each of these numerical methods is "pasted" in a dedicated appendix (cf. *Appendix 5* and *Appendix 6*).

Each of the three analytic methods is characterized by a set of propagation constants $[k_{z;n,j}]$, that form bands. The EXAC and SVEA bands formations are discussed each in a dedicated appendix (cf. *Appendix 1* and *Appendix 2*). The CMT (single) band will be discussed below in the following "comparison of bands" section.

### 3. Comparison of bands

As discussed in *Appendix 1* and in *Appendix 2*, Bloch-functions solve the EXAC and SVEA equations (equation *(59)* and equation *(76)* respectively). The restrictions imposed on these solving functions [cf. equation *(62)*], dictate bands of the propagation constants $(k_{z;n,j})$.

In the case of the CMT, the set of eigenvalues of matrix $(B)$ [cf. equation *(40)*] form a "half band", allowing comparison of such band with the first band calculated in the EXAC case.

In this comparison of bands section, we want to compare these propagation constant bands, starting with SVEA vs. EXAC.

### 3.1. SVEA vs. EXAC

Stated already above is the fact that SVEA propagation constant bands depend on the choice of the reference index [$n_{ref}$ – cf. equation *(5)*] and are generally different when compared with the EXAC bands.

The curves of *Figure 13* show $k_z(EXAC)$ and $k_z(SVEA)$ for the first six bands with $n_{ref} = n_1$.

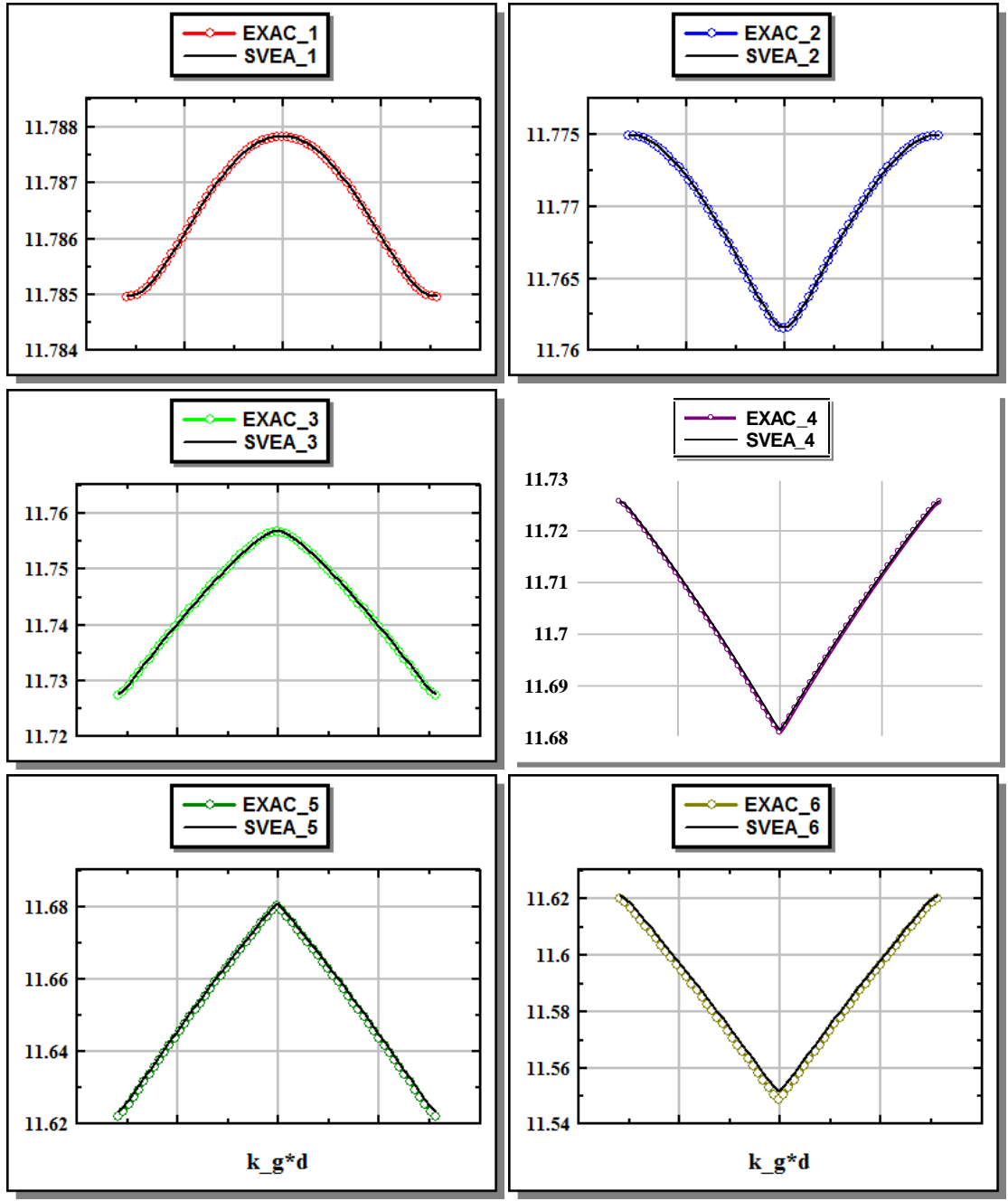

*Figure 13:* *First six bands – EXAC vs. SVEA for $n_{ref} = n_1$. As shown, the bands are quite similar (cf.* ***Figure 14*** *to see the differences). Vertical axis of these six charts is $k_z$ in units of $1/\mu m$.*

Overall the bands shown in *Figure 13* are rather similar. A closer look however, enabled by explicit difference calculations, reveals the magnitude of the differences.

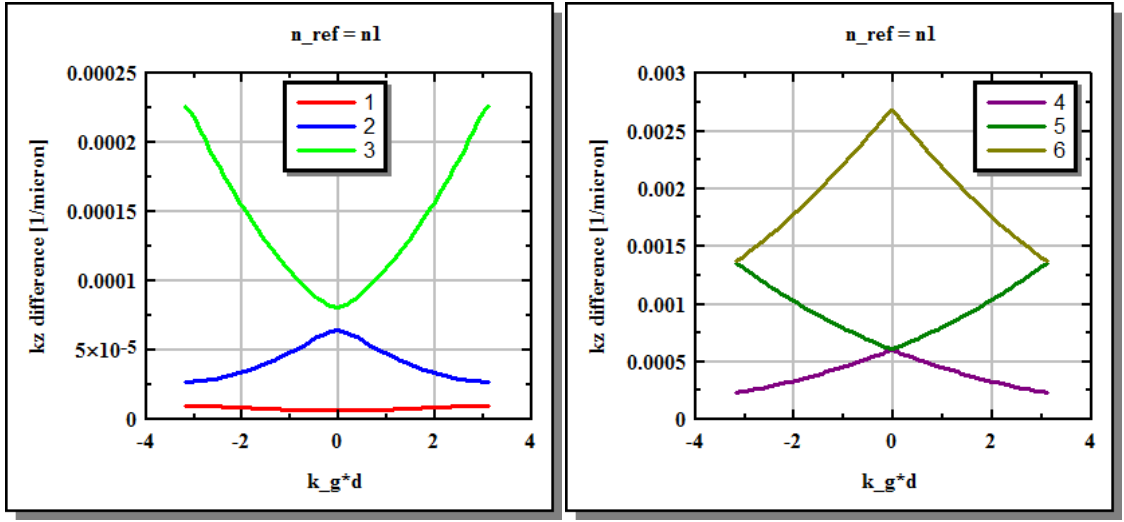

*Figure 14:* *Differences in propagation bands $[k_z(difference) \equiv k_z(SVEA) - k_z(EXAC)]$ for $n_{ref} = n_1$. If higher order bands (right) are excited, then calculated differences in electrical field distributions will show up already at a propagation distance of 1mm or so ($\Delta k_z \cdot z \approx 1$).*

*Figure 14* shows the calculated differences $[k_z(difference) \equiv k_z(SVEA) - k_z(EXAC)]$ for $n_{ref} = n_1$. The first three bands (left of *Figure 14*) exhibit differences on the order of $10^{-5}$ to $10^{-4}$ $\mu m^{-1}$ [for the specific case at hand – cf. list *(22)*]. Proceeding to the next three bands (right of *Figure 14*), the differences are larger by an order of magnitude. If these higher bands are excited (cf. section 4), electrical field distribution differences (SVEA vs. EXAC) will already show up at $z$ distances of $1 mm$ or so.

For $n_{ref} = n_1$, as is the case displayed by *Figure 13* and by *Figure 14*, the error $[k_z(SVEA) - k_z(EXAC)]$ increases with band number. But such monotonic increase in error with band number is NOT a general rule.

In fact, comparing equation *(59)* to equation *(76)*, with the definition *(77)*, we realize that

$$K_e^2 = \beta_{ref}^2 \cdot (1 - 2 \cdot q)$$

(52)

or

$$K_e = \beta_{ref} \cdot \left[1 - \frac{1}{2} \cdot (2 \cdot q) - \left(\frac{1}{8}\right) \cdot (2 \cdot q)^2 - \left(\frac{1}{16}\right) \cdot (2 \cdot q)^3 - \left(\frac{5}{128}\right) \cdot (2 \cdot q)^4 - \cdots \right]$$

(53)

Given that $k_z(EXAC) = K_e$ [equation *(70)*] and $k_z(SVEA)] = \beta_{ref} \cdot (1 - q) \equiv k_0 \cdot n_{ref} \cdot (1 - q)$ [equation *(85)*], we find for the difference -

$$k_{z;n,j}(SVEA) - k_{z;n,j}(EXAC) \cong k_0 \cdot n_{ref} \cdot [(1/2) \cdot q_{n,j}^2 + (1/2) \cdot q_{n,j}^3]$$
*(54)*

And since the "eigenvalues" ($q_{n,j}$) are also functions of the reference index:

$$q_{n,j} = q_{n,j}(n_{ref})$$
*(55)*

**it is possible to select a reference index so as to minimize the error $[k_{z;n,j}(SVEA) - k_{z;n,j}(EXAC)]$ for any specific band**.

Consult *Figure 15*. As the reference index is continuously lowered, the bands continuously "slide" to the left. *Figure 15* shows two snap-shots. The one on the left for a high reference index ($n_{ref} = n_1$) where (the $q$ values of) all bands are positive, and one on the right for a low reference index ($n_{ref} = n_2 - 0.0016$) where the entire band number one and part of band number two are negative.

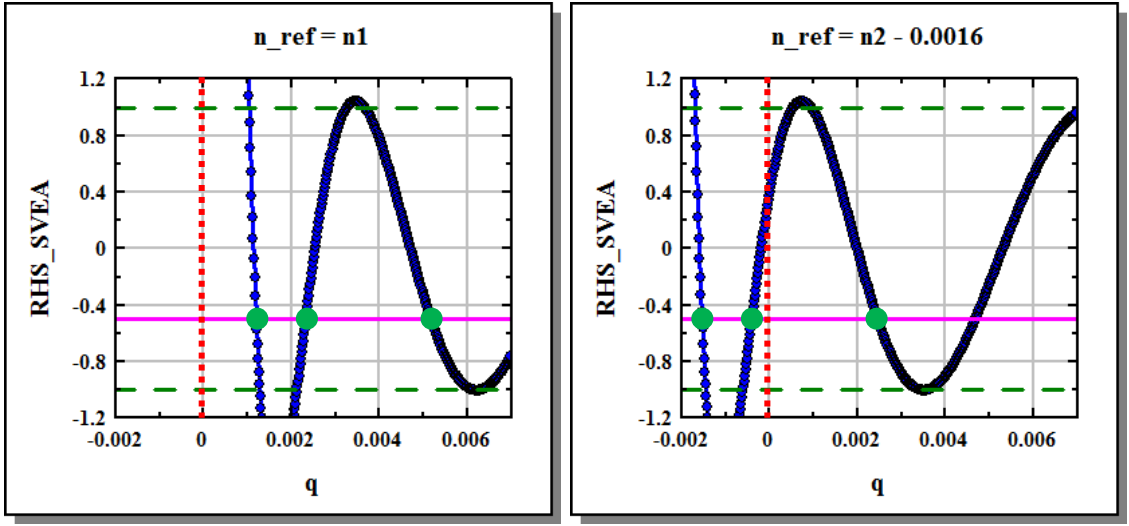

*Figure 15: Change of SVEA band position with value of the reference index. Left ( $n_{ref} = n_1$) – all bands are positive. Errors are smallest for band number one and are increasing with band number (cf. Figure 14). Right ( $n_{ref} = n_2 - 0.0016$) – band number one and part of band number two are negative. Minimum errors $[k_{z;n,j}(SVEA) - k_{z;n,j}(EXAC)]$ are now for band number two (cf. equation (54) and see Figure 16).*

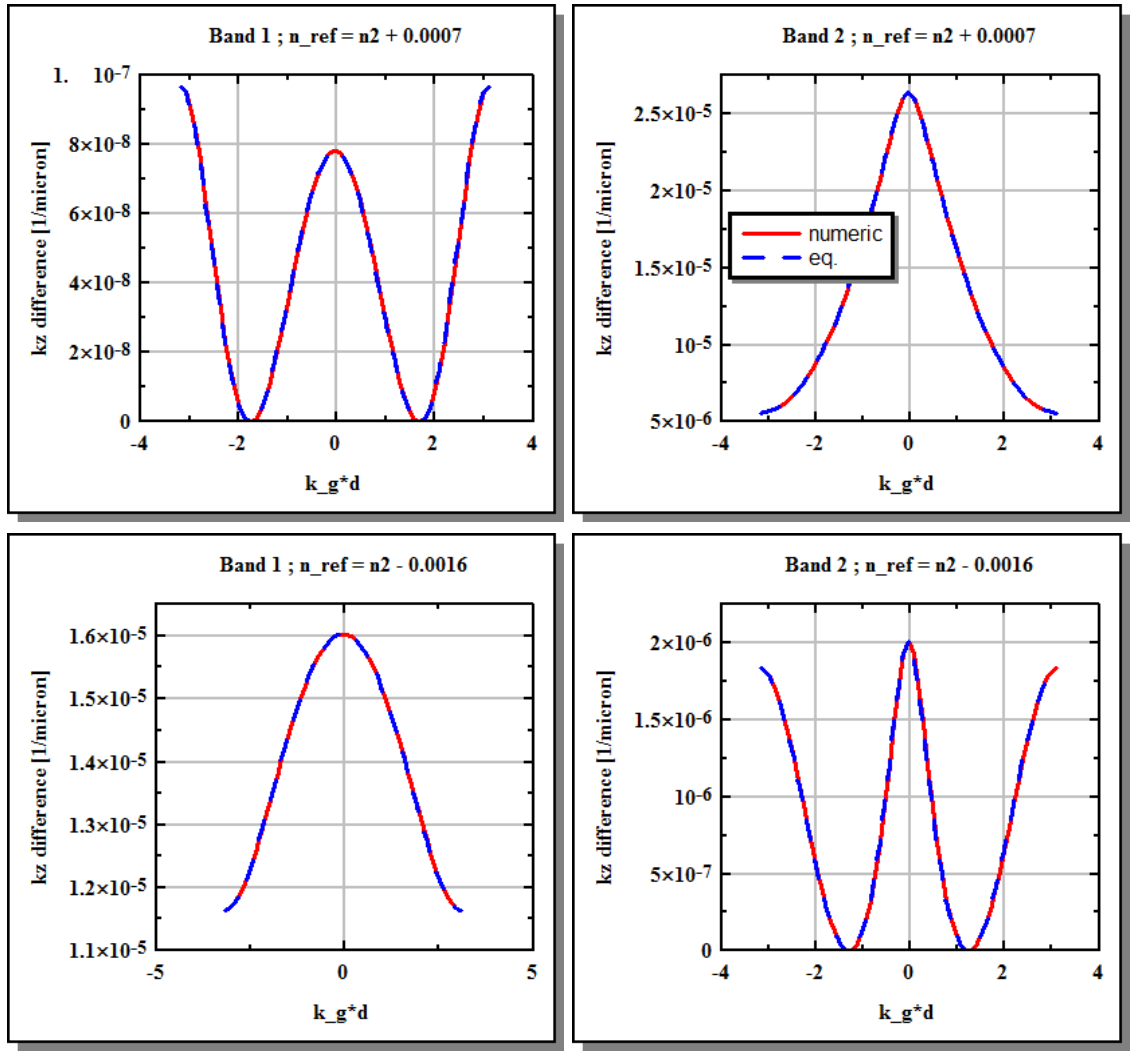

*Figure 16:* *Errors in propagation constant bands $[k_{z;n,j}(SVEA) - k_{z;n,j}(EXAC)]$ for different values of the reference index. Top ($n_{ref} = n_2 + 0.0007$) – minimum errors for band number one. Bottom ($n_{ref} = n_2 - 0.0016$) - minimum errors for band number two (see **Figure 15**). Continuous red line – numerically calculated (exact) differences. Blue dashed line – analytic approximation [equation **(54)**].*

Let's take now a closer look at the values of the reference index such that the errors for band number one are minimized. First zero crossing of the eigenvalues of band number one ($q_{1,j}$) (as the value of $n\_ref$ is lowered from $n_1$) is shown by the red curve on the left of *Figure 17*. Last zero crossing is shown by the blue curve. Inside the "zero-crossing zone" (between the two lines), two $q_{1,j}$ values are zeroed and therefore the errors for the respective propagation constants $[k_{z;1,j}(SVEA) - k_{z;1,j}(EXAC)]$ are zeroed two. The width of the calculated zero-crossing zone is shown by the curve on the right of *Figure 17*. Below the zero-crossing zone (grey area, left of *Figure 17*), the entire set of eigenvalues calculated for band number one ($q_{1,j}$) is negative. Further lowering of the reference index will minimize the error for band number two $[k_{z;2,j}(SVEA) - k_{z;2,j}(EXAC)]$, etc.

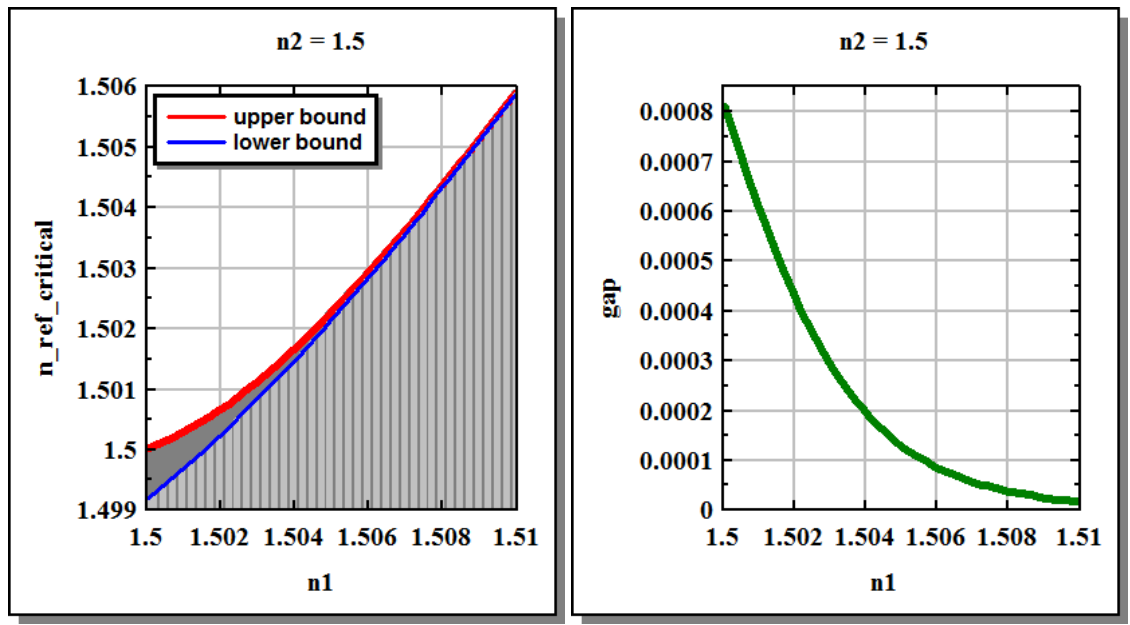

*Figure 17: Left – "zero-crossing zone". Inside the zero-crossing zone (between the red curve and the blue curve), two eigenvalues of band number one ($q_{1,j}$) are zeroed, thereby the errors of the respective propagation constants are zeroed too [top of Figure 16 and equation (54)]. Further below the zero-crossing zone (grey area) another eigenvalue band will cross zero, and thus the errors for the band's propagation constants are minimized [bottom of Figure 16 and equation (54)].*
*Right – gap of the zero-crossing zone (distance between the upper bound and the lower bound on the left).*
*Note: the lowest $n_1$ value for the shown curves is $n_1 = 1.5001$.*

Note again that while the SVEA Bloch functions are identical to the corresponding EXAC Bloch functions, the overall calculated electrical field distributions are NOT (see section *4.1.1*), since the propagation constant bands (SVEA vs. EXAC) are not identical.

Note also that the propagation constants in the SVEA case are always (for all values of the reference index and for all bands) greater than their EXAC "partners" (cf. equation *(54)* with $q_{n,j} < 1$).

In summary for the $k_z$ bands comparison – the error $[k_{z;n,j}(SVEA) - k_{z;n,j}(EXAC)]$, to first order, is proportional to $q_{n,j}^2$ [equation *(54)*]. $q_{n,j}$'s drift to lower values with lowering of $n_{ref}$ (from $n_1$) and even cross zero to negative values. By selecting the reference index, calculated errors for a specific propagation constants band can be minimized.

These comments conclude the EXAC vs. SVEA bands comparison section.

The discrete Coupled Mode Theory yields a half first band. In the following section, this CMT half first band is compared with the EXAC first band.

## 3.2. CMT vs. EXAC

Half bands calculated according to the CMT are shown by *Figure 18*. The shown bands are actually the eigenvalues of matrix $(B)$ [cf. equation *(41)*], calculated assuming nearest-neighbors coupling only (top) or adding next-nearest-neighbors coupling (bottom).

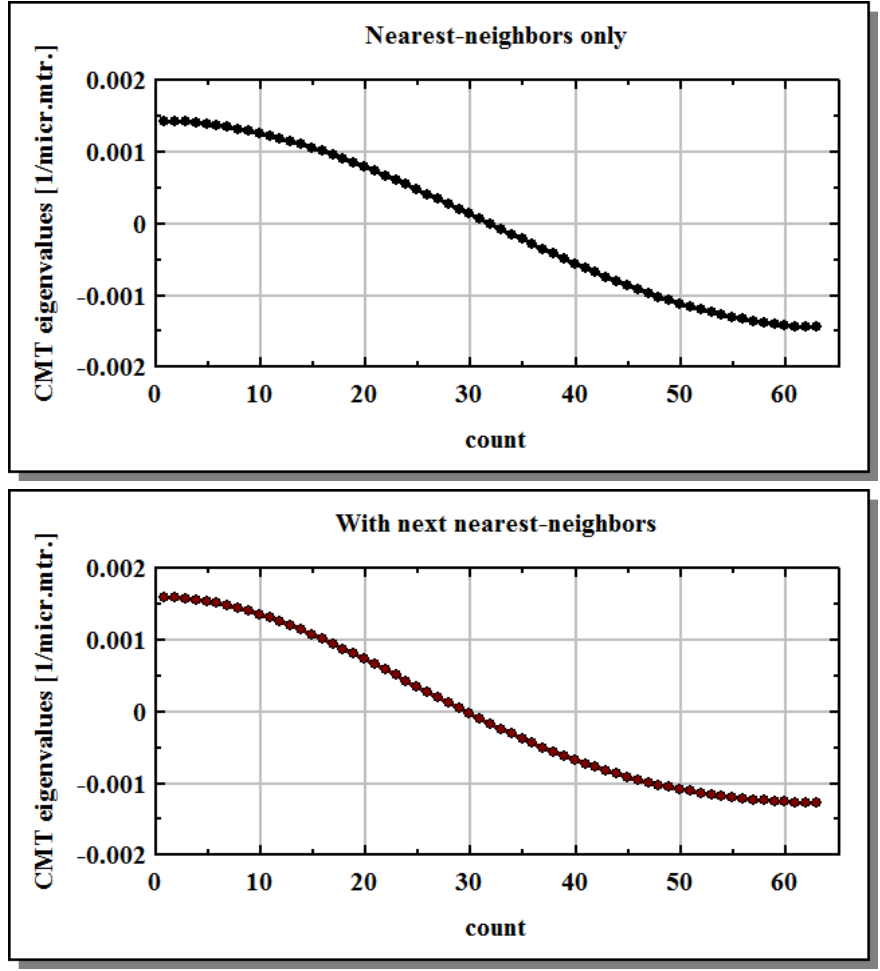

*Figure 18:* *CMT-calculated eigenvalues (cf. section 2.3), forming a half band. Top – nearest-neighbors coupling only. Bottom – adding next-nearest-neighbors coupling.*

The calculated CMT-eigenvalues shown in *Figure 18* form the "envelope propagation constants" [equation *(36)*]. To compare their values to the values of the EXAC $k_z$ propagation constants of the first band – left of *Figure 19*, the half CMT band was mirrored and its top was "locked" to the top of the first EXAC band.

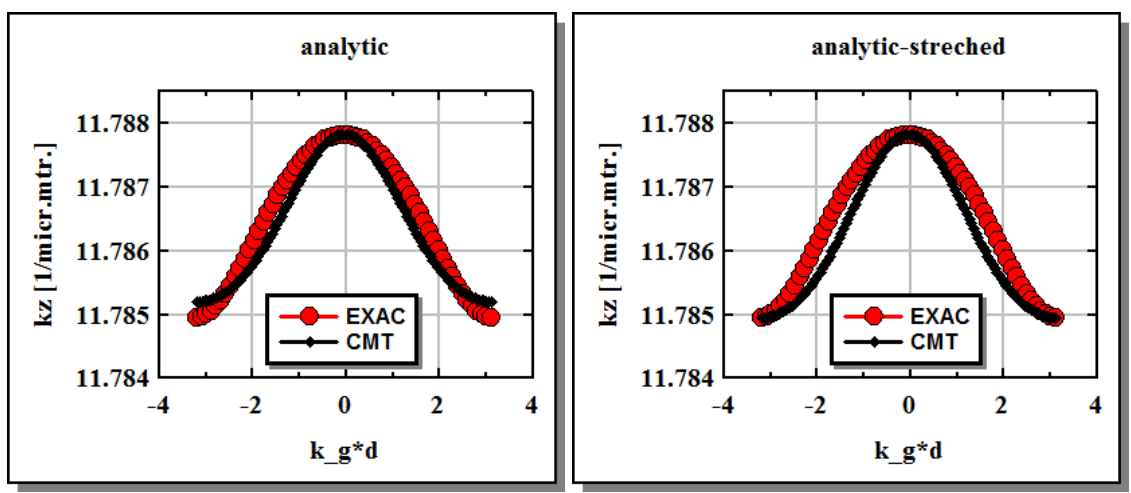

***Figure 19:*** *CMT band and EXAC first band. The top of the CMT band is "locked" to the top of the EXAC first band. Left – CMT band as calculated. Right – CMT band stretched (by about x1.1 for the case at hand).*

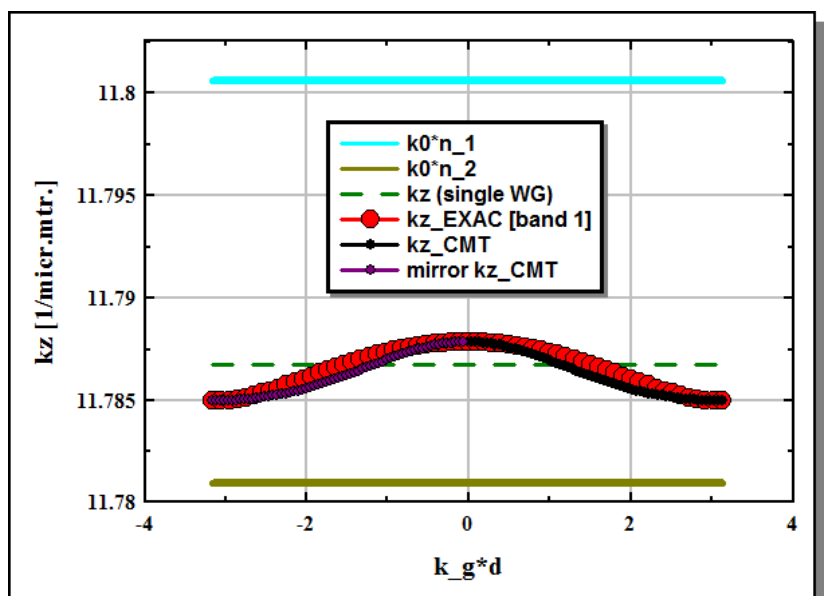

***Figure 20:*** *Zoom out, showing the bands in between a $k_0 \cdot n_1$ line and a $k_0 \cdot n_2$ line. The dashed green line represent the propagation constant ($k_z$) calculated for an isolated single WG (of width and "height" identical to the width and height of all WGs in the array).*

As shown (left of ***Figure 19***), the calculated CMT band [based on analytically calculated coupling coefficients (cf. for section *2.3*)] fits the EXAC band fairly well. Yet, taking a closer look, the CMT band needs a small "stretch" (by a factor of about 1.1 in the example at hand) to better fit the EXAC first band – right of *Figure 19*). The need for the small scaling of the calculated CMT band calls for a further study.

A zoom-out view, looking at the EXAC first band and the (stretched two-halves) CMT band along with $k_0 \cdot n_1$, $k_0 \cdot n_2$, and $k_z$- single WG lines, is shown by *Figure 20*.

Theoretically, the eigenvalues of a tridiagonal matrix (nearest neighbors coupling only) with constant coefficients, strictly follow a *cosine* curve [20]. However, the EXAC first band does NOT strictly follow a *cosine* curve (independent of the number of periods). Therefore, even with the small correction of next-nearest neighbors coupling, a small gap is seen between the CMT band and the EXAC first band, with or without stretching the CMT band (right and left of *Figure 19*).

To summarize the CMT vs. EXAC $k_z$ band comparison – while overall similarity is established (based on analytically calculated coupling coefficients), the calculated CMT band needs some stretching (to better fit the EXAC first band) and even after stretching, differences are observed, particularly at the center of each half band.

Now, having compared the propagation bands calculated by each of the three analytic methods, we want to compare electrical field distributions calculated by all five methods. In fact, in the next section we compare the four approximately-calculated distributions to the EXAC calculated distribution (that we consider to be exact and hence treat it as a reference distribution) and compute a "distance" by which the four methods are graded.

## 4. Comparison of distributions

Comparison of electrical field distributions is divided into two sub-sections. The first sub-section (section *4.1*) treats field distributions calculated by all five methods showing distribution maps and showing field cross-sections for essentially small angles of excitation. In this first sub-section the focus is on comparing the distributions and on ranking the distribution-calculating methods.
The second sub-section (section 4.1) is about band excitation with growing tilt angles of the input field.

### 4.1. Five methods – small angles of excitation

Five different sets of distributions are discussed in this section, each due to a specific set of parameters ("case") as summarized in *Table 1* (see equation *(23)* for the first three columns).

| Case # | Case description | $\sigma$ [$\mu m$] | Tilt | Shift | $n_{ref}$ | $z_{end}$ |
|---|---|---|---|---|---|---|
| I | Discrete diffraction | 2 | No | No | $n_1$ | $\lesssim 8.2mm$ |
| II | On-axis wide Gaussian | 10 | No | No | $n_1$ | $\lesssim 8.2mm$ |
| III | Off-axis wide Gaussian | 10 | Yes | No | $n_1$ | $\lesssim 8.2mm$ |
| IV | Shifted narrow Gaussian | 2 | No | $-d/2$ | $n_1$ | $5mm$ |
| V | $n_{ref}$ below the zero-crossing zone | 2 | No | No | $n_2$ | $\lesssim 8.2mm$ |

*Table 1: Parameters for the five electrical field distributions discussed in section 4.1 [see also the fixed parameter list (22)].*

### 4.1.1. Case I - discrete diffraction

The first set of maps is related to the frequently presented case of "discrete (or ballistic) diffraction" **[21]**,**[22]**. Namely – essentially a single center WG is excited and the light "beam" spreads linearly with distance - *Figure 21*.

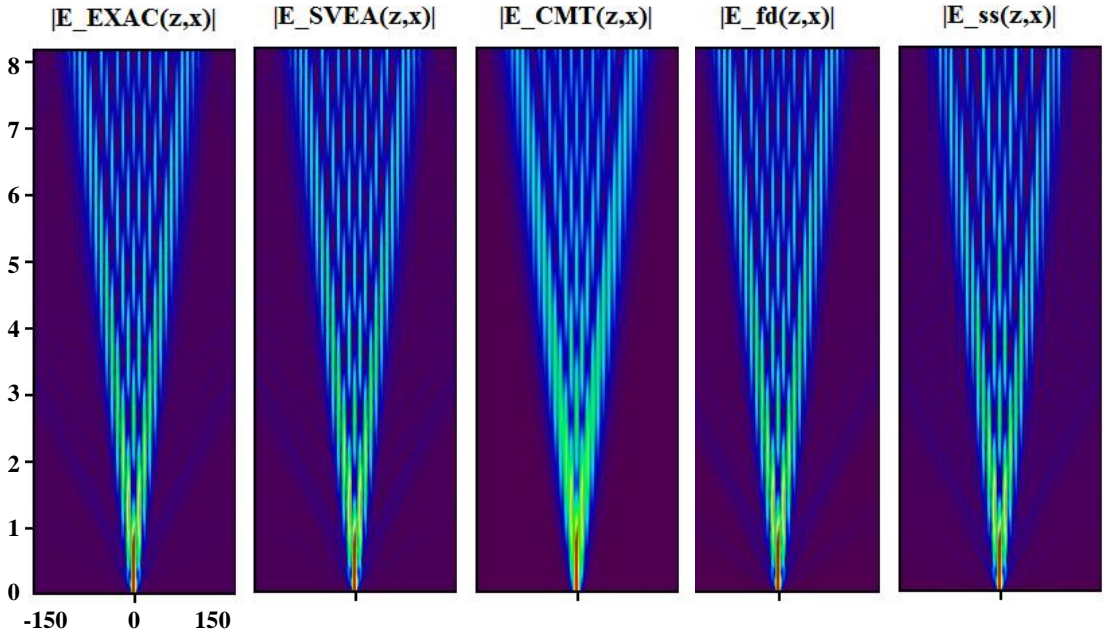

*Figure 21: Discrete diffraction maps as calculated by five different methods. For accuracy evaluation, each of the four maps on the right is compared to the "reference map" on the left (Figure 23 and Figure 24). Color scale is amplified for better visibility.*

Cross-sections showing the calculated electrical field intensities at $z \gtrsim 8mm$ are shown by *Figure 22*. Top – full view. Bottom (as the cross-section functions are all symmetric) – half view for higher resolution.

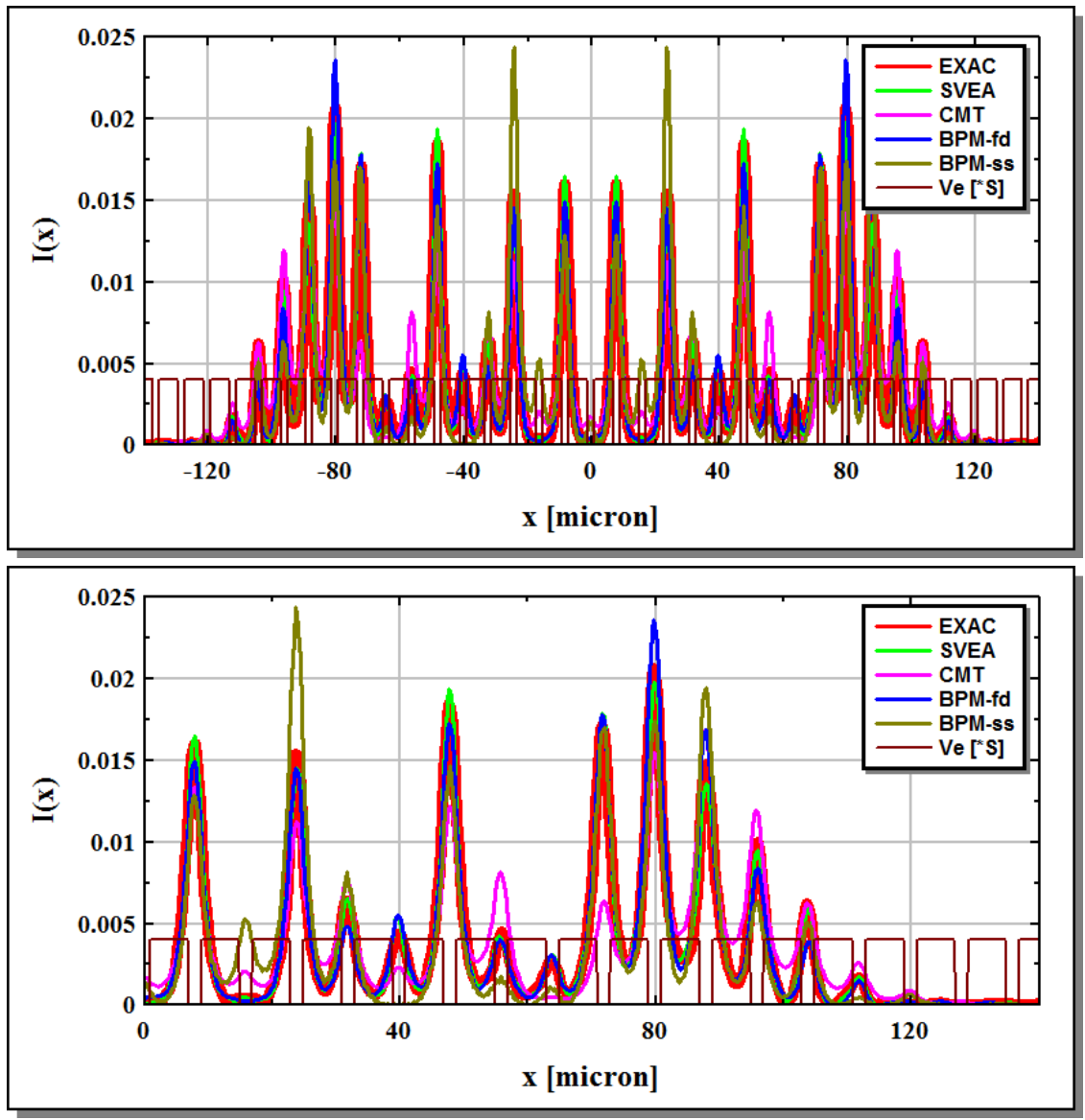

*Figure 22:* *Intensity cross-sections at $\gtrsim 8mm$ . Top – full view. Bottom (as the cross-section functions are all symmetric) – half view for better visibility. Comparisons in pairs are shown by* ***Figure 23*** *and calculated "distances" are shown by the bar charts of* ***Figure 24***.

Comparisons in pairs are shown by *Figure 23*. The figure shows four pairs of curves. Top of each pair – the EXAC field intensity and a second field intensity as marked. Bottom of each pair – normalized difference (between the EXAC intensity and the other intensity). The definition we adopted for "difference" between two functions is stated by equation *(56)*.

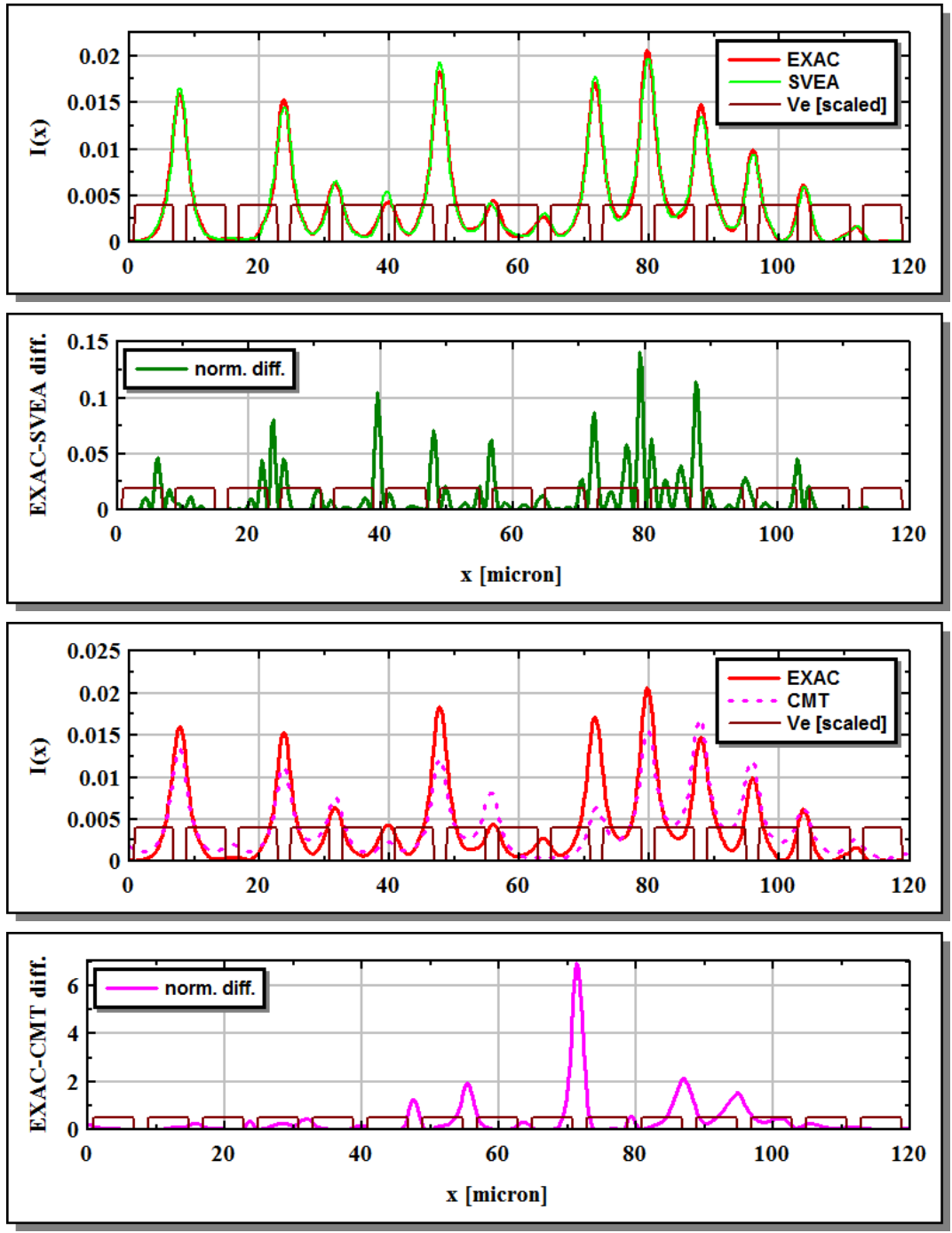

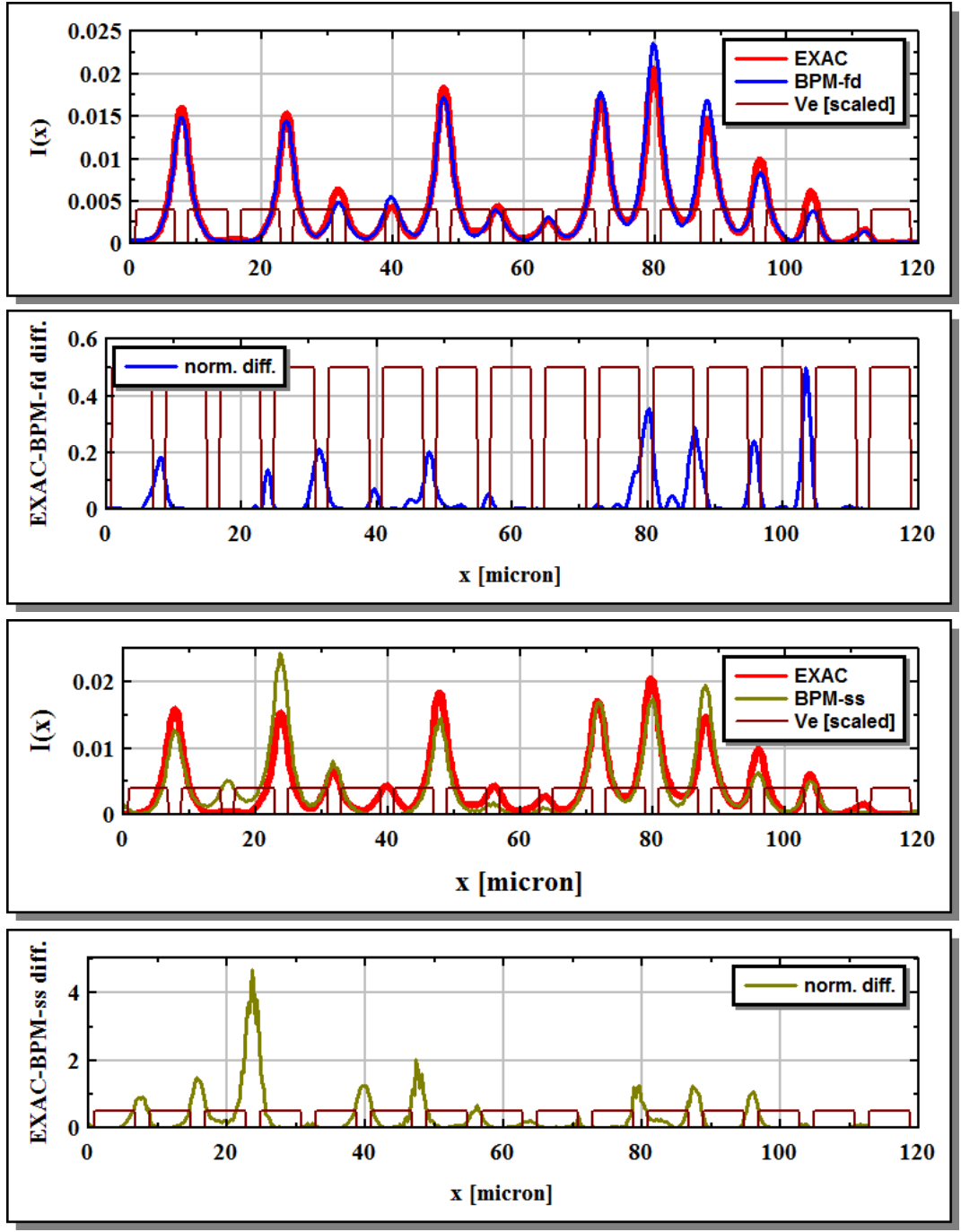

*Figure 23:* Intensity differences for discrete diffraction at $z \gtrsim 8mm$. Note that the differences are "normalized" by the standard deviations [cf. equation (56)]. Smallest differences are found for SVEA. Next are the two numeric methods with BPM-fd coming before (more accurate than) BPM-ss. Largest differences are calculated for the CMT intensity cross-section. Reference index for the BPM methods is $n_{ref} = n_1$.

For the purpose of quantitative comparison between the EXAC distribution and each of the four distribution calculated by the four approximate methods, we adopted the following $difference[p(x), q(x)]$ and $distance(p, q)$ definitions **[23]**:

$$difference[p(x), q(x)] \equiv \left[\frac{p(x)}{stdev[p(x)]} - \frac{q(x)}{stdev[q(x)]}\right]^2$$

$$distance(p, q) \equiv \left[\sum_{i=1}^{N} difference[p(x_i), q(x_i)]\right]^{1/2}$$

(56)

Amplitude and intensity distances according to definition *(56)* (with $n_{ref} = n_1$ for the BMP methods) are shown by the bar charts of *Figure 24*.

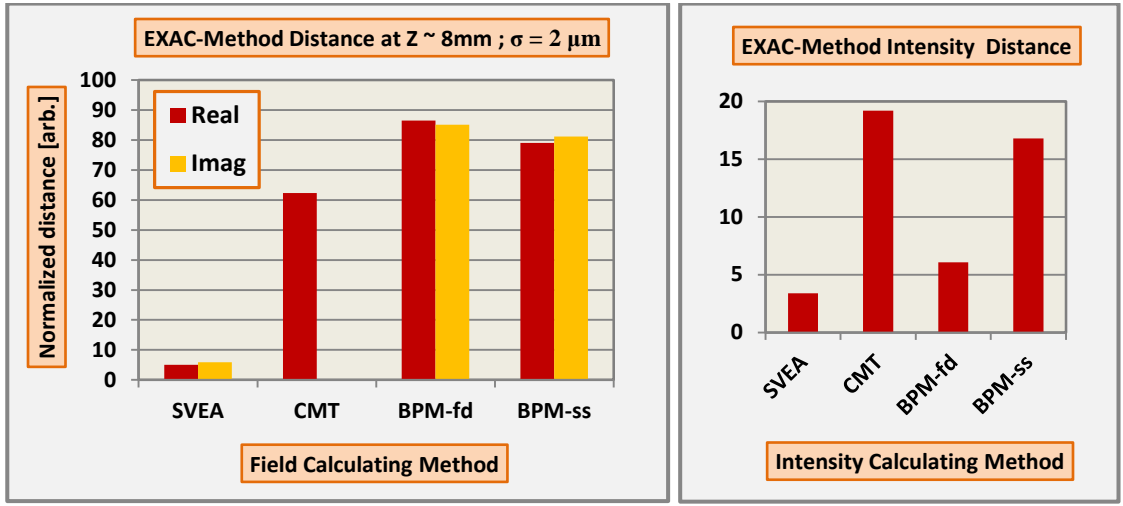

*Figure 24: EXAC vs. other method distances. The distances on the left are for the real and imaginary cross-sections (the associated functions themselves are not shown by dedicated figures). Intensity distances are shown on the right (and the functional differences are shown by the four pairs of Figure 23). Not surprising, the distances of the real and imaginary cross-sections are much larger than the distances of the corresponding intensity cross-sections.*
*Method ranking according to the calculated intensity differences for the discrete diffraction case (best to worst): SVEA, BPM-fd, BPM-ss, CMT.*

The next four cases are not quantitatively analyzed and are depicted here for visual inspection only. However, it becomes clear by mere visual inspection that the accuracy (or inaccuracy) order found by analyzing the discrete diffraction distributions holds.

### 4.1.2. Case II – on-axis wide Gaussian

The second set of distributions we look at, results from an on-axis "wide" Gaussian ($\sigma = 10\mu m$) as the input field.

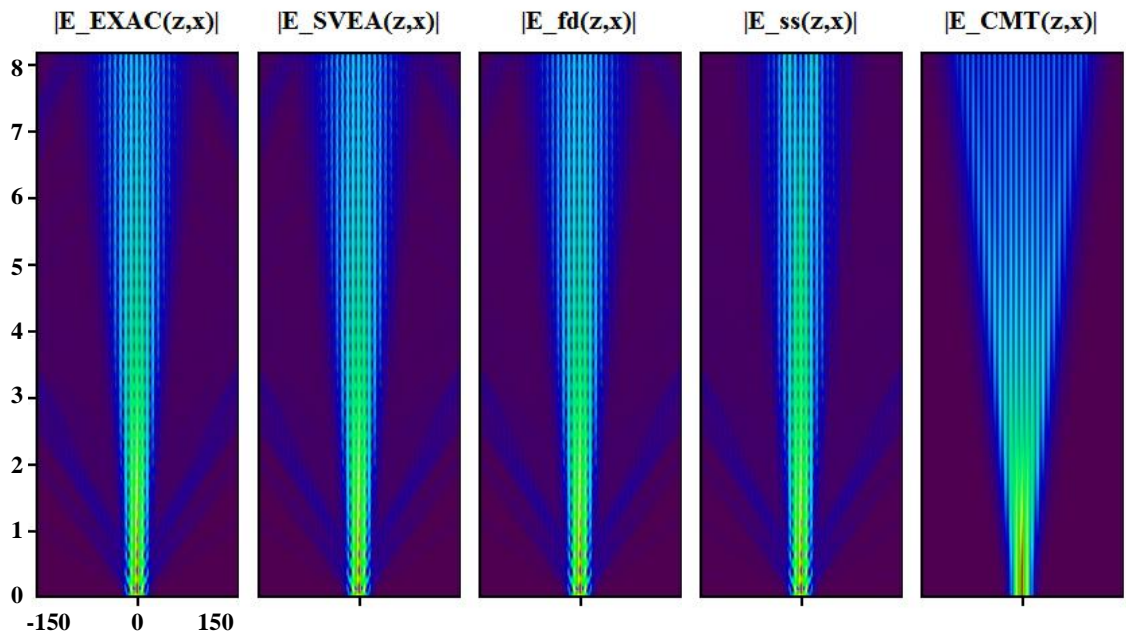

*Figure 25: On-axis wide Gaussian excitation. Weakly excited higher-order bands are seen in all four maps to the left. Color scale is amplified.*

The four maps on the left of *Figure 25* show that higher bands are weakly excited (cf. section *4.2* for more on band excitation), and appear rather similar. Inspection of the intensity cross-sections of *Figure 26* reveals the differences. Color scales for these maps and for all other maps in this section are amplified.

The noticeably different CMT map, on the right of *Figure 25*, results from a first band excitation only. According to the CMT and the excitation scheme detailed in the related theoretical section (section *2.3*), only first band is excited under all excitation conditions (as seen also by all maps that follow).

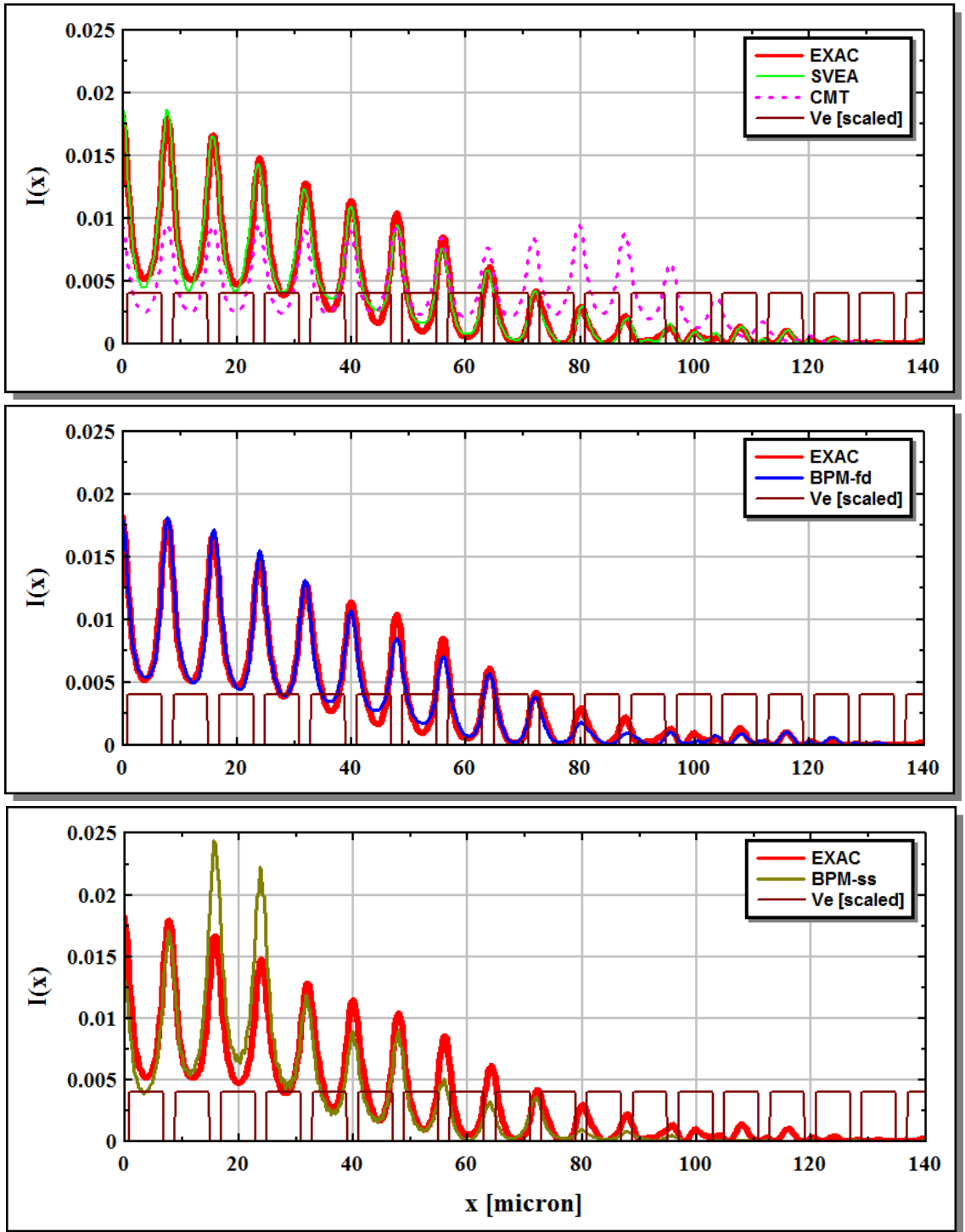

***Figure 26:*** *Intensity cross-sections at z ≳ 8mm for an on-axis wide Gaussian excitation. SVEA and BPM-fd are fairly on target. BPM-ss cross-section is less accurate and more so for the CMT cross-section.*

### 4.1.3. Case III – off-axis wide Gaussian

Distributions and "last *z*" cross-sections for an off-axis wide Gaussian excitation are shown by *Figure 27* and by *Figure 28*.

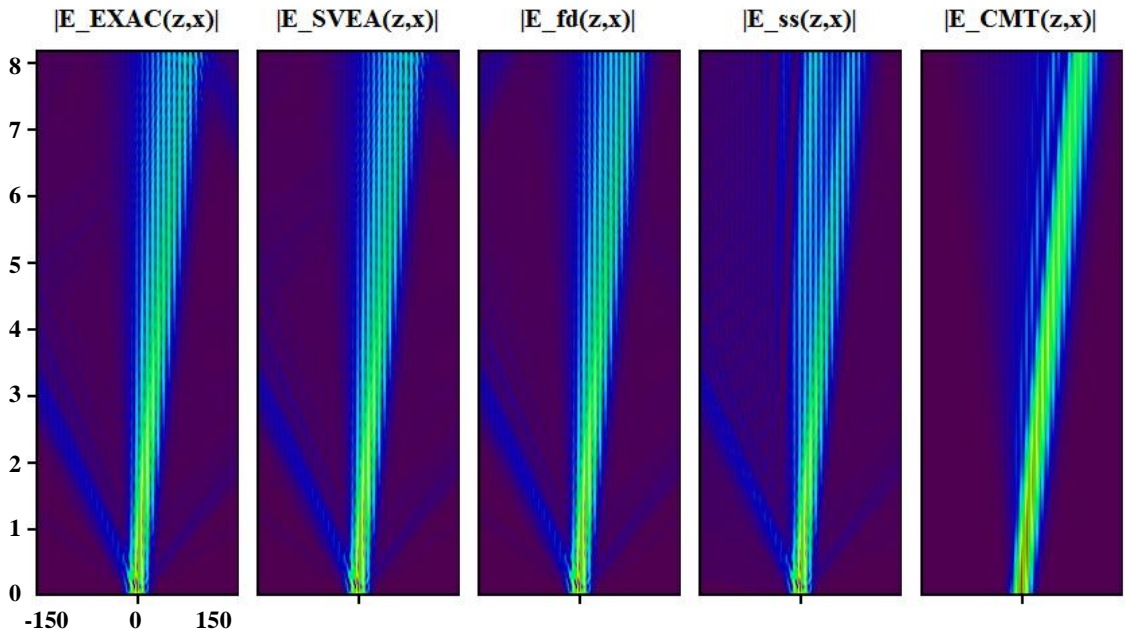

*Figure 27:* *Off-axis wide Gaussian excitation. Map accuracy degrades going from left to right. Color scale is amplified.*

Second and third band excitation are clearly seen in all four maps at the left of *Figure 27*. The weak band at the top left of the BPM-fd map (central map) is a result of "reflection" from the finite array and should be ignored.

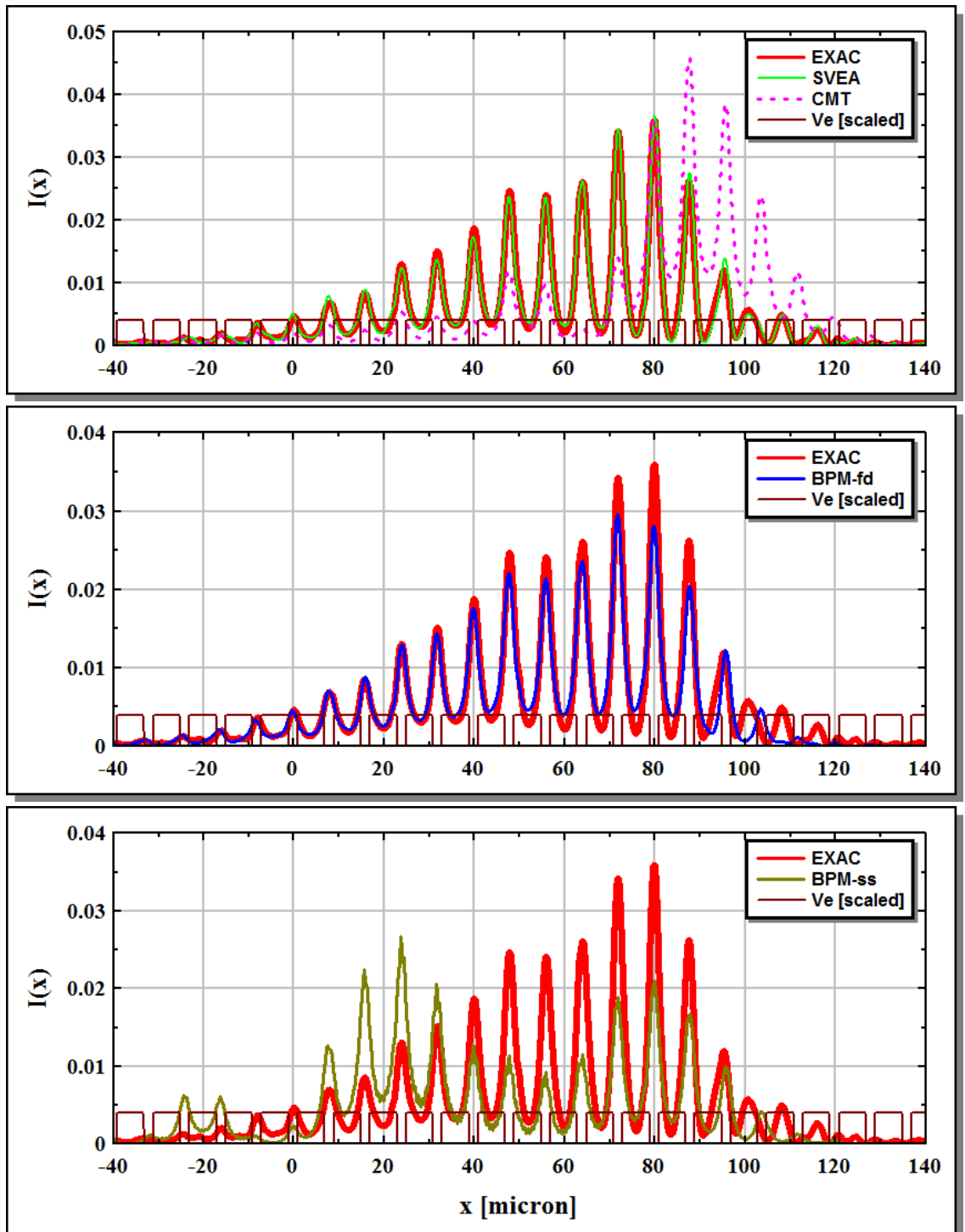

***Figure 28:*** *Intensity cross-sections at z ≳ 8mm for an off-axis wide Gaussian excitation. Map accuracy (cf. **Figure 27**) degrades going from left to right.*

### 4.1.4. Case IV - shifted narrow Gaussian

The maps of *Figure 29*, case number IV, were calculated for a shifted narrow Gaussian (in between two neighboring waveguides). The shifted Gaussian, as already discussed in section *2.1.2*, strongly excites the second band along with the first band and even the third band (cf. *Figure 8*). The resulted distributions (*Figure 29*) are very different from the discrete diffraction distributions (*Figure 21*) that are launched by the same narrow Gaussian only half period to the right.

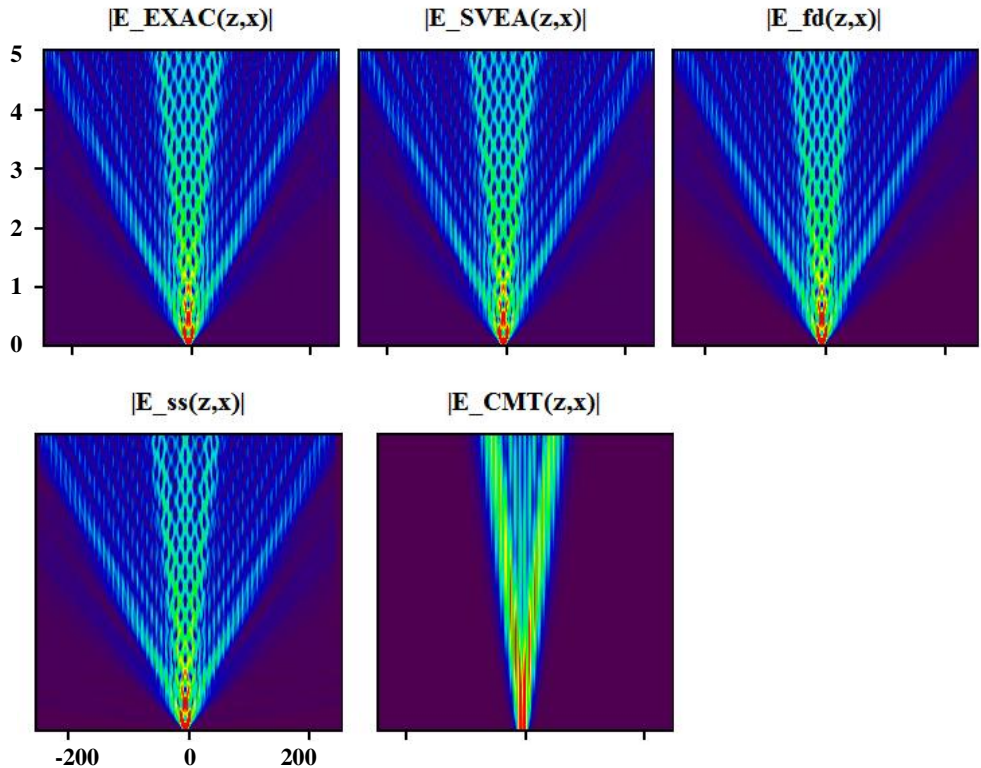

*Figure 29:* Intensity distributions for shifted narrow Gaussian excitation (in between two neighboring waveguides). For this case of strongly excited higher bands, the CMT-predicted distribution is completely off target. Color scale for these maps is amplified.

*Figure 29* clearly indicates that CMT completely fails in predicting electrical field distributions every time bands higher than the first one are significantly excited (see also the angled excitation discussed in section *4.2*).

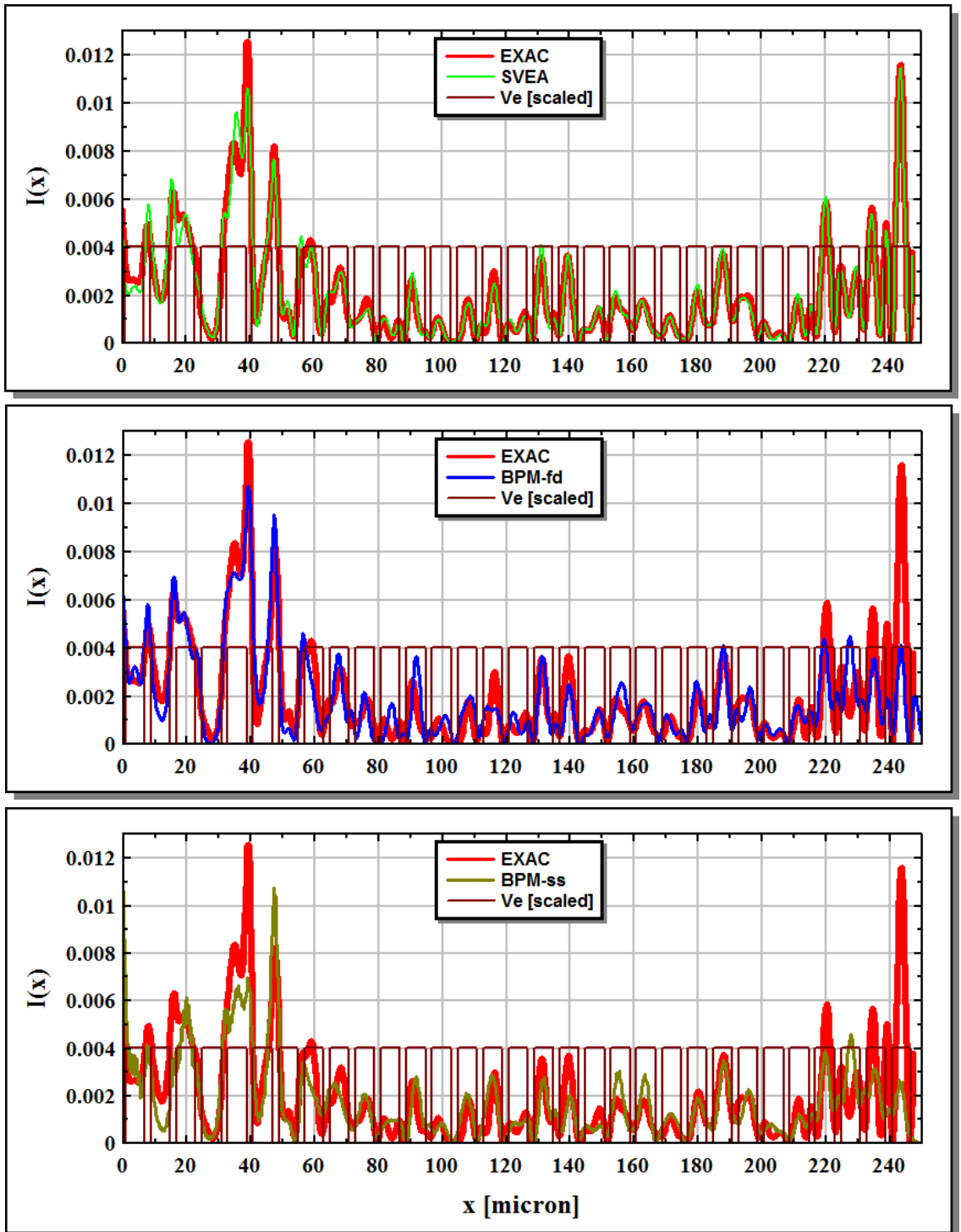

***Figure 30:*** *Intensity cross-sections at z = 5mm for shifted narrow Gaussian excitation.*

### 4.1.5. Case V - $n_{ref}$ below the zero crossing zone

The last case in this sub-section, case V, is the case of reference index below the zero crossing zone (cf. *Figure 17*). Excitation is again the discrete diffraction excitation (as in case I. See also *Figure 21*).

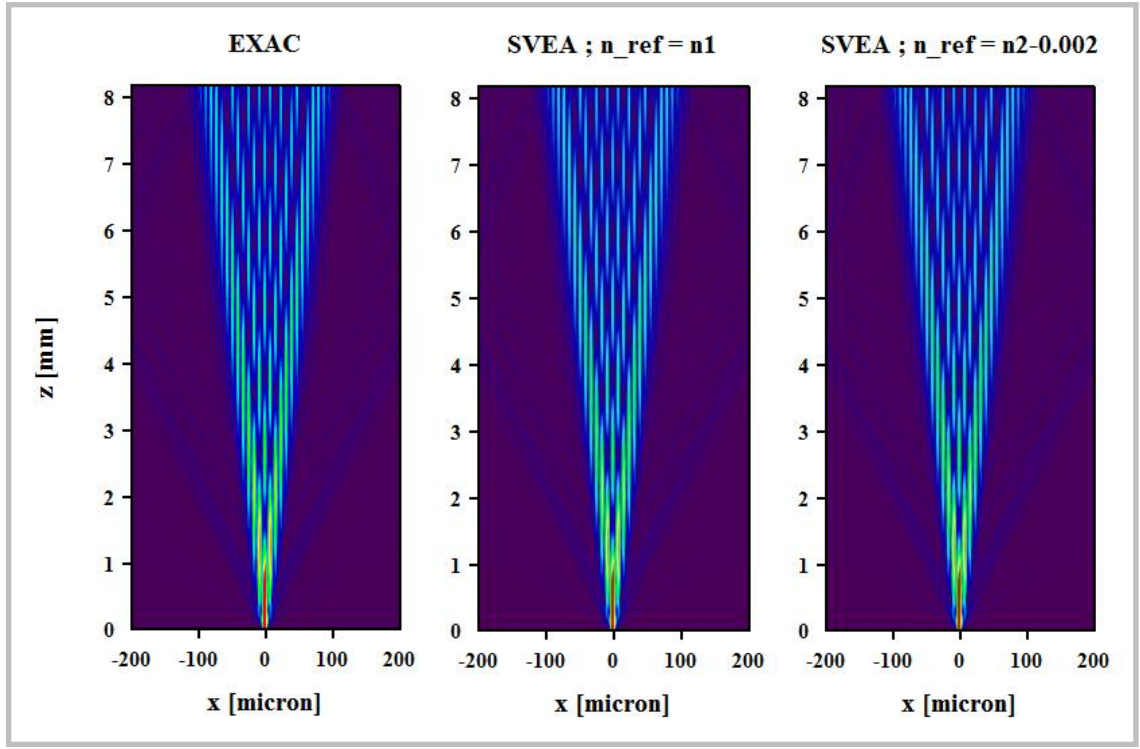

*Figure 31:* Field intensity distribution for discrete diffraction with reference index above and below the zero crossing zone (cf. *Figure 17*). The two SVEA maps are similar to each other and are similar to the EXAC map. To see the differences, call for *Figure 32*.

As discussed in section *3.1* above, the choice of the reference index affects the bands of propagation constants and thus the distributions of the electrical fields.

For the shown discrete diffraction distributions, choice of $n_{ref} = n_2 - 0.002$ results in a calculated SVEA field intensity distribution slightly more accurate than the SVEA field intensity distribution calculated with $n_{ref} = n_1$. At a propagation length of ~8mm, the distances [equation *(56)*] between the EXAC cross-sectional intensity and the SVEA cross-sectional intensities are 2.55 and 3.38 respectively (*Figure 32*).

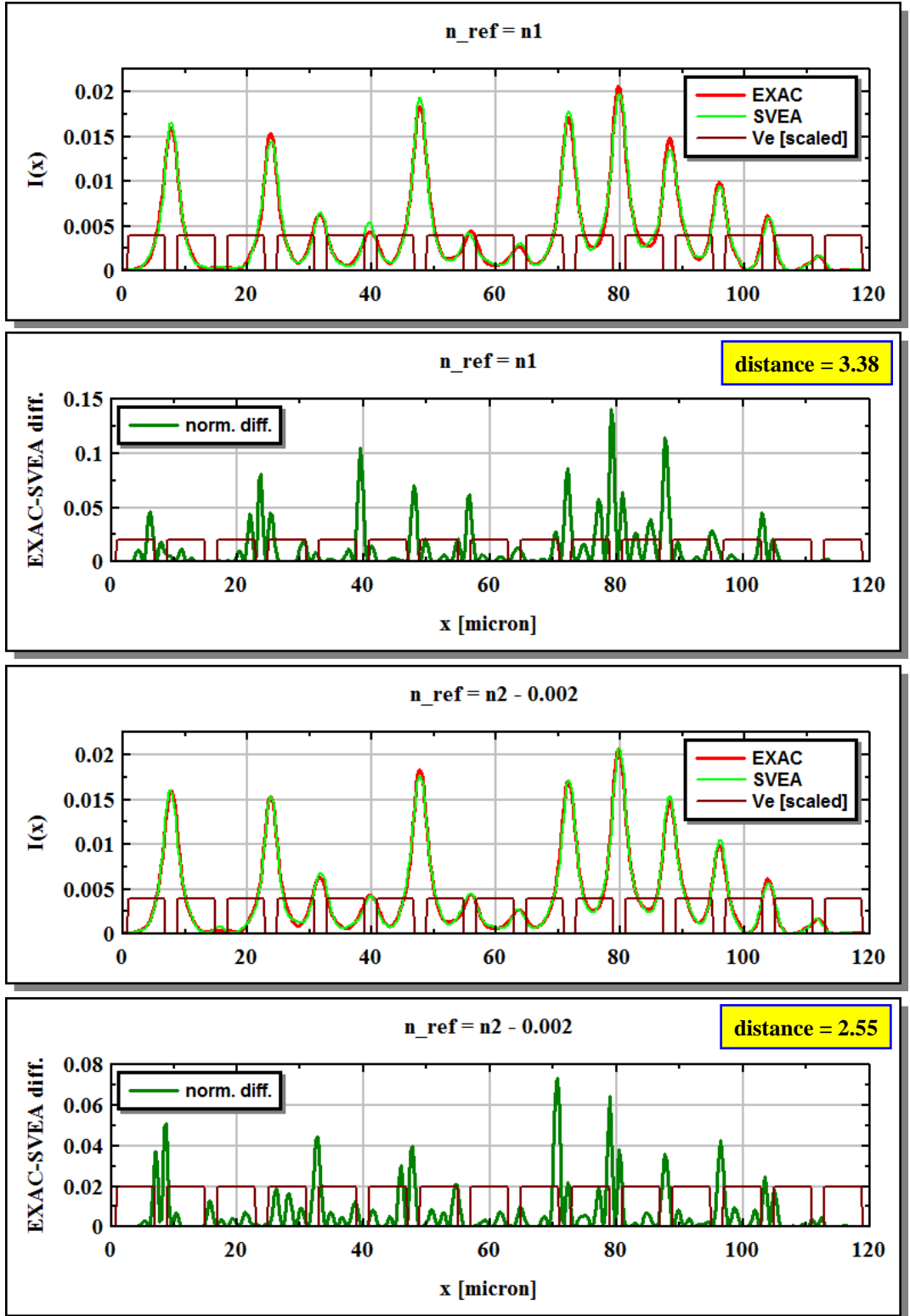

*Figure 32:* *Intensity differences. SVEA maps differ only slightly from the EXAC map. The EXAC-SVEA intensity difference depends on the choice of the reference index. For the shown discrete diffraction case (**Figure 31**), the map with $n_{ref} = n_2 - 0.002$ is more accurate than the map with $n_{ref} = n_1$ [cf. equation (56)].*

To summarize the first part of distributions comparison – we can now rank the four methods (of calculating electrical field distribution along WG arrays) in order of higher-to-lower accuracy:

  A. SVEA
  B. BPM-fd
  C. BPM-ss
  D. CMT

Note that the above ranking holds for the "discrete diffraction" case. For the angled input case of *Appendix 5* and *Appendix 6*, BPM-fd and BMP-ss switch places (shorter distances between EXAC and BPM-ss cross-sections).

The CMT approximate distributions hold only for input fields such that only the first band is significantly excited.

Now to the last section of distributions comparison, with emphasis on band excitation.

## 4.2. Three methods – large angles of excitation

In this last distributions-comparison section we look at seven cases with emphasis on band excitation. For all seven cases we keep the same width of the (wide) exciting Gaussian ($\sigma = 30\mu m$), and for all cases we look at central excitation only [$x_{shift} = 0$, cf. equation *(23)*].

| Case # | $F_{bz}$ | θ [⁰] | Dominating band |
|---|---|---|---|
| VI | 0.0 | 0.00 | 1 |
| VII | 0.5 | 1.43 | 1 |
| VIII | 1.0 | 2.87 | 2,1 |
| IX | 1.5 | 4.30 | 2 |
| X | 2.0 | 5.74 | 3,2 |
| XI | 2.5 | 7.18 | 3 |
| XII | 5.5 | 16.0 | 6 |

*Table 2: Angle of excitation and dominating band (strongest band excited) for the seven cases discussed below.*

The parameter changed from case to case in this section is angular tilt of the input field (*Table 2*). For a systematic increase in tilt angle we define a "fraction of the Brillouin Zone" ($F_{bz}$) to which the tilt angle and the tilt vector [equation *(23)*] are related through equation *(57)*:

$$\theta = \text{asin}\left[F_{bz} \cdot \frac{\lambda_0}{2 \cdot d}\right] \; ; \; k_{tilt} = k_0 \cdot \sin(\theta)$$

(57)

With increasing angle of excitation, higher bands are excited. The strongest excited band (the "dominating band") for each tilt angle is also listed in *Table 2*.

Throughout this section we left aside the analytic SVEA method and the numeric BPM-ss method and look only at electrical field distributions calculated by the remaining three methods – EXAC, BPM-fd, and CMT (reference index for the BPM-fd, for all cases, is taken as $n_{ref} = n_1$).

### 4.2.1.  $F_{bz} = 0.0 \; \{\Theta = 0^0\}$

Exciting field for case VI [$F_{bz} = 0.0$] is shown by *Figure 33* along with the calculated bands of expansion coefficients [$C_{e;n,j}$, cf. equation *(20)*].

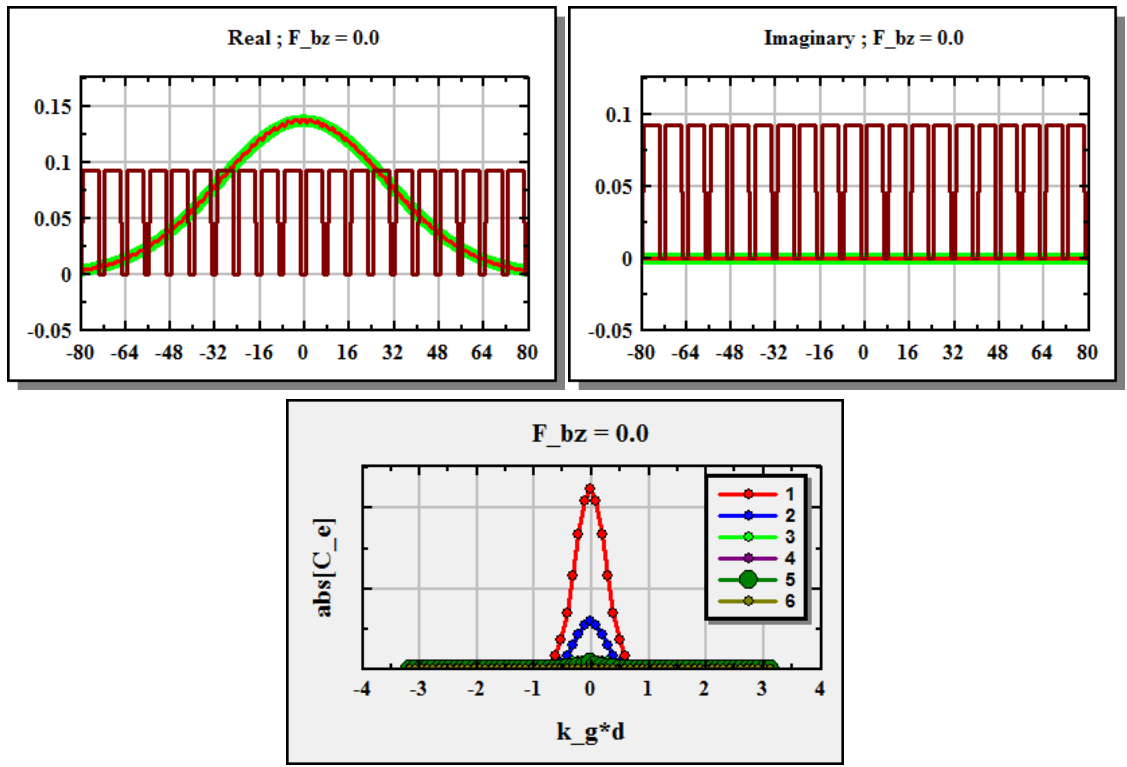

*Figure 33:* *Top - exciting field [green - external field, red - $E_e(x, 0)$]. Bottom – expansion coefficients. Lower chart curves indicate strongest excitation of band one but "visible" contribution from band two also.*

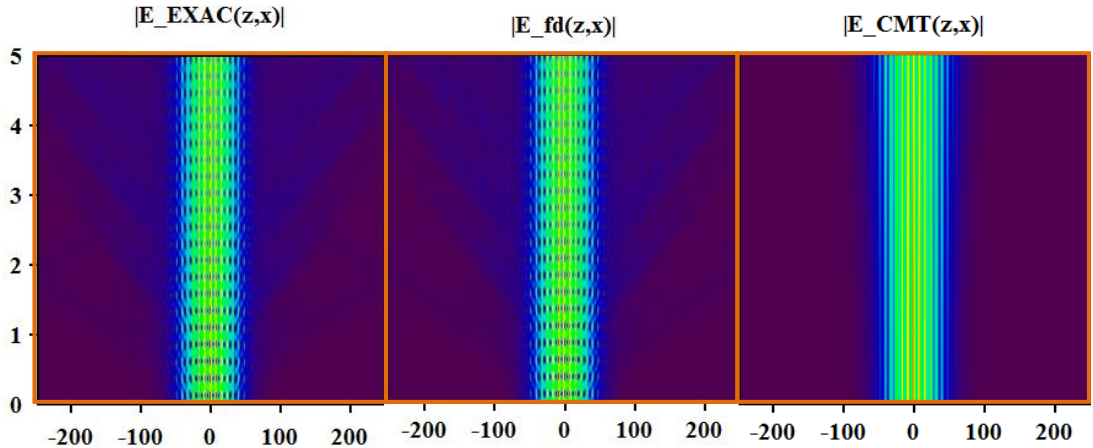

*Figure 34:* Distribution maps for case VI ($F_{bz} = 0.0$). The two maps on the left show "beating" of the two excited bands (band one and band two). CMT map – no beating since (always) only the first band is excited.

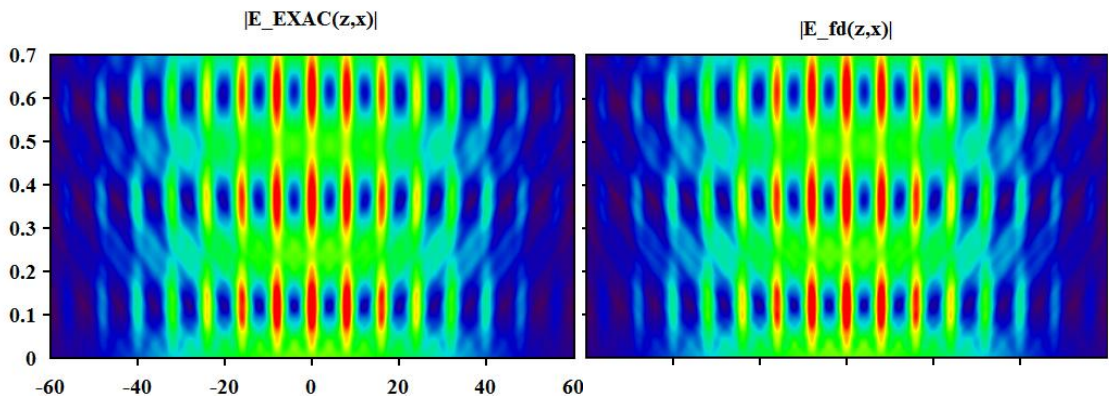

*Figure 35:* Zoom to the first 700 micrometers of propagation to see the beating in higher resolution and to realize the similarity of the two maps. Horizontal axis - cross-sectional distance in microns.

*Figure 34* – maps for case VI. *Figure 35* – zoom to see beating (between the two excited bands) and to examine the similarity of the analytic EXAC map (left of *Figure 34*) and the numeric BPM-fd map (center of *Figure 34*). *Figure 36* – maps by individual bands. As the center of each band is (symmetrically) excited, the individual band distributions are symmetric and equally spread to the left and to the right (cf. *Figure 46*).

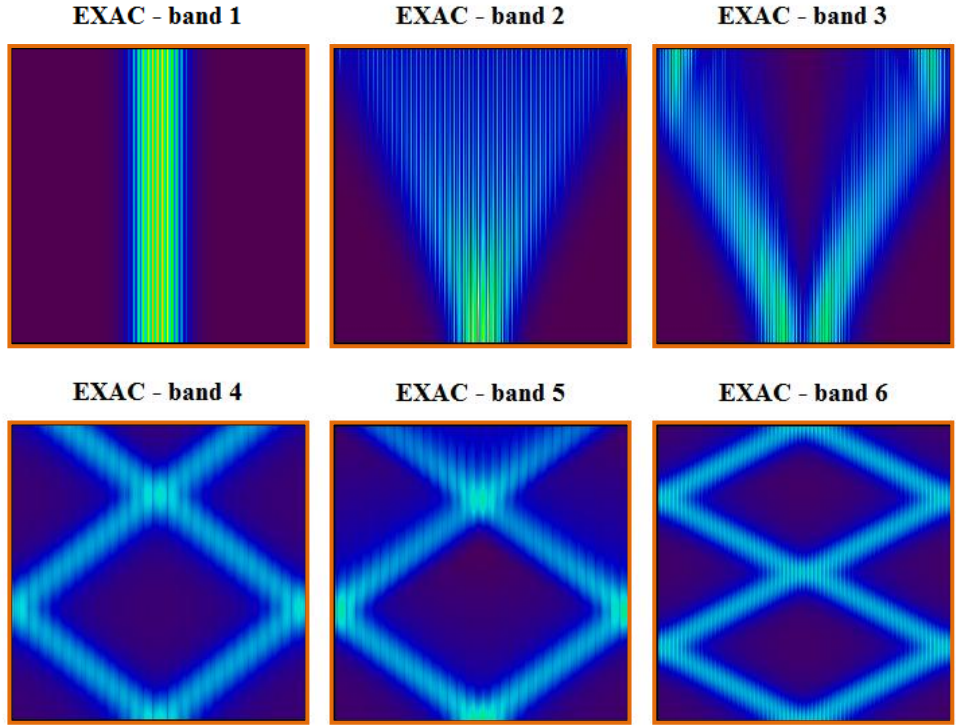

*Figure 36: Maps by individual bands for case VI. The relative strength of the bands is NOT preserved (each band is individually renormalized).*

### 4.2.2. $F_{bz} = 0.5$ $\{\Theta = 1.4^0\}$

Exciting field for case VII ($F_{bz} = 0.5$) is shown by *Figure 37* along with the calculated bands of expansion coefficients. Again band one dominates with visible contribution from band two. This time however, the bands are excited in the middle of one-half of the Brillouin Zone (*Figure 37*) and propagate at the corresponding angles (cf. *Figure 46*) completely breaking the symmetry of the maps.

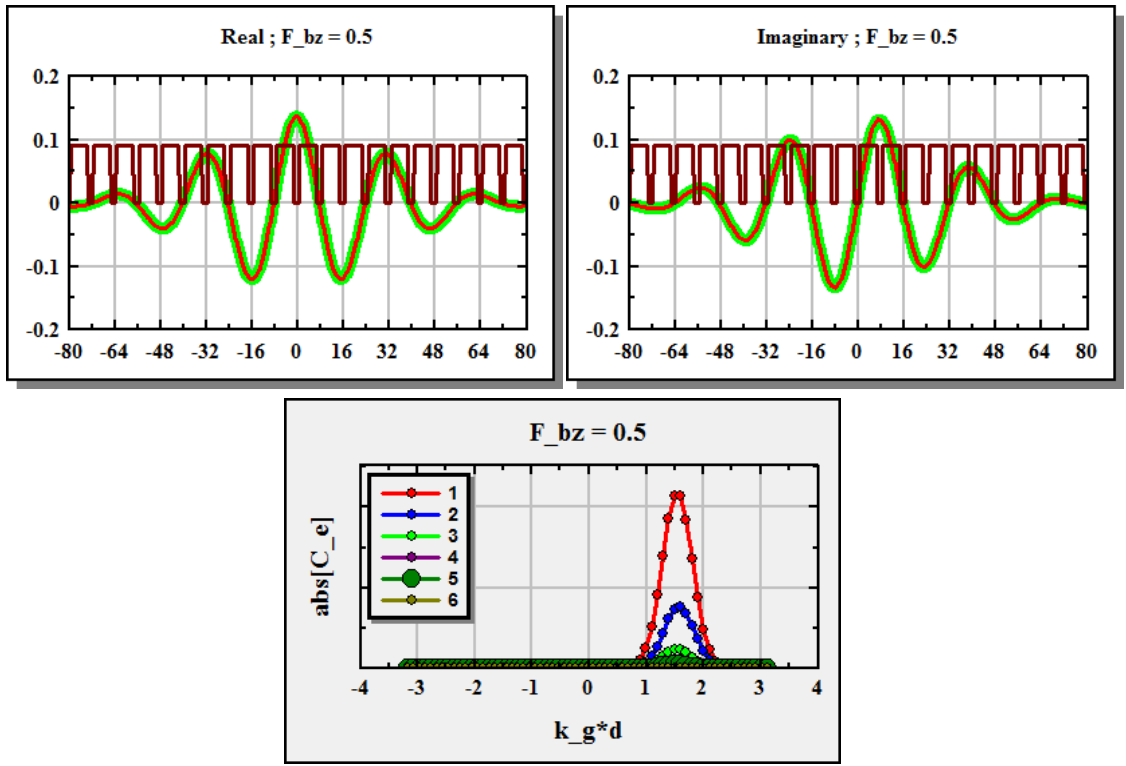

*Figure 37:* Input field reconstruction and band expansion coefficients for case VII. Bands are excited in the middle of the positive half of the Brillouin Zone, mandating angled propagation (cf. **Figure 46**).

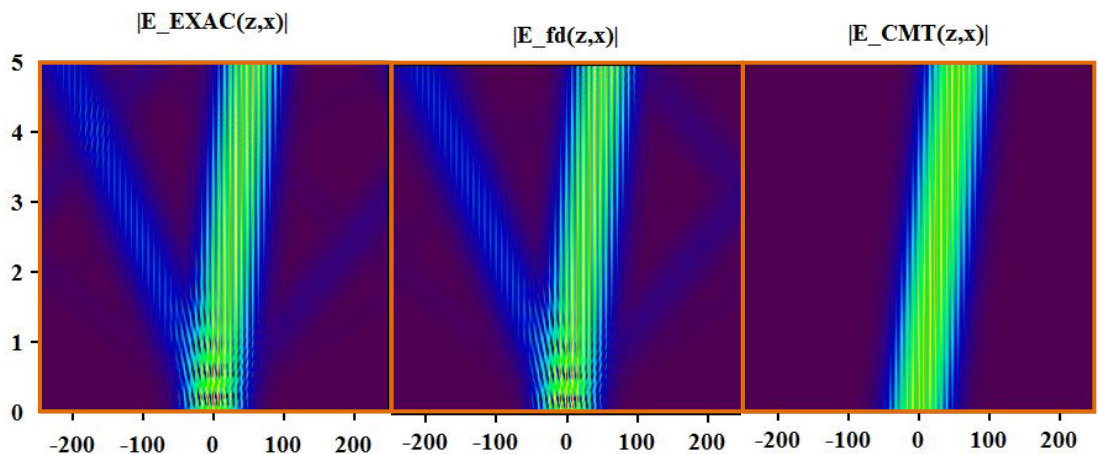

*Figure 38:* Maps for case VII ($F_{bz} = 0.5$). Contribution from band two clearly seen (left and center) and even contribution from band three is visible (weak beam propagating to the right). Map colors are amplified. For the CMT map (right) band one only is contributing.

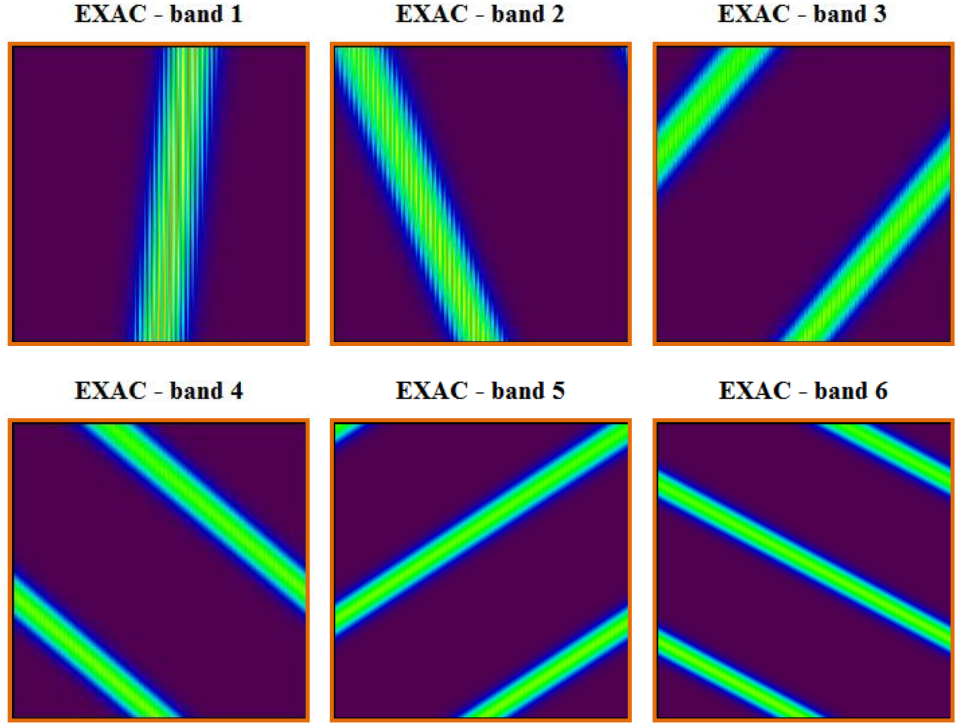

*Figure 39:* *EXAC maps by individual bands for case VII. The four high-order bands show folding of the "beams" (reappearance on the opposite side), as a result of the assumed finite array and the imposed cyclic boundary conditions (equation (62) of Appendix 1).*

### 4.2.3. $F_{bz} = 1.0 \;\; \{\Theta = 2.9^0\}$

Exciting field for case VIII ($F_{bz} = 1.0$) is shown by *Figure 40* along with the calculated bands of expansion coefficients. Band edges are now excited (bottom of *Figure 40*). Propagation angle (for all bands) at the edge of the Brillouin Zone is zero (on axis) but for band three and higher the propagation angle quickly rotates to larger values as the excited part of the band gets further away to either side of the band edge (*Figure 48*), giving rise to the shown case VIII maps (*Figure 41* and *Figure 42*).

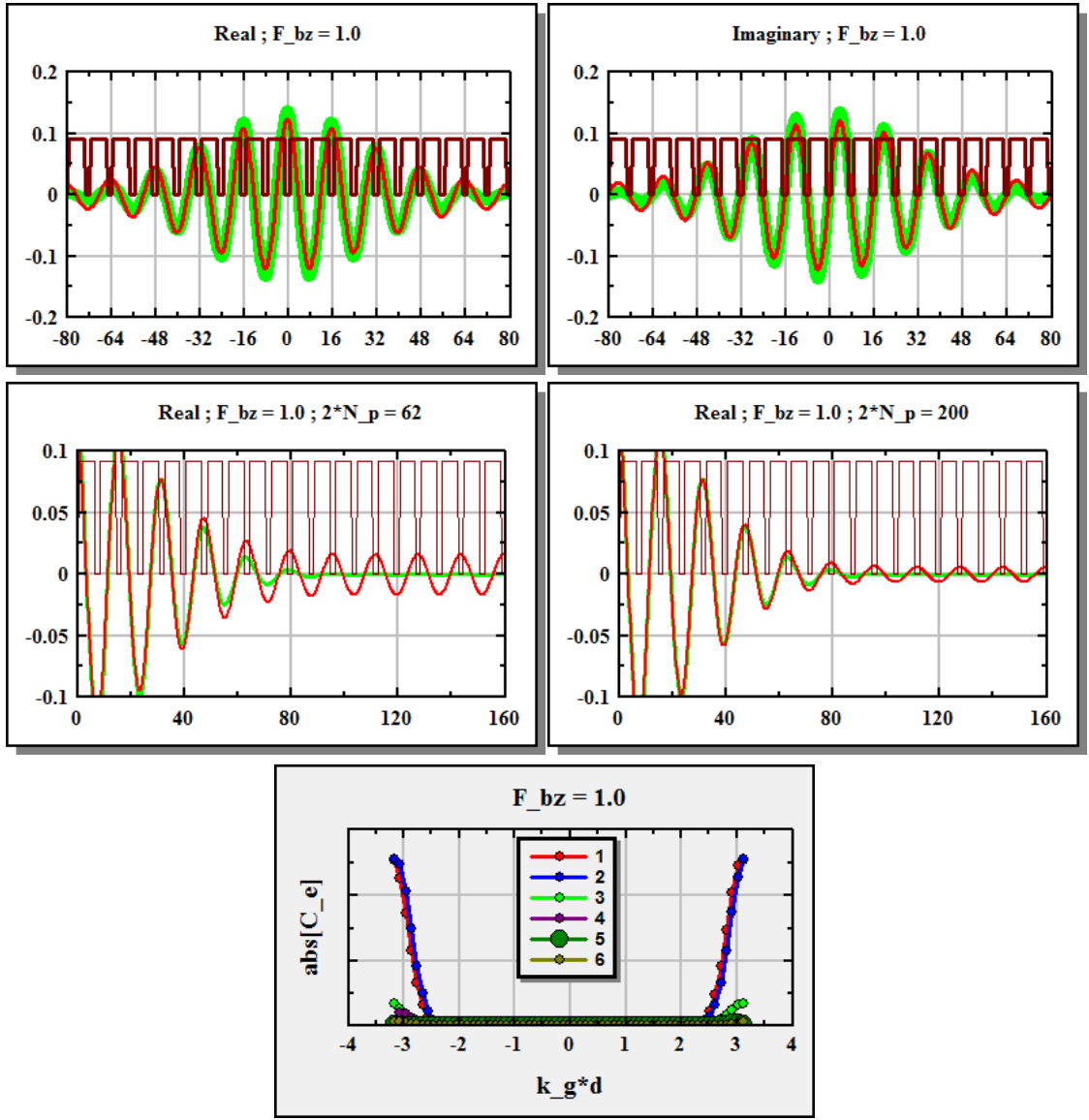

*Figure 40:* *Input field reconstruction and band expansion coefficients for case VIII. Bands are excited at the edges of the Brillouin Zone, bringing the propagation direction back along the z axis ("north").*
*The graphs in the center row show inadequate "reconstruction" (red) of the input Gaussian (green), as a result of assembling Bloch functions from only sixty-two periods (left) or only two-hundred periods (right). The EXAC distribution maps (**Figure 41** and **Figure 42**) thus incorrectly display a bluish background.*

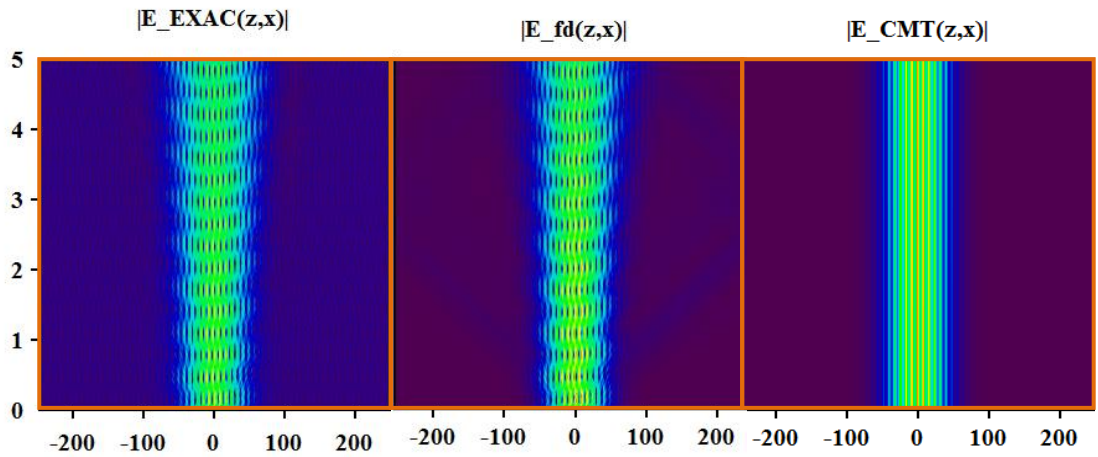

*Figure 41:* Maps for case VIII ($F_{bz} = 1.0$). Band one and band two strongly (and nearly equally) excited (*Figure 40*). Beam propagation is essentially back on axis (despite the tilted excitation [25]). Note the bluish background in the EXAC map (left), contributed by inadequate reconstruction of the tilted input Gaussian beam by adding Bloch functions from only sixty-two periods (see *Figure 40*). The numerical SVEA map does not show such (erroneous) background.

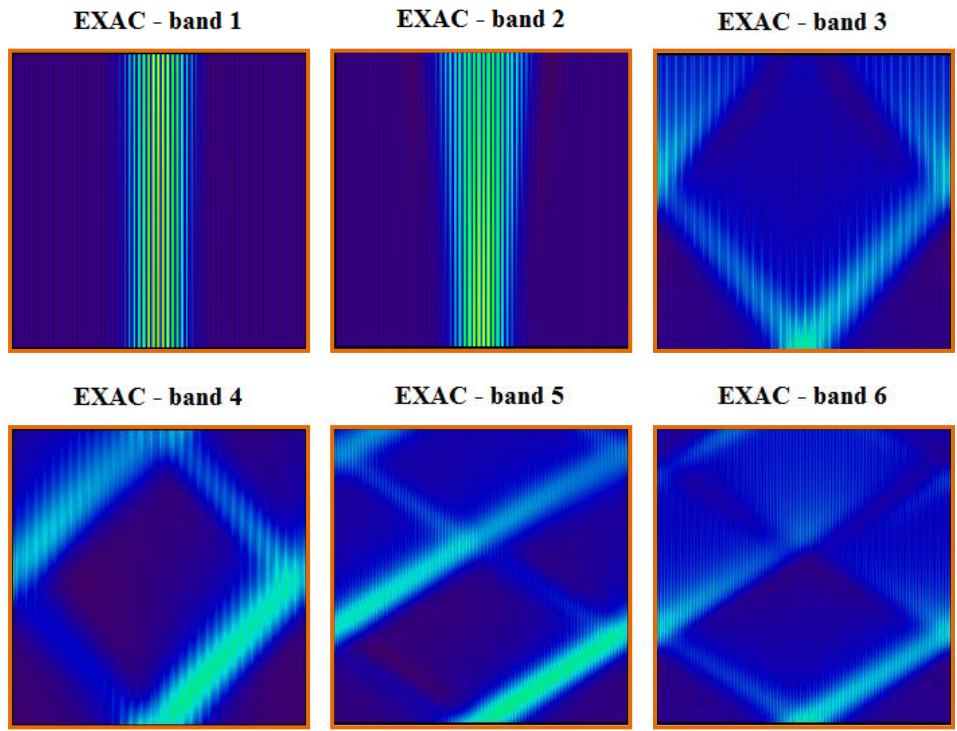

*Figure 42:* EXAC maps by individual bands for case VIII.

### 4.2.4. $F_{bz} = 1.5$ {Θ = $4.3^0$}

At $F_{bz} = 1.5$ (case IX), bands are again (vs. case VII with $F_{bz} = 0.5$) excited in the middle of one half of the Brillouin Zone, this time the negative half (*Figure 43*). The input Gaussian is adequately reconstructed (even with only 62 periods). The dominating band, however, is band two with band one, band three and even band four "visibly" excited (*Figure 44* and mainly *Figure 47*).

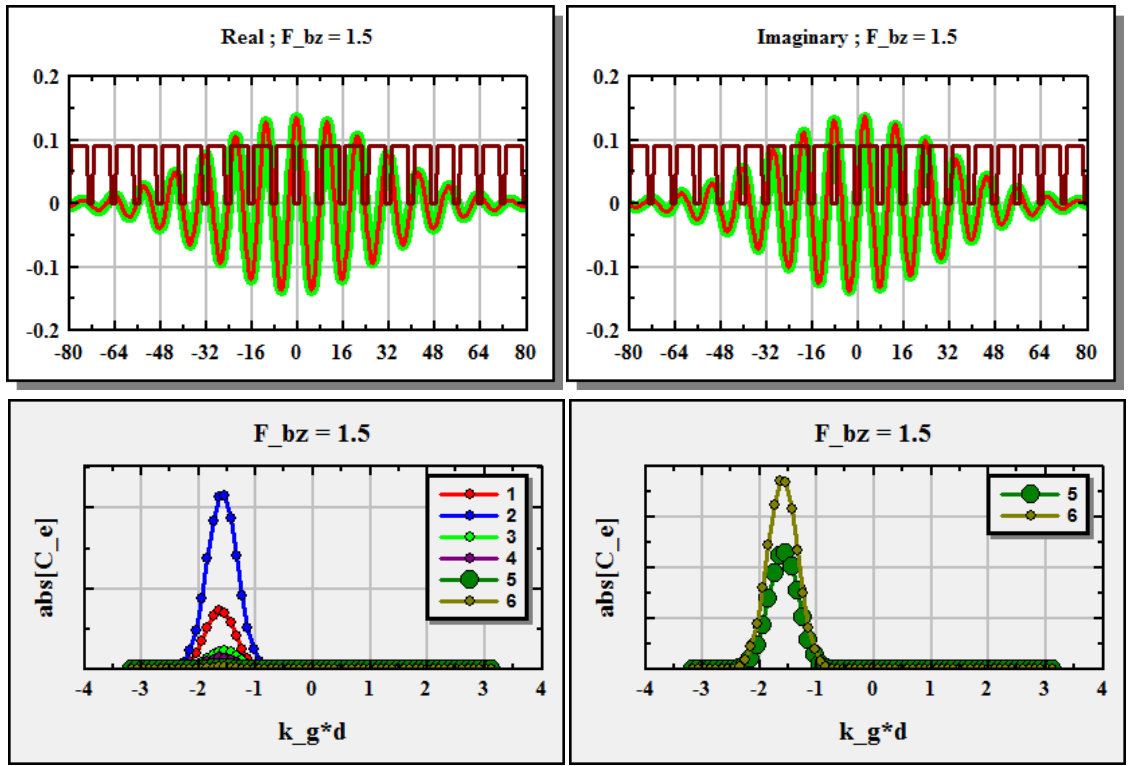

*Figure 43:* *Input field reconstruction and band expansion coefficients for case IX. Bands are excited at the center of the negative half of the Brillouin Zone, resulting in angled propagation at well-defined angles for every band (Figure 46 and Figure 47). Bottom right – zoom to see excitation coefficients for band five and band six. Band six is seen to be excited stronger (vs. band five) as can also be seen by carefully exploring the color-amplified map of Figure 47.*

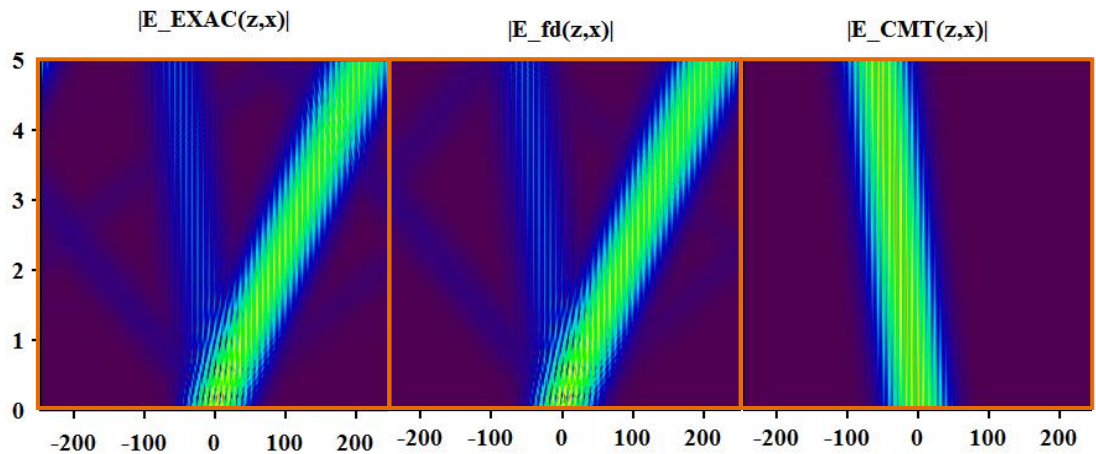

*Figure 44:* Maps for case IX ($F_{bz} = 1.5$). Dominating is band number two but band one, band three and even band four are visible (for both the analytic EXAC map and the numeric BPM-fd map). Note the CMT map showing the first band only (and thus wrong propagation direction altogether).

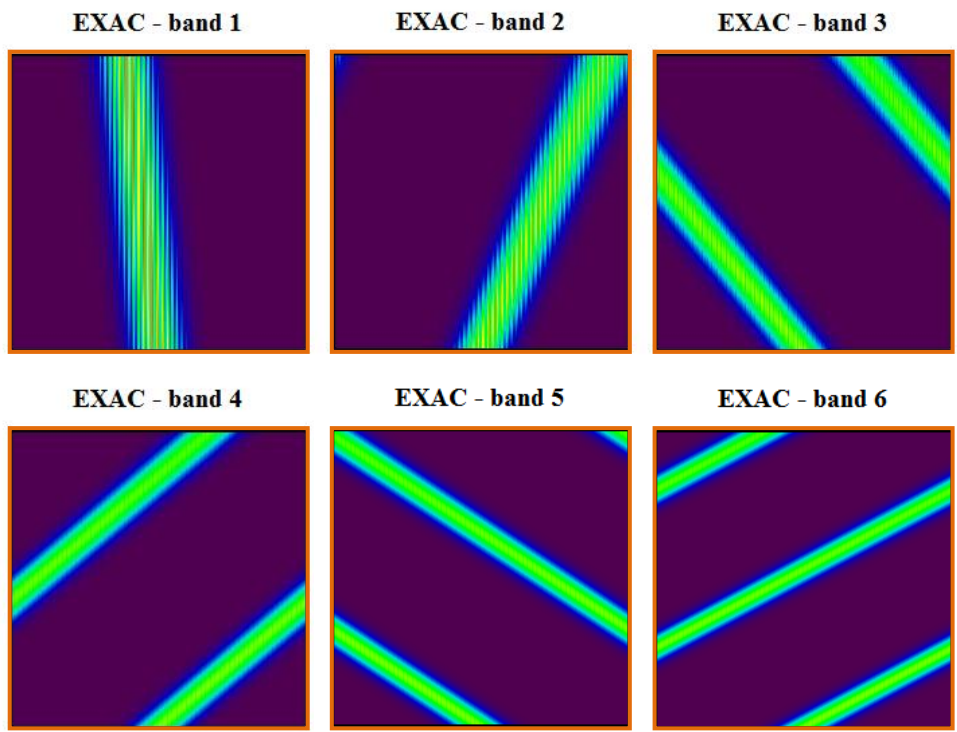

*Figure 45:* EXAC maps by individual bands for case IX.

Given the propagation bands ($k_{ze;n}$), the group velocity direction of a Bloch function is "normal to the curve in wave-vector space" [24] or "normal to the diffraction curve - $k_z(k_x)$" [25] and points "outwards" (towards increasing $k_z$) [24],[25],[26]. The propagation angle ($\theta$) of a Bloch function [$\psi_{e;n,j}$], measured from the z axis, is thus calculated for a one dimensional WG array [$\epsilon(x)$] by the derivative [25] -

$$\tan(\theta_{e;n,j}) = -\left.\frac{dk_{ze;n}(k_g)}{dk_g}\right|_{k_{gj}}$$

(58)

The curves in *Figure 46* show the calculated propagation angles of the respective Bloch functions [equation *(58)*] for the first six propagation bands. The figure shows, as expected, higher angles (typically) for higher band count.

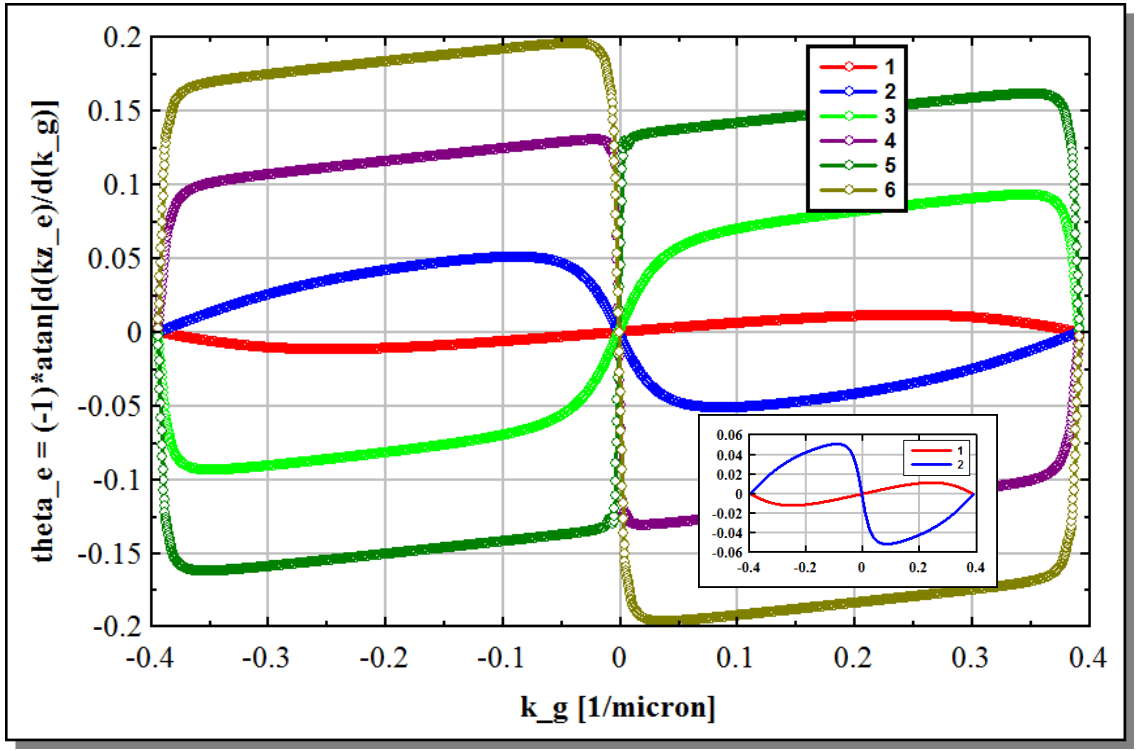

*Figure 46:* *Propagation angles of Bloch functions (with respect to the z axis) – equation (58). Inset – band one and band two only. All curves continuously cross zero at the center and edges of the first Brillouin Zone. Note that zero (spatial) dispersion (zeros of the next derivative) occur "somewhere" in the Brillouin Zone even for the first band, and not at quarter or three-quarters of the Brillouin Zone as predicted (for the first band) by nearest-neighbors-coupling CMT.*

For all bands, angles switch sign at the origin and angles of all bands continuously cross zero at the center and at the edge of the Brillouin Zone. That is – a group of Bloch functions with propagation constants near the exact center or exactly at the edge of the Brillouin Zone will propagate in the z direction, irrespective of band count. Note however, that for high order bands, the propagation angle will "quickly" rotate (away from zero) to its high (and nearly constant) value, and stay there across most of the Brillouin Zone.

Propagation directions for case IX ($F_{bz} = 1.5$) are emphasized by the map of *Figure 47*, with its highly amplified colors. The arrows on the chart to the right indicate the propagation directions of Bloch functions of the first six bands ("normal to the curve in wave-vector space" or "perpendicular to the diffraction curve" [24],[25]).

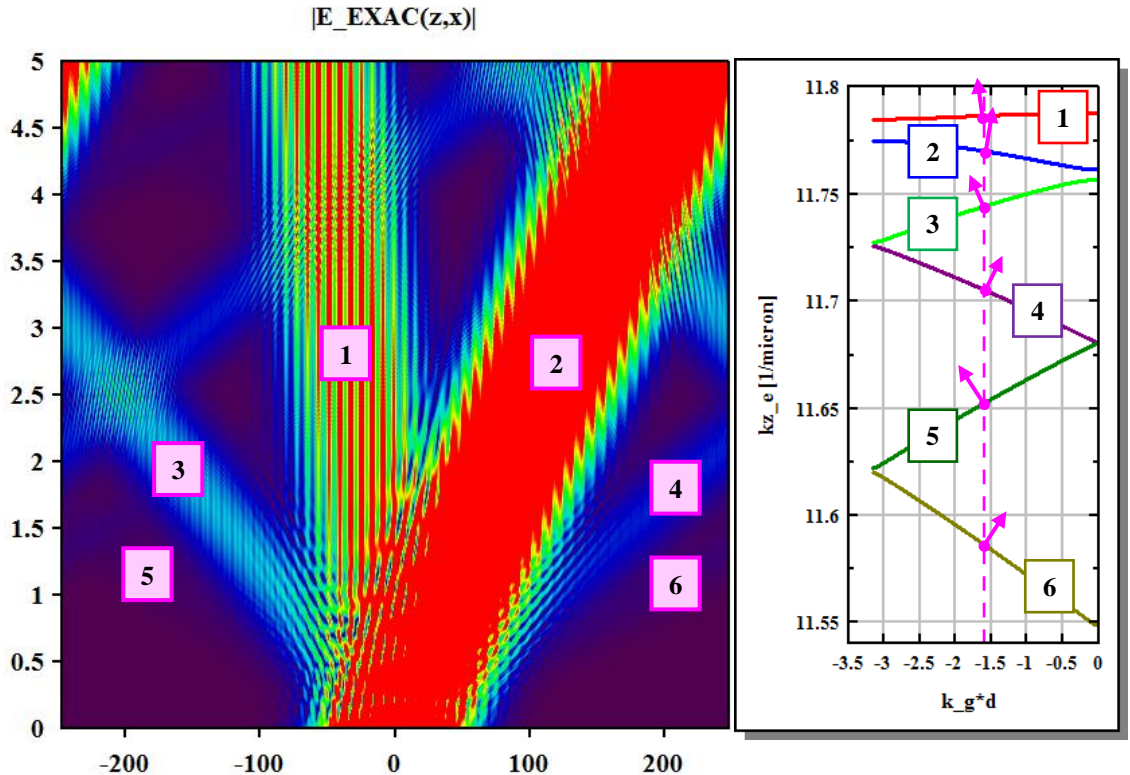

*Figure 47:* Left - propagation of Bloch functions for case IX on a color-amplified map. Contributions from the various bands are designated by their numbers (compare with the individual maps of *Figure 45*). Right – arrows on the diffraction bands [25] pointing to the direction of propagation (of the respective Bloch function). Compare the angles on the map to the calculated angles (*Figure 46*).

### 4.2.5. $F_{bz} = 2.0$ {$\Theta = 5.7^0$}

Case X ($F_{bz} = 2.0$) relates to a relatively large tilt angle of the input field where the phase of the input field differs by full $2 \cdot \pi$ going from one WG to the next (across eight microns in our case, see *Figure 48*). Band two and band three are nearly equally excited, giving rise to the electrical field distribution maps of *Figure 49*.

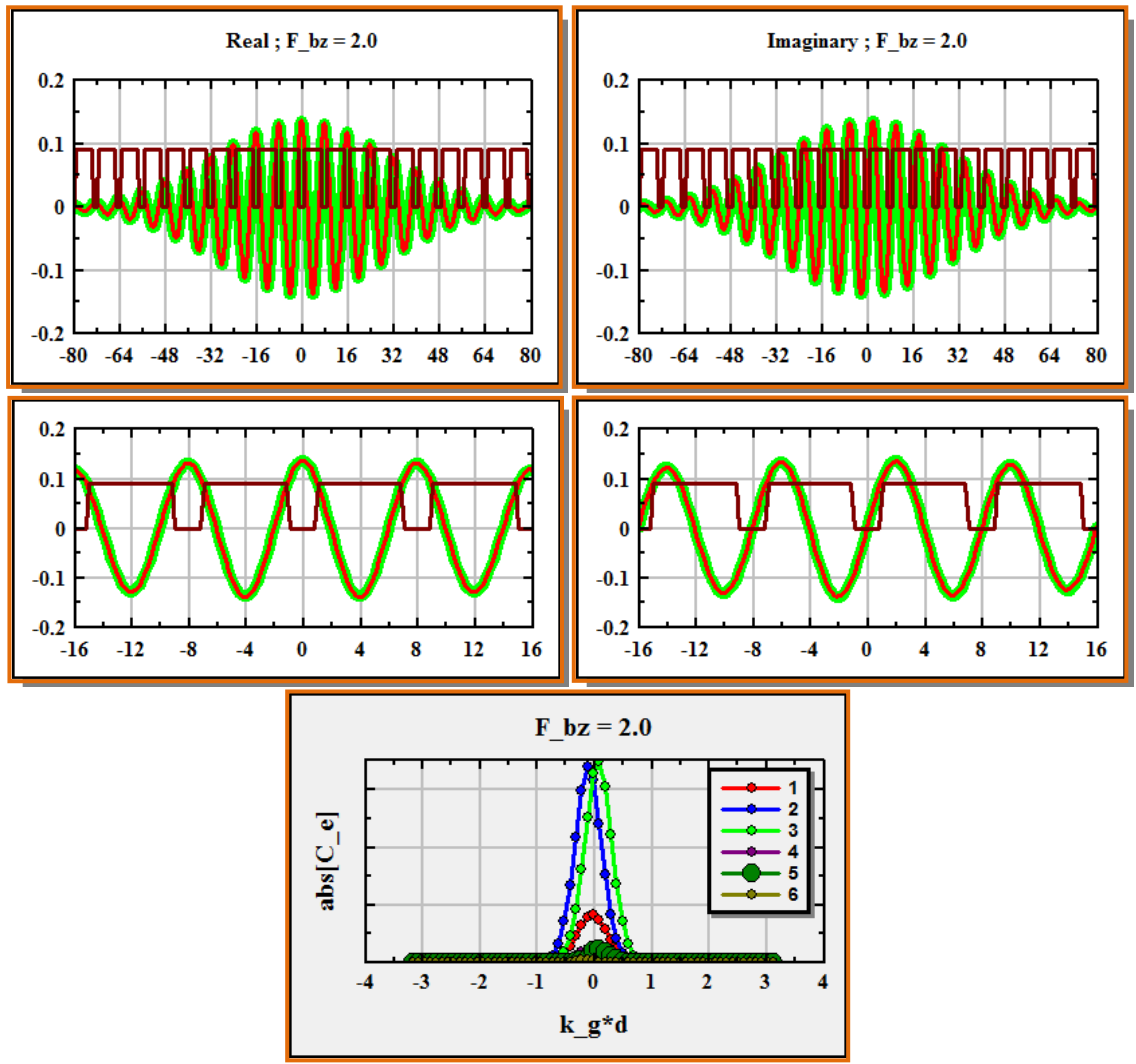

*Figure 48: Input field reconstruction and band expansion coefficients for case X. The real and imaginary parts of the input field are well reproduced by the Bloch functions. Band two and band three are strongly excited.*

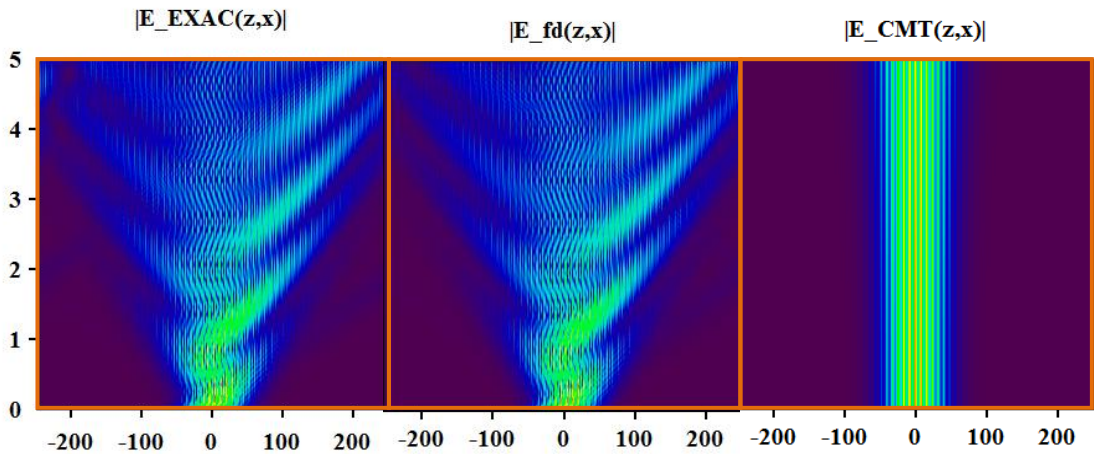

*Figure 49:* Maps for case X ($F_{bz} = 2.0$).

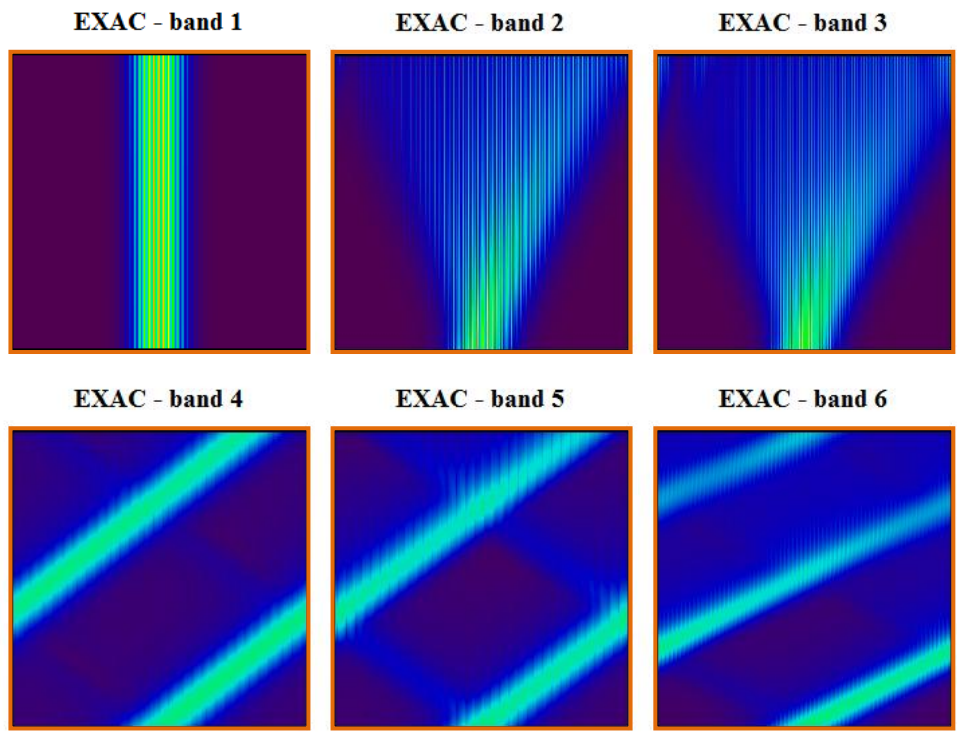

*Figure 50:* EXAC maps by individual bands for case X.

### 4.2.6. $F_{bz} = 2.5 \ \{\Theta = 7.2^0\}$

Case XI ($F_{bz} = 2.5$) – band three (solely) strongly excited (*Figure 51*). The distribution maps indeed show essentially band three propagation (*Figure 52* and *Figure 53*).

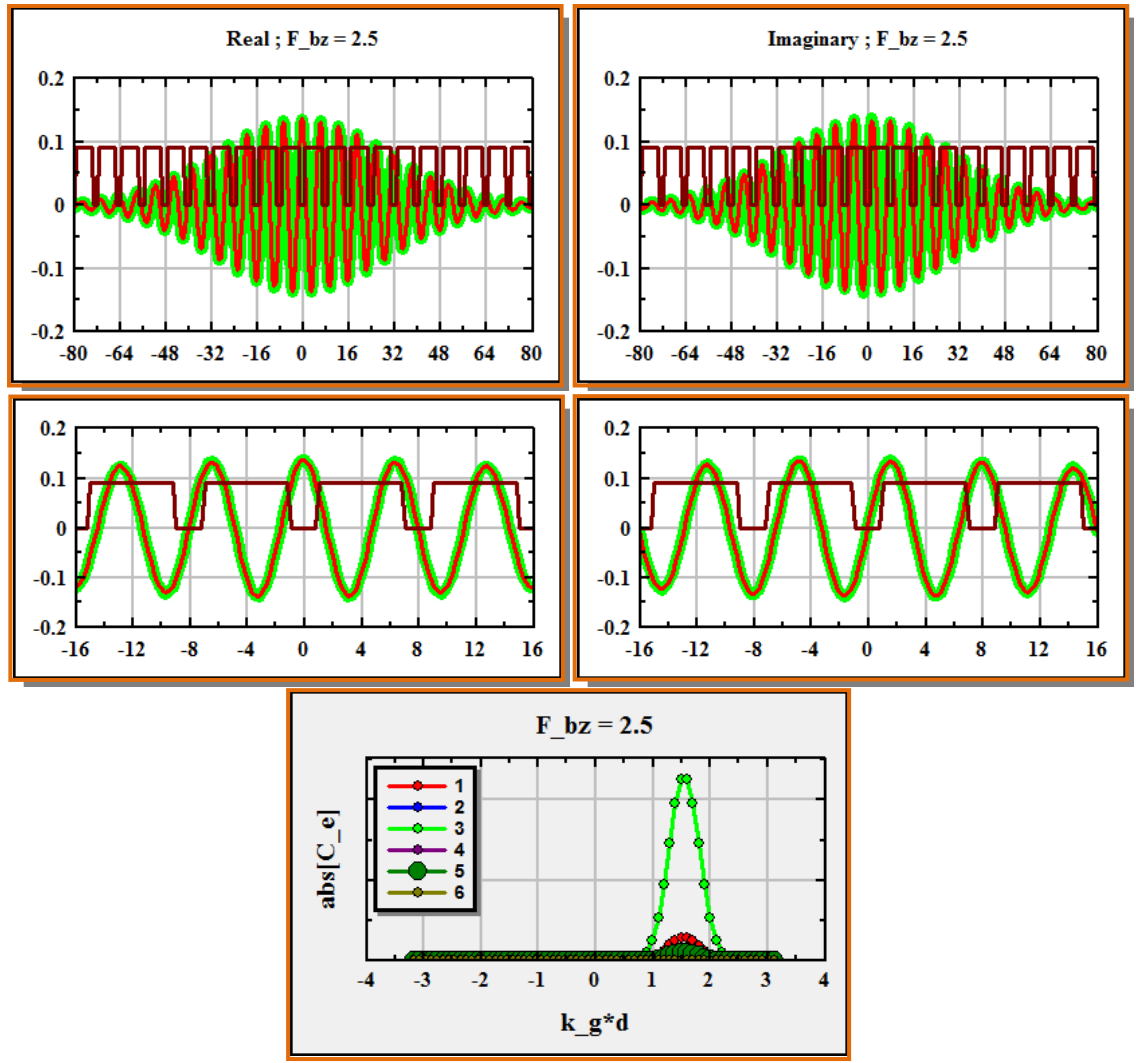

*Figure 51: Input field reconstruction and band expansion coefficients for case XI. Band three dominates.*

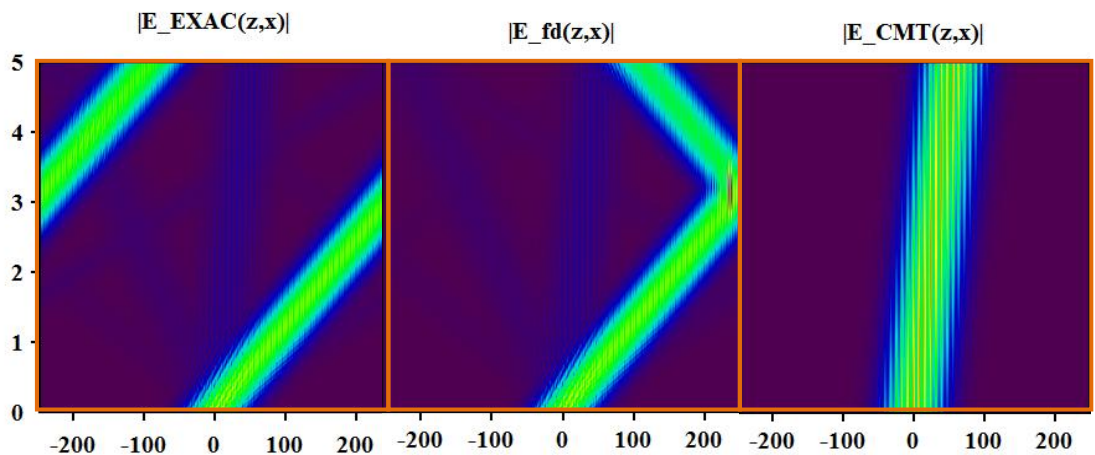

*Figure 52:* *Maps for case XI. Compare the EXAC map (left) to the EXAC band-three map of Figure 53. In the central map, generated by BPM-fd, the beam going right to left at the top of the map is just "reflection" from the boundary (no prevention provisions in the code – see Appendix 5) and should be ignored.*

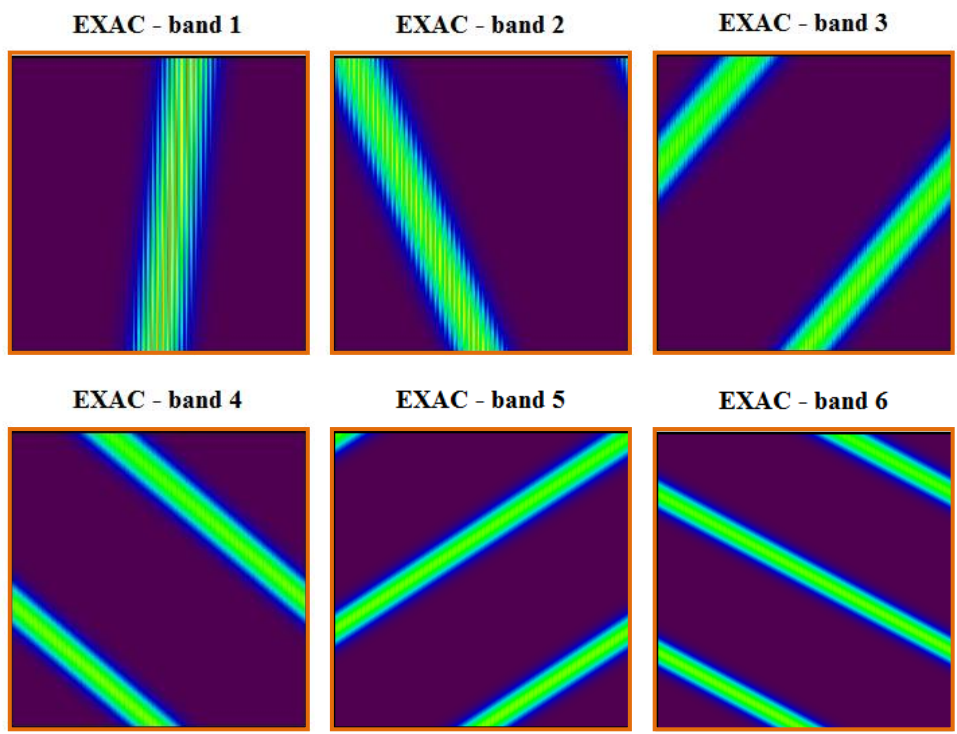

*Figure 53:* *EXAC maps by individual bands for case XI.*

### 4.2.7. $F_{bz} = 5.5$ {$\Theta = 16^0$}

At $F_{bz} = 5.5$, case XII, essentially only band six is excited - *Figure 54*. The input Gaussian is well reconstructed (with only sixty-two periods and with Bloch functions from only six bands - *Figure 54*). The maps of *Figure 55* and *Figure 56* indeed indicate band-six propagation only.

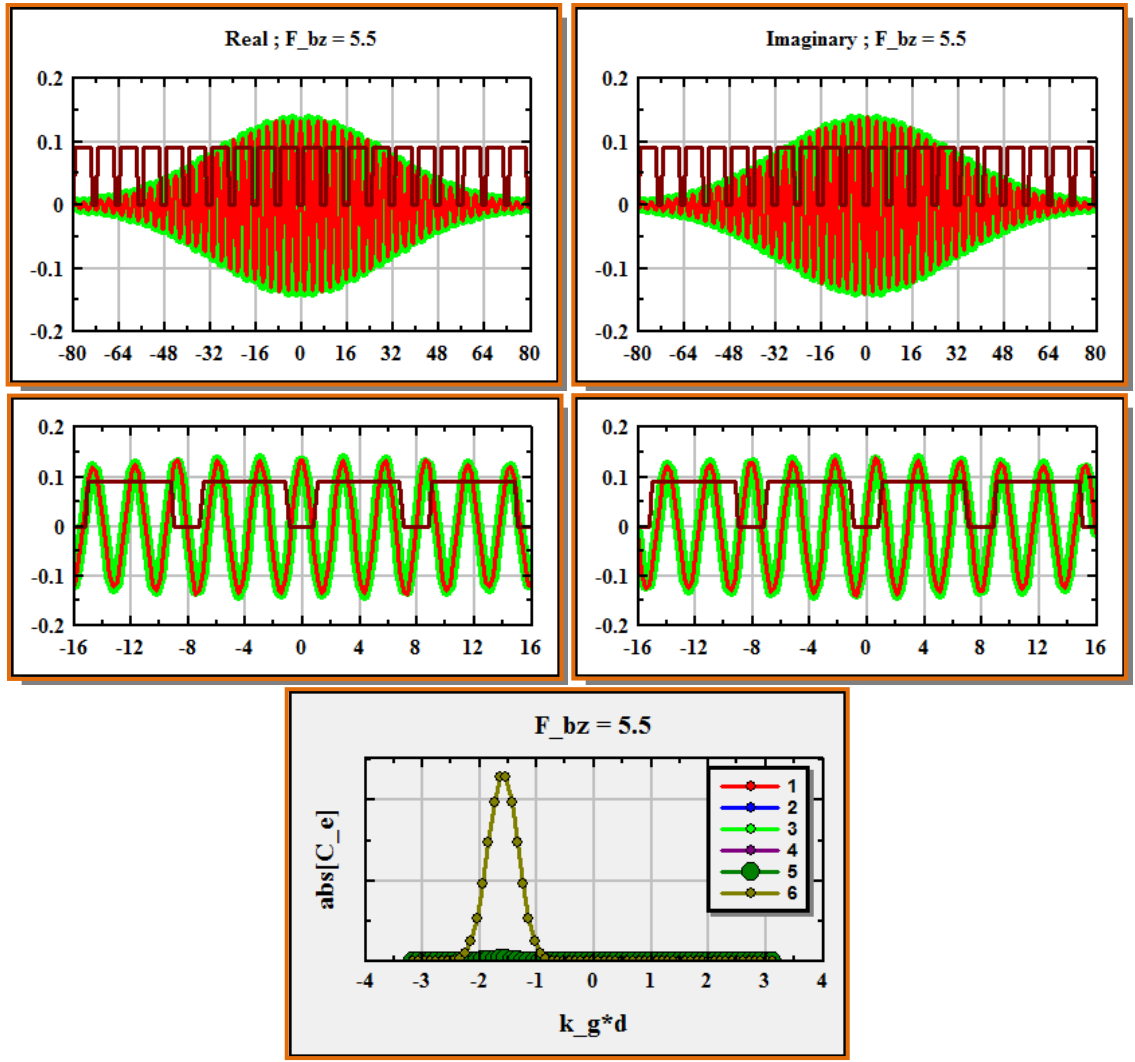

*Figure 54:* Input field reconstruction and band expansion coefficients for case XII. With $F_{bz} = 5.5$ band six is very purely excited (next is band four, down by a factor of about thirty). The Input Gaussian is very closely reproduced (essentially by band six Bloch functions only).

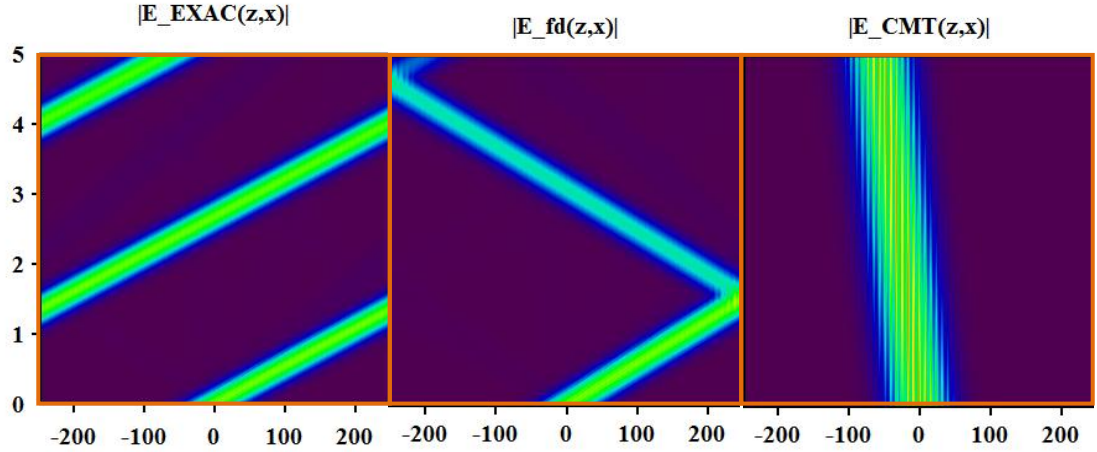

*Figure 55: Maps for case XII. The EXAC map (left) and the BPM-fd map (center) indeed show propagation of Bloch-functions from band-six only (compare with the right-down map of Figure 56).*

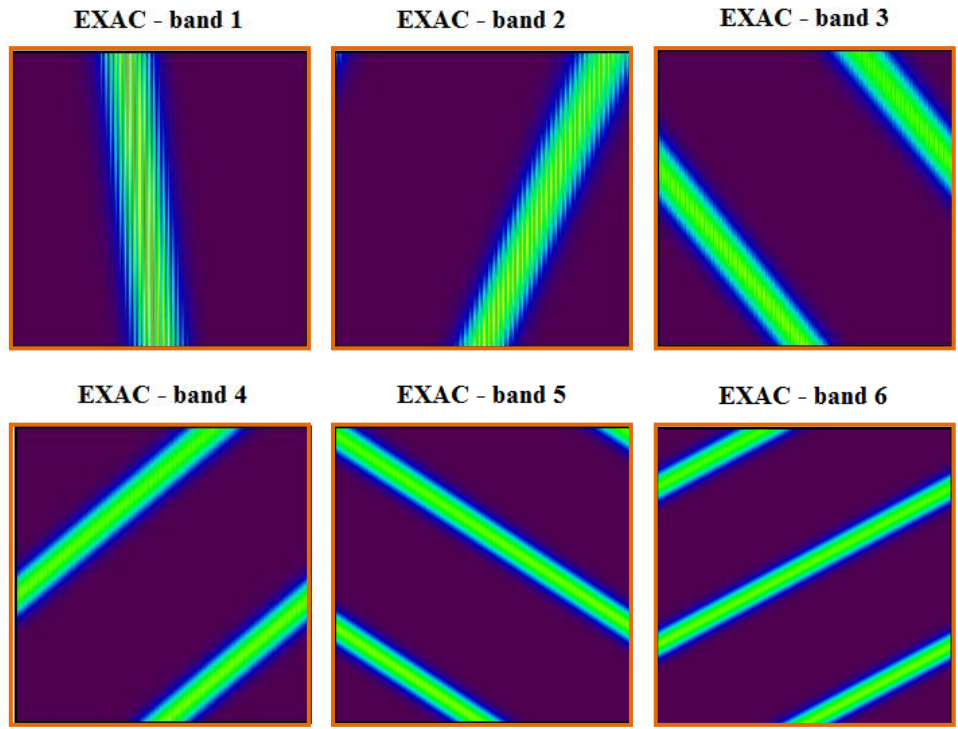

*Figure 56: EXAC maps by individual bands for case XII.*

To summarize this band excitation section – a tilted wide Gaussian will excite a group of Bloch-functions associated with a confined range of associated lattice wave-vectors ($k\_g$). The small set of excited Bloch functions from every band will typically have a rather well defined direction of propagation (will form a "beam" of light), particularly if a single high order band is excited and the associated confined group of lattice wave-vectors is not at the center or edge of the Brillouin Zone.

The maps generated by the EXAC analytical solution and by the BPM-fd numeric solution (of the SVEA equation) are fairly similar to each other and correctly represent the electrical field distributions. The map generated by CMT is associated with band one excitation only. If other, higher order bands are excited to any significant level, the CMT map fails to show the correct electrical field distribution.

## 5. Summary

The five methods studied in this document can be divided into two groups: analytic group and numeric group. The three methods of the analytic group – EXAC, SVEA, CMT are actually only two distinct methods since EXAC and SVEA can be considered one. First, because they yield very similar distributions. Second, and more importantly – if the permittivity depends on $x$ only, the SVEA equation can be "upgraded" back to coincide with the EXAC equation. So we are left with only two distinct analytic methods – EXAC and CMT.

EXAC and CMT indeed differ significantly. The CMT method only treats amplitudes of "guided" fields. Guided field amplitudes can be associated with band one Bloch functions only - *Figure 57*. If higher bands are excited, CMT will yield completely erroneous distributions. Even with band-one-only excitation, differences will still result (between CMT-calculated distributions and the EXAC-calculated distributions) since the propagation constants calculated by each of the two methods don't match (cf. *Figure 19*).

As for the two methods in the numeric group – BMP-fd and BMP-ss, we see some differences in calculated distribution (vs. the EXAC-calculated distributions), even if the selected value of the reference index is "rational".

Between the two methods (BMP-fd and BMP-ss), we found that their relative accuracy is "case dependent" (for a fixed pixel size).

Note that while the BPM-calculated **intensities** are fairly close to the EXAC-calculated intensities, the calculated **fields**, both the real part as well as the imaginary part, are very different from their EXAC counterparts (cf. *Figure 24*). Meaning – if amplitude distributions are of interest, for example in interference experiments, predictions by the BPM methods should be carefully reviewed.

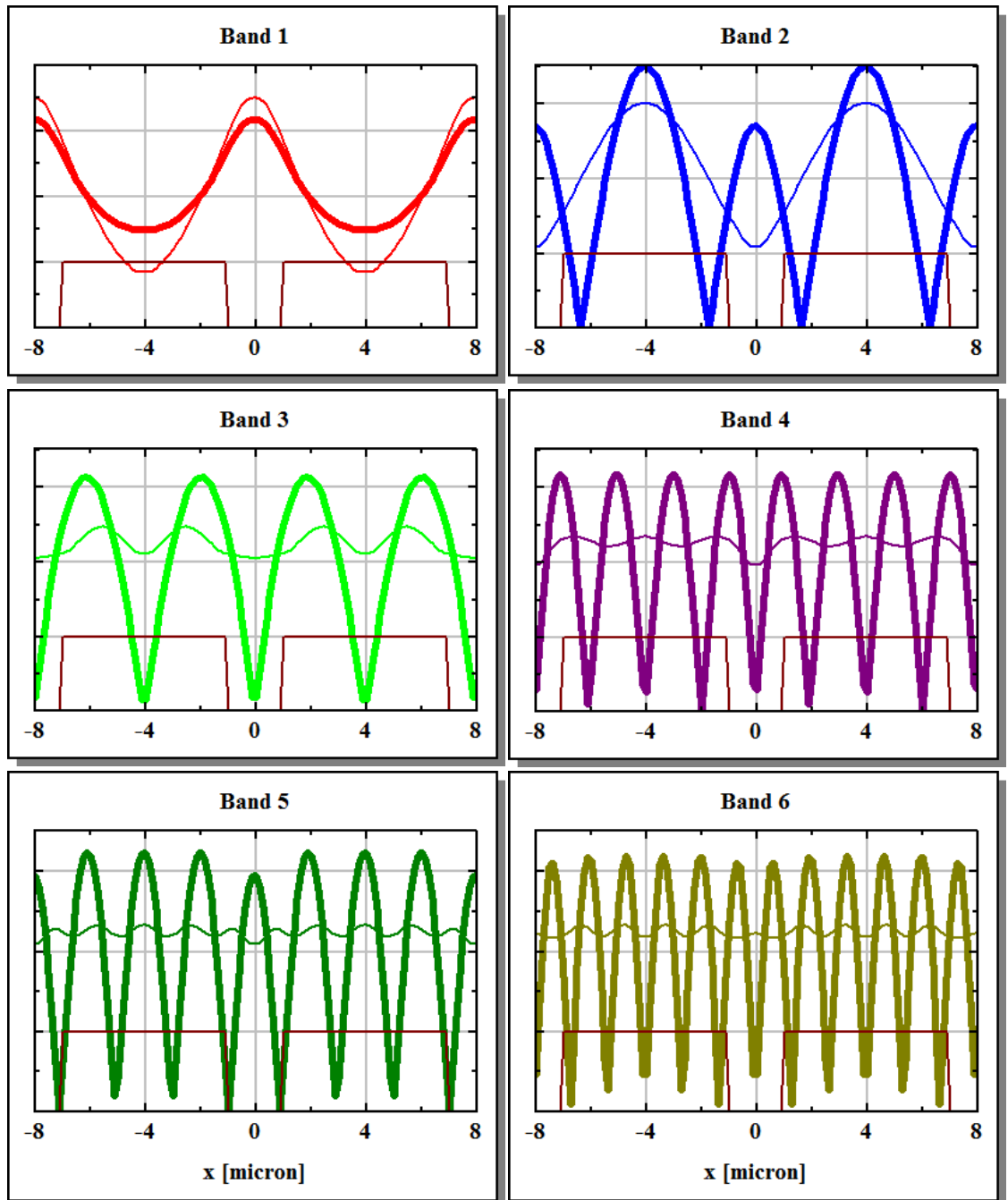

*Figure 57:* Band intensities of Bloch functions. Thick line - band intensities for center-of-band excitation ($j = 32$ ; $k_{g;32} = 0$). Thin line - band intensities for near-edge excitation ($j = 55$ ; $k_{g;55} = 0.29\ [\mu m]^{-1}$ ; Total number of periods – 62). The curves show that for all bands but the first one, Bloch functions "carry" high intensities in between the WGs (i.e. in the regions of low refractive index). Thus, the CMT, treating the amplitude evolution of **guided** fields, works for band one excitation only.

Of-course, the great advantage of the BPM methods is their applicability to the (large variety of) cases where analytic solutions are hard to find or simply do not exist.

## *Appendix 1:* **"EXAC" Solution of equation** *(16)*

In this appendix we solve the Full (scalar) Helmholtz Equation [equation *(16)*] for $\psi_e(x)$ [the subscript "e" stands for "EXAC"] –

$$\frac{d^2\psi_e(x)}{dx^2} + (k_0^2 \cdot \epsilon(x) - K^2) \cdot \psi_e(x) = 0$$

*(59)*

with $\epsilon(x)$ as "square" 1D finite periodic structure (*Figure A1*):

$$\sqrt{\epsilon(x)} = n(x) = \begin{cases} n_2 & j \cdot d < x \leq j \cdot d + a \\ n_1 & j \cdot d - b < x \leq j \cdot d \\ j = -N_p, -(N_p - 1), \dots 1, 0, 1, \dots (N_p - 1) \end{cases}$$

*(60)*

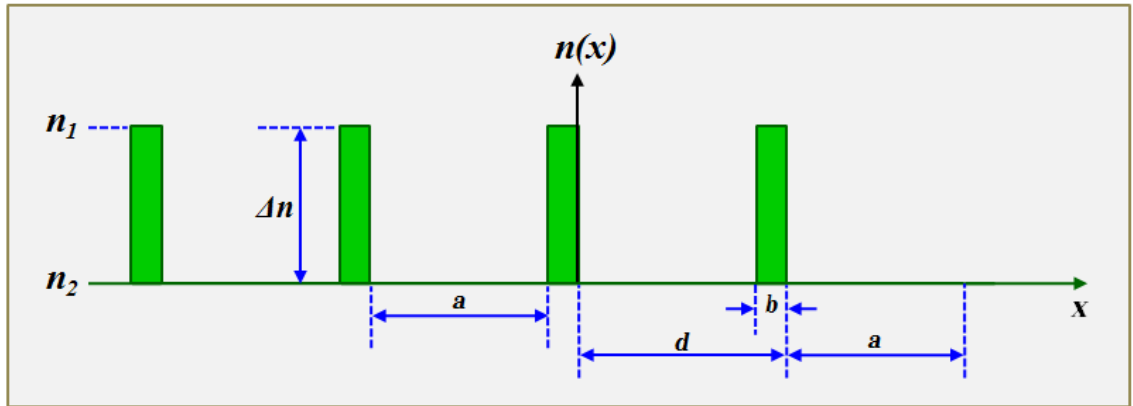

*Figure A1:* Geometry and parameters of the waveguide array in question [equation *(60)*].

The general solution of equation *(59)*, for the finite periodic structure *(60)* follows the Floquet-Bloch Theorem (FBT) **[12]**:

$$\psi_e(x) = e^{i \cdot k_g \cdot x} \cdot u_e(x)$$

*(61)*

For the scalar electrical field to be continuous and smooth, the function $\psi_e(x)$ must be continuous and smooth (continuous first derivative). In addition, according to the FBT, the "cell function" [$u_e(x)$] of equation *(61)* must have a cell periodicity, and, for a large number of periods, we want to impose the cyclic Born-von Karman boundary conditions **[12]**.

These requirements translate into the following four restrictions imposed on $\psi_e(x)$ for any $x$:

$$\text{I.} \quad \psi_e(x)]\!]_{x^-} = \psi_e(x)]\!]_{x^+}$$

$$\text{II.} \quad \frac{d\psi_e(x)}{dx}\bigg]\!\bigg]_{x^-} = \frac{d\psi_e(x)}{dx}\bigg]\!\bigg]_{x^+}$$

$$\text{III.} \quad u_e(x) = u_e(x+d)$$

$$\text{IV.} \quad \psi_e(x) = \psi_e(x+L) \; ; \; L \equiv 2 \cdot N_p \cdot d$$

*(62)*

The cyclic boundary condition [condition IV in equation *(62)*], requires quantization of the "lattice wave-vector" $k_g$ in the solution *(61)*:

$$k_{g;j} = \frac{j}{N_p} \cdot BZ \; ; \; BZ \equiv \frac{\pi}{d} \; ; \; j = -N_p, \ldots 0, \ldots, N_p$$

*(63)*

At this point, the above formulation fits the (simplest case of the) Kronig–Penney model developed around 1930 in the context of quantum mechanics (idealizing the atomic potential in a periodic 1D lattice to square wells).

For the sake of completeness, we follow here the steps of Kronig and Penney, with emphasis on the resulting bands of the propagation constant $[k_z \equiv K$, cf. equation *(15)*$]$.

Let's first define two distinct "wave-numbers" - $\delta_e$ and $\gamma_e$ for the two structure regions – region "H" for the $n_1$ region and region "L" for the $n_2$ region:

$$\text{H:} \quad \delta_e^2 \equiv k_0^2 \cdot n_1^2(x) - K^2$$

$$\text{L:} \quad \gamma_e^2 \equiv k_0^2 \cdot n_2^2(x) - K^2$$

*(64)*

The solution to equation *(59)*, given the periodic permittivity *(60)*, and following the FBT form of equation *(61)* becomes:

$$u_{eH}(x) = A \cdot e^{i \cdot (\delta_e - k_g) \cdot x} + A' \cdot e^{-i \cdot (\delta_e + k_g) \cdot x}$$
$$u_{eL}(x) = B \cdot e^{i \cdot (\gamma_e - k_g) \cdot x} + B' \cdot e^{-i \cdot (\gamma_e + k_g) \cdot x}$$
$$\psi_e(x) = e^{i \cdot k_g \cdot x} \cdot u_e(x)$$

*(65)*

where, for the time-being, we omitted the index $j$ from $k_g$ [cf. equation *(63)*].

The cell periodicity and the continuity requirements [I, II, and III in equation *(62)*] result in a four-by-four matrix equation of the form -

$$\begin{bmatrix} a_1, a_2, a_3, a_4 \\ b_1, b_2, b_3, b_4 \\ c_1, c_2, c_3, c_4 \\ d_1, d_2, d_3, d_4 \end{bmatrix} \cdot \begin{bmatrix} A \\ A' \\ B \\ B' \end{bmatrix} = 0$$

*(66)*

A non-trivial solution to equation *(66)* exists if and only if the matrix is singular, i.e. if and only if its determinant is zero. The zero-determinant requirement leads to the quantization of $K$ and thus to the formation of bands for the propagation constants - $k_z$ [$= K$, cf. equation *(15)*].

We show below, that for every $j$ [cf. equation *(63)*], a set of solutions to equation *(66)* ($K_{n,j}^2$) exists. The four coefficients in equation *(66)* [$A, A', B, B'$] are thus enumerated too and the set of solutions $\psi_{e;n,j}(x)$ we are after [cf. equation *(59)*] reads:

$$u_{e;n,j}(x) = \begin{cases} A_{n,j} \cdot e^{i \cdot (\delta_{e;n,j} - k_{g;j}) \cdot x} + A'_{n,j} \cdot e^{-i \cdot (\delta_{e;n,j} + k_{g;j}) \cdot x} \; ; \; H \; region \\ B_{n,j} \cdot e^{i \cdot (\gamma_{e;n,j} - k_{g;j}) \cdot x} + B'_{n,j} \cdot e^{-i \cdot (\gamma_{e;n,j} + k_{g;j}) \cdot x} \; ; \; L \; region \end{cases}$$

$$\psi_{e;n,j}(x) = e^{i \cdot k_{g;j} \cdot x} \cdot u_{e;n,j}(x)$$

*(67)*

The set of Bloch functions [equation *(67)*] is the exact solution of equation *(16)* as appears in the body of this paper [and of equation *(59)* of this appendix].

Back to equation *(66)*, the sixteen elements of the matrix can be found in **[13]**, or, more conveniently, in **[12]**, and are also listed below in ***Appendix 3***.

Let's write the det = 0 equation in the following way:

$$LHS(k_g) \equiv \cos(k_g \cdot d)$$

$$RHS_e(K^2) \equiv \cos(\delta_e \cdot b) \cdot \cos(\gamma_e \cdot a) - \frac{\delta_e^2 + \gamma_e^2}{2 \cdot \delta_e \cdot \gamma_e} \cdot \sin(\delta_e \cdot b) \cdot \sin(\gamma_e \cdot a)$$

$$LHS(k_g) = RHS_e(K^2)$$

*(68)*

with $K^2$ "buried" in $\delta_e$ and in $\gamma_e$ [cf. equation *(64)*].

Now, since we already know that for any $k_{g;j}$ there are several $K^2$ values (formally several tens at most if we restrict $K^2$ to be positive, i.e. require that $k_z$ be real, representing a propagating wave) that solve equation *(68)*, we rewrite equation *(68)* as -

$$LHS(k_{g;j}) = RHS_e(K_{n,j}^2)$$

*(69)*

The bands of propagation constants $[k_{ze;n,j}]$ for the EXAC solutions are given by [cf. equation *(15)*]:

$$k_{ze;n,j} = K_{n,j}$$

*(70)*

$RHS_e(K^2)$ curve and $LHS(k_{g;j})$ line (for a particular $j$) are shown in ***Figure A1***. The $K^2$ solutions to equation *(69)* are indicated by the green circles. Scanning the value of $k_{g;j}$ across the first Brillouin Zone will create the set of propagation bands [equation *(70)* and ***Figure A3***]. ***Figure A4*** is a zoom showing individual propagation bands.

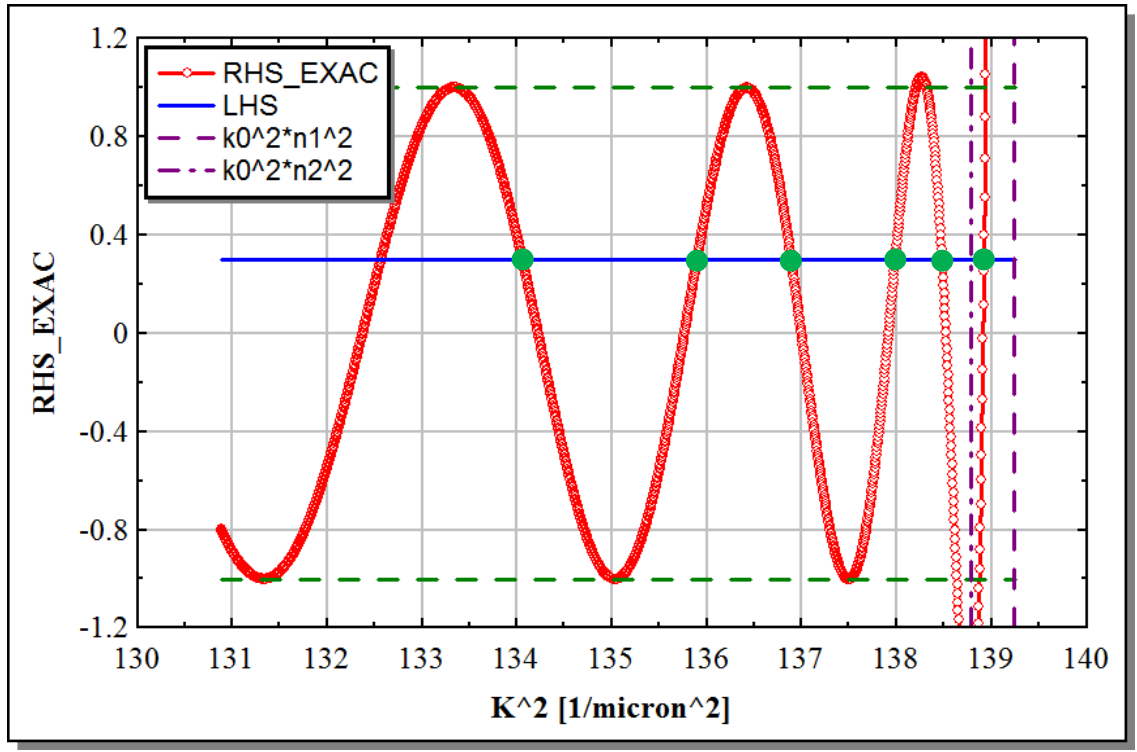

***Figure A2***: $RHS_e(K^2)$ *(red) and* $LHS(k_{g;j})$ *(blue). The green circles show the first six $K^2$ values that solve equation **(69)** for the particular $k_{g;j}$ of the LHS line. "Scanning" $k_{g;j}$ values across the first Brillouin Zone (so that the LHS line scans the minus one to plus one range and back) will create the first six $K^2$ bands (and thus the first six bands of propagation constants - $k_{ze;n,j} = K_{n,j}$ [equation **(70)**]. See the tips for numerical calculations in **Appendix 7**.*

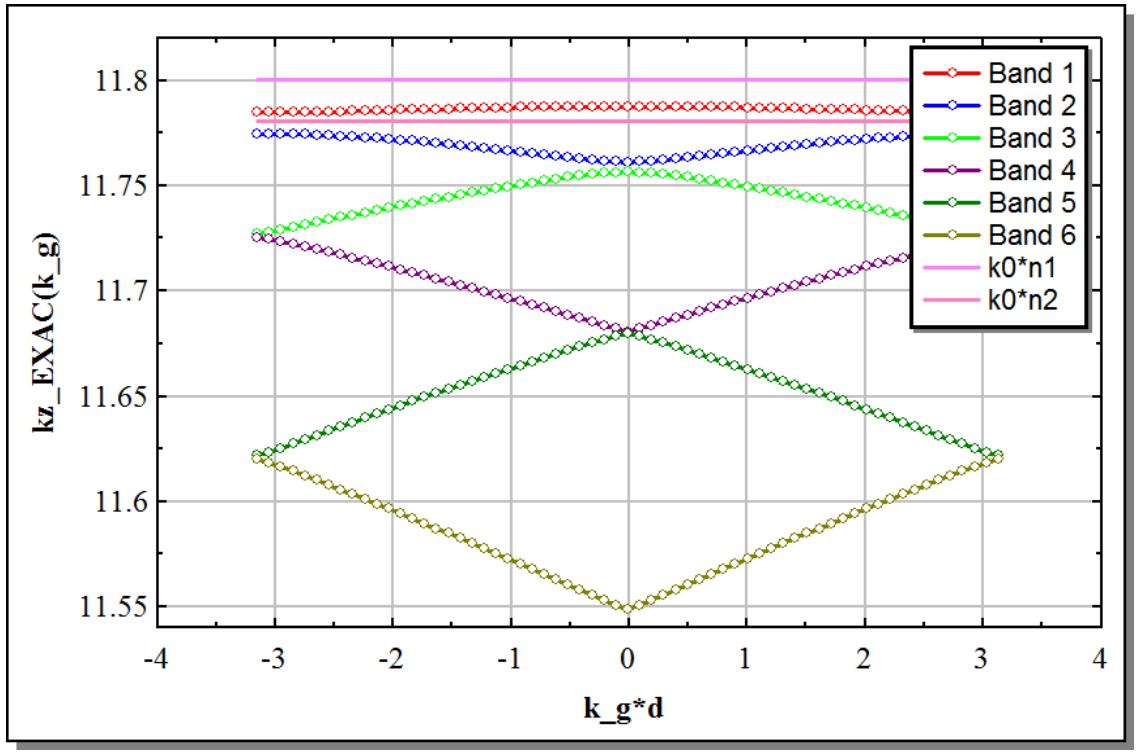

*Figure A3*: *First six EXAC propagation bands as calculated by solving equation (69). The two pink lines indicate the position of the propagation constant for a homogeneous bulk with refractive index $n_1$ (high pink line) or $n_2$ (lower pink line). The first band of EXAC propagation constants (and only the first band) lies entirely between the two pink lines.*

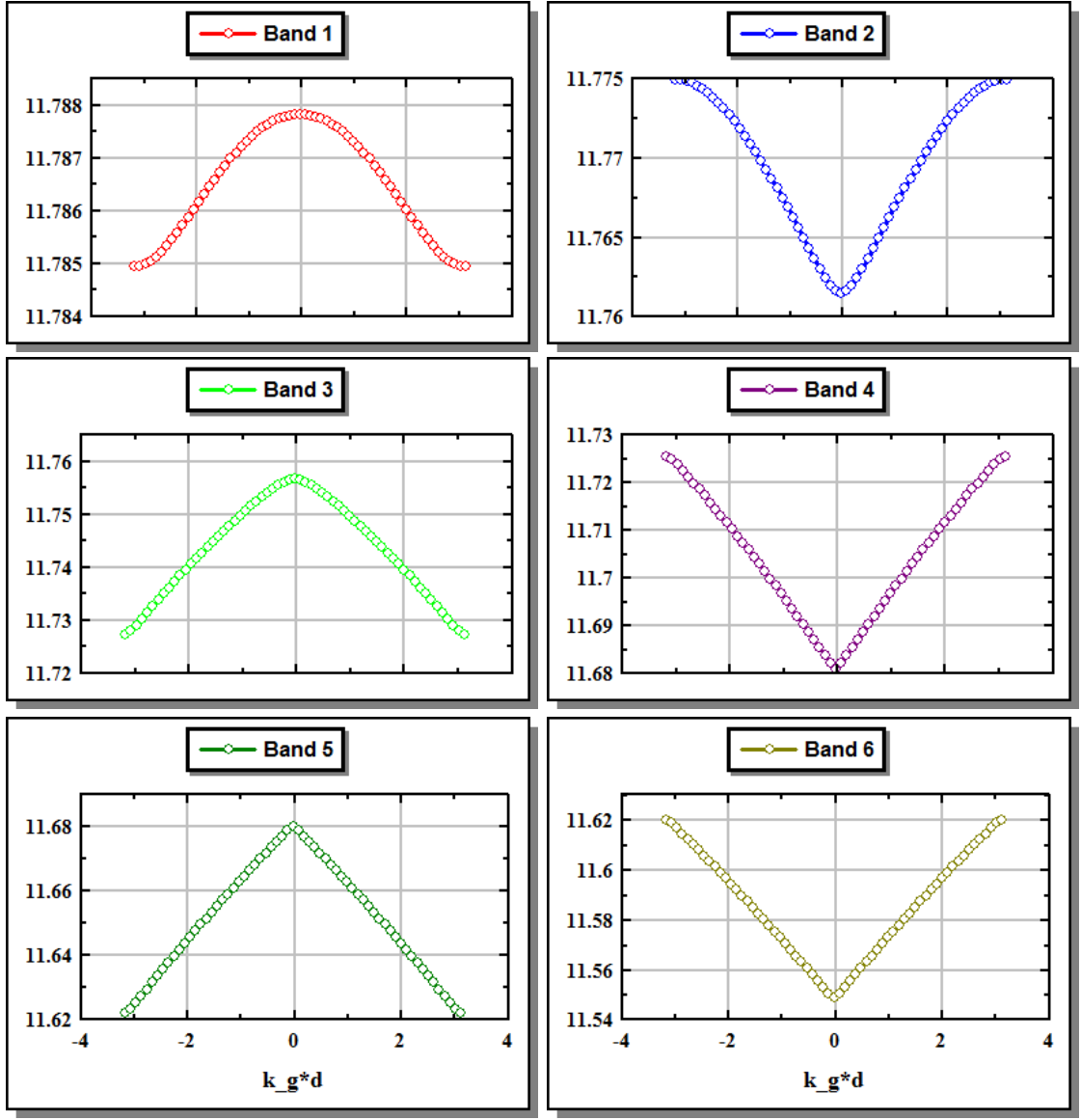

***Figure A4***: *Zoom to individual EXAC propagation bands. Numbers on the vertical axis are $k_z$ in units of 1/micrometer. First derivative (with respect to $k_g$) of all bands continuously crosses zero at the center and edges of the Brillouin Zone (not obvious by looking at the shown curves, but consult **Figure 46** in the body of the paper).*

Once the propagation bands are found, the coefficients $[A, A', B, B']$ [equation *(66)*] are calculated (*Appendix 3* and *Appendix 4*) and the sought-for Bloch functions [equation *(67)*] explicitely computed.

## *Appendix 2*: "SVEA" - Solution of equation *(7)*

In this appendix we solve the SVEA Helmholtz Equation [equation *(7)*], for a 1D periodic array as described by equation *(60)* (and cf. **Figure A1**). For the given 1D array [$\epsilon(x, y, z) \Rightarrow \epsilon(x)$] (and for y-independent initial conditions), equation *(7)* is simplified to [cf. equation *(24)* or cf. equation *(30)*]:

$$i \cdot \frac{\partial U_s(x,z)}{\partial z} = -\frac{1}{2 \cdot k_0 \cdot n_{ref}} \cdot \frac{\partial^2 U_s(x,z)}{\partial x^2} - \frac{k_0^2}{2 \cdot k_0 \cdot n_{ref}} \cdot [\epsilon(x) - n_{ref}^2] \cdot U_s(x,z)$$

*(71)*

To solve equation *(71)* we resort to the separation of variables method and write:

$$U_s(x,z) \equiv \psi_s(x) \cdot g_s(z)$$

*(72)*

For writing simplification let's define a reference propagation constant $\beta_{ref}$ as –

$$\beta_{ref} \equiv k_0 \cdot n_{ref}$$

*(73)*

and get for equation *(71)*:

$$i \cdot \frac{1}{g_s(z)} \cdot \frac{\partial g_s(z)}{\partial z} = \frac{1}{\psi_s(x)} \cdot \left\{ -\frac{1}{2 \cdot \beta_{ref}} \cdot \frac{\partial^2 \psi_s(x)}{\partial x^2} - \frac{1}{2 \cdot \beta_{ref}} \cdot [k_0^2 \cdot \epsilon(x) - \beta_{ref}^2] \cdot \psi_s(x) \right\} \equiv K_s$$

*(74)*

For $g_s(z)$ we get [the constant for $g_s(z=0)$ is "absorbed" by $\psi_s(x)$]

$$g_s(z) = e^{-i \cdot K_s \cdot z}$$

*(75)*

For the x-dependence [$\psi_s(x)$] we need to solve

$$\frac{d^2\psi_s(x)}{dx^2} = -\left[k_0^2 \cdot \epsilon(x) - \beta_{ref}^2 + 2 \cdot \beta_{ref} \cdot K_s\right] \cdot \psi_s(x)$$

*(76)*

To convert equation *(76)* into a more familiar form, we replace the constant $K$ with another (dimension-less) constant – $q$:

$$K_s \equiv \beta_{ref} \cdot q$$

*(77)*

Now let's rewrite equation *(76)* as

$$\frac{\partial^2\psi_s(x)}{\partial x^2} = -2 \cdot \beta_{ref}^2 \cdot \left[\frac{k_0^2 \cdot \epsilon(x) - \beta_{ref}^2}{2 \cdot \beta_{ref}^2} + q\right] \cdot \psi_s(x)$$

*(78)*

Now insert the "optical potential" $V_s$ [equation *(33)*] and write equation *(78)* in its final form -

$$\frac{\partial^2\psi_s(x)}{\partial x^2} = -2 \cdot \beta_{ref}^2 \cdot [q - V_s(x)] \cdot \psi_s(x)$$

*(79)*

Note that with the definition *(77)*, the electrical field is given by [cf. equations *(5)*,*(73)*,*(75)*] -

$$E_s(x,z) = \psi_s(x) \cdot e^{i \cdot \beta_{ref} \cdot (1-q) \cdot z} \ .$$

*(80)*

Let's first define two distinct "wave-numbers" - $\delta_s$ and $\gamma_s$ for the two structure regions – region "H" for the $n_1$ region and region "L" for the $n_2$ region:

$$\text{H:} \quad \delta_s^2 \equiv 2 \cdot \beta_{ref}^2 \cdot [q - V_{s1}(x)]$$

$$\text{L:} \quad \gamma_s^2 \equiv 2 \cdot \beta_{ref}^2 \cdot [q - V_{s2}(x)]$$

$$V_{s1}(x) \equiv -\frac{1}{2} \cdot \left[\frac{n_1^2(x) - n_{ref}^2}{n_{ref}^2}\right] \; ; \; V_{s2}(x) \equiv -\frac{1}{2} \cdot \left[\frac{n_2^2(x) - n_{ref}^2}{n_{ref}^2}\right]$$

*(81)*

The solution to equation *(79)*, given the periodic permittivity *(60)*, is again (like in the case of the Full Helmholtz Equation) a set of Bloch functions –

$$\psi_s(x) = e^{i \cdot k_g \cdot x} \cdot u_s(x)$$

*(82)*

From here we just follow the procedure outlined in ***Appendix 1***, replace "$e$" by "$s$", replace $K^2$ by $q$ and recall the distinct "wave-numbers" - $\delta_s$ and $\gamma_s$ as defined by equation *(81)*.

For reference convenience let's just restate the eigenvalue equation (equation *(69)* with the necessary modifications):

$$LHS(k_{g;j}) = RHS_s(q)$$

*(83)*

So for the analytic solution of the SVEA Helmholtz Equation [equation *(71)*], given the finite 1D periodic array as described by equation *(60)*, we find again a set of Bloch functions of the form:

$$u_{s;n,j}(x) = \begin{cases} A_{n,j} \cdot e^{i \cdot (\delta_{s;n,j} - k_{g;j}) \cdot x} + A'_{n,j} \cdot e^{-i \cdot (\delta_{s;n,j} + k_{g;j}) \cdot x} \; ; \; H \; region \\ B_{n,j} \cdot e^{i \cdot (\gamma_{s;n,j} - k_{g;j}) \cdot x} + B'_{n,j} \cdot e^{-i \cdot (\gamma_{s;n,j} + k_{g;j}) \cdot x} \; ; \; L \; region \end{cases}$$

$$\psi_{s;n,j}(x) = e^{i \cdot k_{g;j} \cdot x} \cdot u_{s;n,j}(x)$$

*(84)*

Note that despite the difference in formal writing of the solutions ("$e$" for the EXAC case - equation *(67)* in ***Appendix 1*** vs. "$s$" here for the SVEA case), the cell functions [$u_{n,j}(x)$] and (necessarily) Bloch functions [$\psi_{n,j}(x)$] are identical (EXAC vs. SVEA – equation *(34)*). The functions are identical because the "energy" values [$K^2_{n,j}$ in the EXAC case and $q_{n,j}$ in the SVEA case] "adjust themselves" to the respective "potential" [$V_e(x)$ in the EXAC case and $V_s(x)$ in the SVEA case] such that the distinct "wave-numbers" - $\delta_{n,j}$ and $\gamma_{n,j}$ that enter the "cell function" [$u_{n,j}(x)$ cf. equation *(67)* and equation *(84)*] are equal.

The overall electrical field distributions (EXAC vs. SVEA) are NOT identical since the propagation constants [$k_{ze;n,j}$ ; $k_{zs;n,j}$] differ.

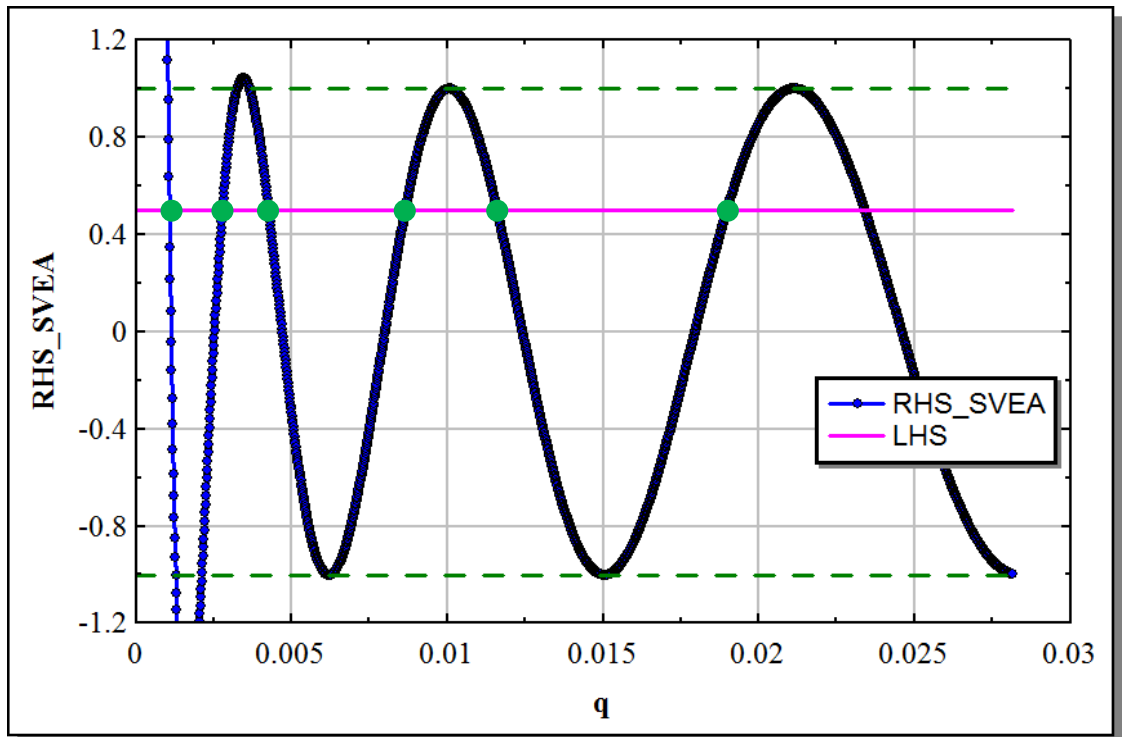

**Figure A5**: *$RHS_s(q)$ (blue) and $LHS(k_{g;j})$ (magenta). The green circles show the first six q values that solve the eigenvalue equation [equation (83)] for the particular $k_{g;j}$ of the LHS line. "Scanning" $k_{g;j}$ values across the first Brillouin Zone (so that the LHS line scans the plus one to minus one range and back) will create the first six q bands (and thus the first six bands of propagation constants - $k_{zs;n,j} = k_0 \cdot n_{ref} \cdot (1 - q_{n,j})$ [equation (85)].*

The bands of propagation constants $[k_{zs;n,j}]$ for the SVEA solutions are given by [cf. equations *(73)* and *(80)* and compare with equation *(70)* for the EXAC solutions]:

$$k_{zs;n,j} = k_0 \cdot n_{ref} \cdot (1 - q_{n,j})$$

*(85)*

$RHS_s(q)$ curve and $LHS(k_{g;j})$ line (for a particular $j$) are sown in *Figure A5*. The $q$ solutions to equation *(83)* are indicated by the green circles. Scanning the value of $k_{g;j}$ across the first Brillouin Zone will create the set of propagation bands [equation *(85)* and *Figure A3*]. *Figure A7* is a zoom showing individual propagation bands.

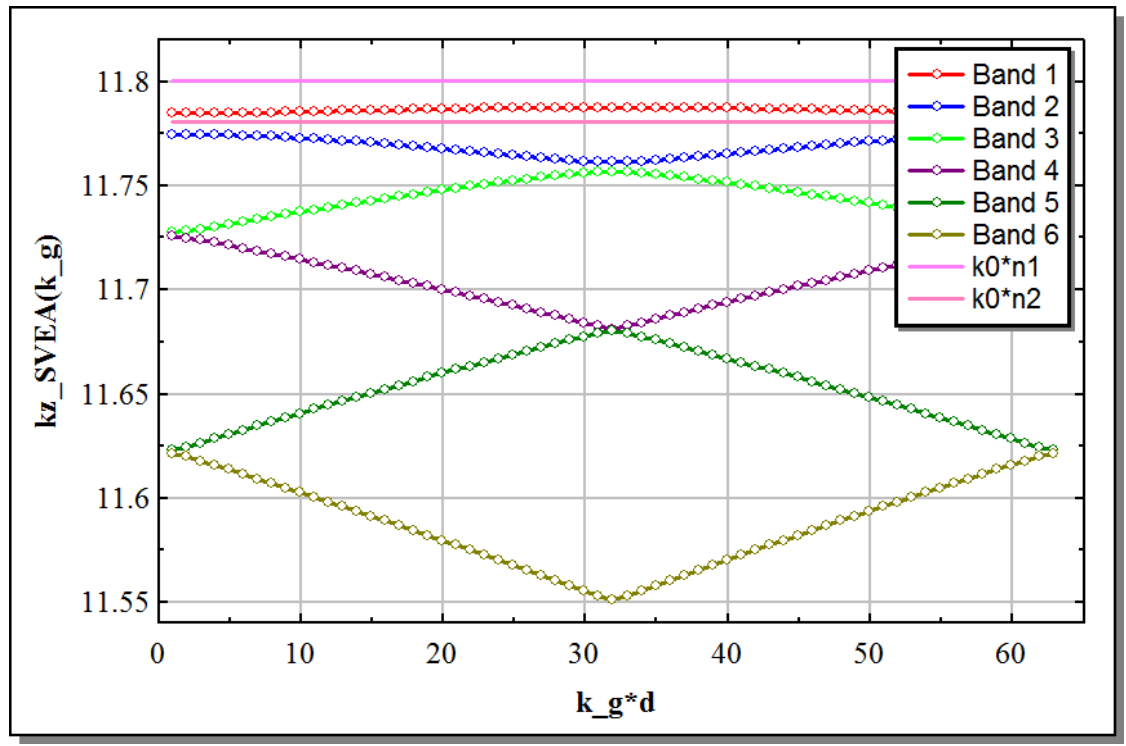

*Figure A6*: First six SVEA propagation bands as calculated by solving equation *(83)*. The two pink lines indicate the position of the propagation constant for a homogeneous bulk with refractive index $n_1$ (high pink line) or $n_2$ (lower pink line). The first band of SVEA propagation constants (and only the first band) lies entirely between the two pink lines.

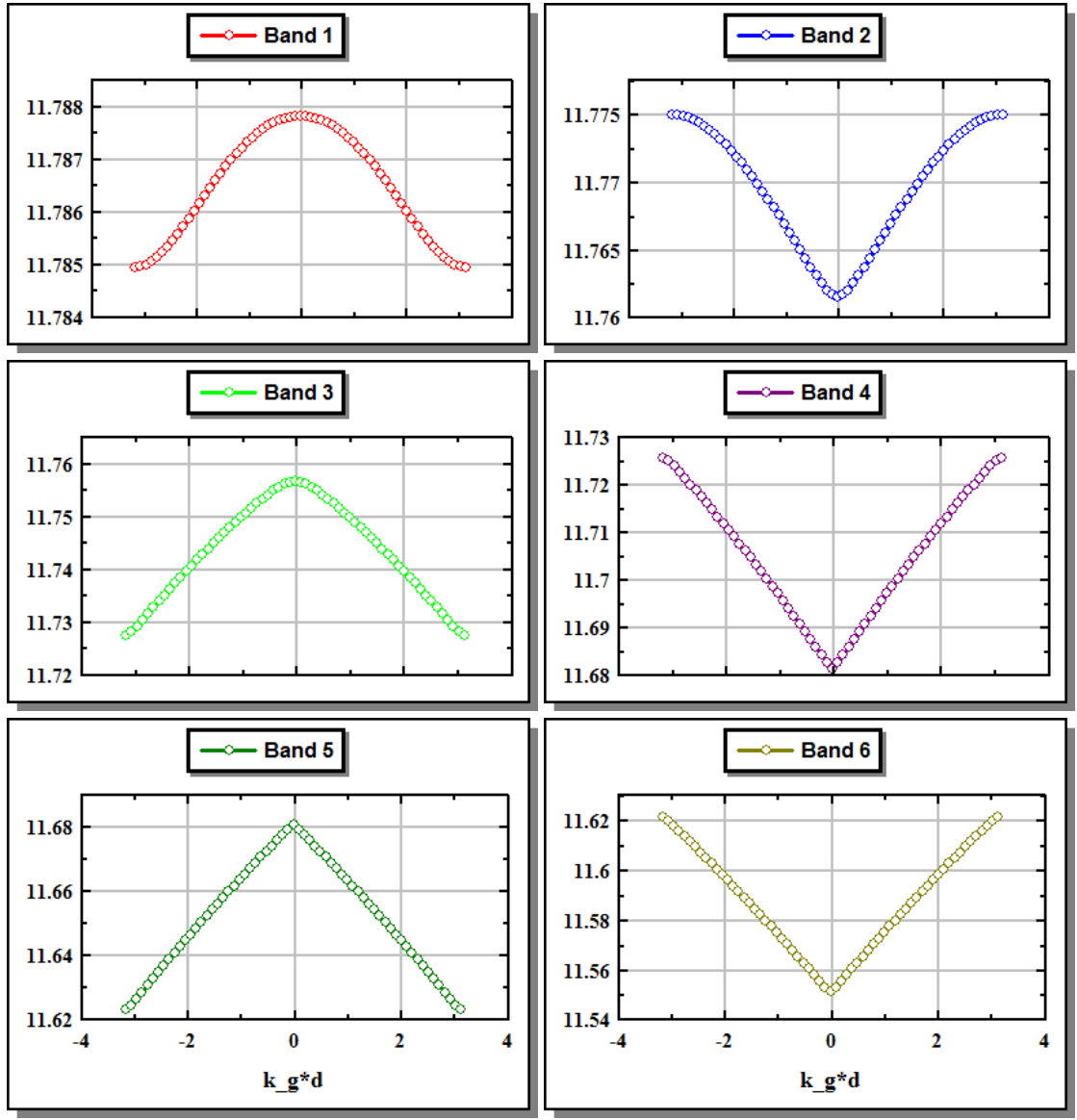

***Figure A7***: *Zoom to individual SVEA propagation bands for $n_{ref} = n_1$. Numbers on the vertical axis are $k_z$ in units of 1/micrometer. For SVEA bands too, the first derivative (with respect to $k_g$) of all bands continuously crosses zero at the center and edges of the Brillouin Zone (not obvious by looking at the shown curves, but consult **Figure 46** in the body of the paper).*

## *Appendix 3:* **Matrix elements for the Kronig–Penney model**

Equation *(66)* for the coefficients of the general solution [$\psi(x)$] states:

$$\begin{bmatrix} a_1, a_2, a_3, a_4 \\ b_1, b_2, b_3, b_4 \\ c_1, c_2, c_3, c_4 \\ d_1, d_2, d_3, d_4 \end{bmatrix} \cdot \begin{bmatrix} A \\ A' \\ B \\ B' \end{bmatrix} = 0$$

*(86)*

For the Kronig-Penney model, the sixteen coefficients of the matrix in equation *(86)* are (cf. equation *(60)* and cf. ***Figure A1***):

$a_1 = 1 \;;\; a_2 = 1 \;;\; a_3 = -1 \;;\; a_4 = -1$

$b_1 = \gamma_e \;;\; b_2 = -\gamma_e \;;\; b_3 = -\delta_e \;;\; b_4 = \delta_e$

$c_1 = exp[i \cdot (\gamma_e - k_g) \cdot a] \;;\; c_2 = exp[-i \cdot (\gamma_e + k_g) \cdot a]$

$c_3 = -exp[-i \cdot (\delta_e - k_g) \cdot b] \;;\; c_4 = -exp[+i \cdot (\delta_e + k_g) \cdot b]$

$d_1 = (\gamma_e - k_g) \cdot exp[i \cdot (\gamma_e - k_g) \cdot a]$

$d_2 = -(\gamma_e + k_g) \cdot exp[-i \cdot (\gamma_e + k_g) \cdot a]$

$d_3 = -(\delta_e - k_g) \cdot exp[-i \cdot (\delta_e - k_g) \cdot b]$

$d_4 = (\delta_e + k_g) \cdot exp[i \cdot (\delta_e + k_g) \cdot b]$

## Appendix 4: Solution of a set of four homogeneous equations

We want to solve

$$\begin{bmatrix} a_1, a_2, a_3, a_4 \\ b_1, b_2, b_3, b_4 \\ c_1, c_2, c_3, c_4 \\ d_1, d_2, d_3, d_4 \end{bmatrix} \cdot \begin{bmatrix} x_1 \\ x_2 \\ x_3 \\ x_4 \end{bmatrix} = 0$$

(87)

given that matrix $[a_1, \ldots, d_4]$ is singular $\{det[a_1, \ldots, d_4] = 0\}$.

Define:

$$A \equiv 1 - \frac{b_4 \cdot d_2}{b_2 \cdot d_4} \qquad B \equiv \frac{b_4 \cdot d_1}{b_2 \cdot d_4} - \frac{b_1}{b_2} \qquad C \equiv \frac{b_4 \cdot d_3}{b_2 \cdot d_4} - \frac{b_3}{b_2}$$

$$D \equiv \frac{c_4 \cdot d_1}{d_4 \cdot c_3} - \frac{c_1}{c_3} \qquad E \equiv \frac{c_4 \cdot d_2}{d_4 \cdot c_3} - \frac{c_2}{c_3} \qquad F \equiv 1 - \frac{c_4 \cdot d_3}{d_4 \cdot c_3}$$

$$G \equiv A - \frac{C \cdot E}{F} \qquad H \equiv B + \frac{C \cdot D}{F}$$

(88)

and write the solution as:

$$x_1 = any\ complex\ number$$

$$x_2 = \frac{H}{G} \cdot x_1$$

$$x_3 = \frac{D}{F} \cdot x_1 + \frac{E}{F} \cdot x_2$$

$$x_4 = -\frac{d_1}{d_4} \cdot x_1 - \frac{d_2}{d_4} \cdot x_2 - \frac{d_3}{d_4} \cdot x_3$$

(89)

## *Appendix 5:*   **Key code lines for BPM-fd**

Following are the code lines for BPM-fd. Names of the variables "speak for themselves". The program is written in NUMERIT [27].

```
func bpm_fd(k0_24,delta_z_24,delta_x_24,ref_index_x_array_24,\
            n_reference_24,k0nref_24,N_of_cells_x_24,\
            N_of_cells_zm_24,E_of_x_z0_24,E_x_z_24)

  ``a_i = lower diagonal  --> a[2-N]    {N-1 elements}
  ``b_i = main diagonal   --> b[1-N]    {N elements}
  ``a_i = upper diagonal  --> a[1-N-1]  {N-1 elements}
  ``d_i(z) = RHS(field(z)) d[1-N]    {N elements}

   d_td_24[N_of_cells_x_24]:0
   E_x_z_24[N_of_cells_zm_24,N_of_cells_x_24]:0
   ``The first row
   E_x_z_24[1,*] = E_of_x_z0_24

  ``counting arrays
   count_i = 2..N_of_cells_x_24 - 1 by 1
   count_i_minus_1 = count_i - 1
   count_i_plus_1 = count_i + 1
   count_1_to_N_minus_1 = 1..N_of_cells_x_24 - 1 by 1

  ``Define a_24,b_24[i],c_24[i]
   a_24 = delta_z_24/(2*delta_x_24^2)
     a_td_24[count_1_to_N_minus_1] = (-1)*a_24
     b_td_24 = delta_z_24/(delta_x_24^2) - \
        (delta_z_24/2)*(k0_24^2)*(ref_index_x_array_24^2 - n_reference_24^2) \
        +1j*2*k0nref_24
   c_24 = (-1)*delta_z_24/(delta_x_24^2) + \
      (delta_z_24/2)*(k0_24^2)*(ref_index_x_array_24^2 - n_reference_24^2) \
      +1j*2*k0nref_24

  `Loop on z (the propagation coordinate)
   for i_z = 2 to N_of_cells_zm_24
     ``Right hand side of the equation  {d_td_24[1] = d_td_24[N_of_cells_x_24] = 0}
      d_td_24[count_i] = (+1)*a_24*E_x_z_24[i_z-1,count_i_minus_1] \
             + c_24[count_i]*E_x_z_24[i_z-1,count_i] \
             + (+1)*a_24*E_x_z_24[i_z-1,count_i_plus_1]

      ``Call a fast tridiagonal algorithm
      solve_tridiagonal_set(N_of_cells_x_24,b_td_24,a_td_24,\
             a_td_24,d_td_24,x_tridiag_solution)

      E_x_z_24[i_z,*] = x_tridiag_solution
```

As an illustration to see the sensitivity to the value of the selected reference index, ***Figure A8*** shows two cases with a large $n_{ref}$ swing - high above $n_1$ and far below $n_2$:

- A.     $n_{ref} = n_1 + 0.025 \; (= 1.5275)$
- B.     $n_{ref} = n_2 - 0.025 \; (= 1.475)$

The input beam is a tilted "wide" Gaussian as for case XI of the main body (section *4.2.6*).

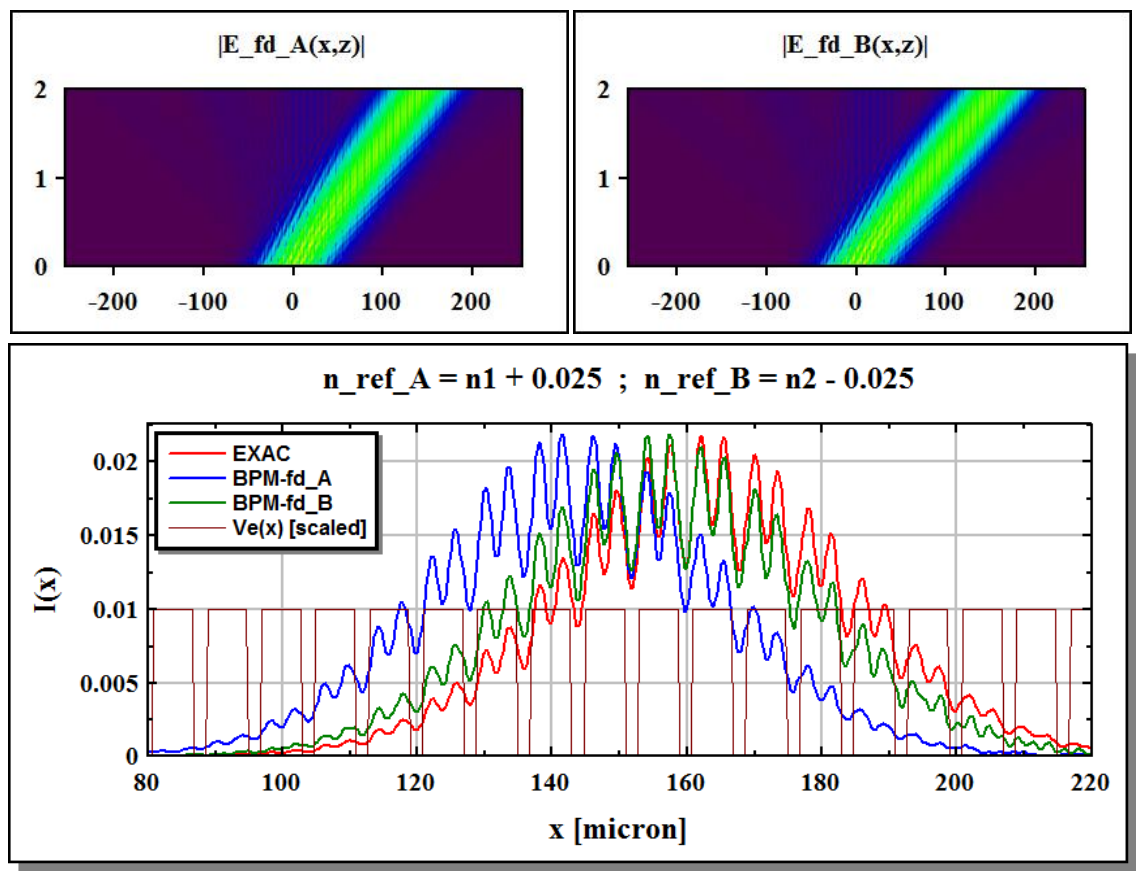

***Figure A8***: *Dependence of BPM-fd calculated electrical field distribution on the selection of the reference index, for a tilted wide Gaussian. See text for the respective reference index values. The map corresponding to high reference index is significantly off target (see **Figure A10** for calculated cross-sectional distances). Note that for a narrow on-axis Gaussian ("discrete diffraction"), BPM-fd scores better (shorter distance to the EXAC cross-section, i.e BPM-fd map is more accurate) than the BPM-ss map (cf. **Figure 24**). Vertical axis of the distribution maps is the propagation coordinate (**z**) in millimeters.*

## *Appendix 6:* **Key code lines for BPM-ss**

Following are the code lines for BPM-ss. Names of the variables "speak for themselves". The program is written in NUMERIT **[27]**.

```
func bpm_ss(`input`k0,n_reference,k0nref,kx,delta_zm,N_of_cells_zm,\
            N_of_cells_x,E_of_x_z0,alfa_x,ref_index_x_array,phase_sign,\
            `output` E_ss_x_z_map,G_of_kx_z,phi_ref_index_x_array,phi_planewaves)
 norm = 1/sqrt(N_of_cells_x)

  phi_ref_index_x_array = (-alfa_x + phase_sign*(+1)*1j*\
                k0^2*(ref_index_x_array^2 - n_reference^2)\
                /(2*k0nref))*delta_zm

 ``Define two maps
  E_ss_x_z_map[N_of_cells_zm,N_of_cells_x]:0
  E_ss_x_z_map[1,*] = E_of_x_z0
  G_of_kx_z[N_of_cells_zm,N_of_cells_x]:0

 ``IMPROVED BPM-ss (delta-z/2 for the planewave phase).
   ``delta_zm/2
   phi_planewaves = phase_sign*(-1j)*(kx^2/(2*k0nref))*(delta_zm/2)

   ``The main loop (two G_of_kx_z steps each with (delta_zm/2)
   for ii = 2 to N_of_cells_zm
     ``with (phi_ref_index_x_array)
       G_of_kx_z[ii-1,*] = norm*fft(E_ss_x_z_map[ii-1,*]*exp(phi_ref_index_x_array))
       E_ss_x_z_map[ii,*] = norm*ifft(G_of_kx_z[ii-1,*]*exp(phi_planewaves))
     ``without (phi_ref_index_x_array)
       G_of_kx_z[ii-1,*] = norm*fft(E_ss_x_z_map[ii,*])
       E_ss_x_z_map[ii,*] = norm*ifft(G_of_kx_z[ii-1,*]*exp(phi_planewaves))

   G_of_kx_z[N_of_cells_zm,*] = norm*\
            fft(E_ss_x_z_map[N_of_cells_zm-1,*]*exp(phi_ref_index_x_array))
```

As done in *Appendix 5* for BPM-fd, we continue here to explore the effect of reference index selection on the electrical field distribution calculated by BPM-ss.

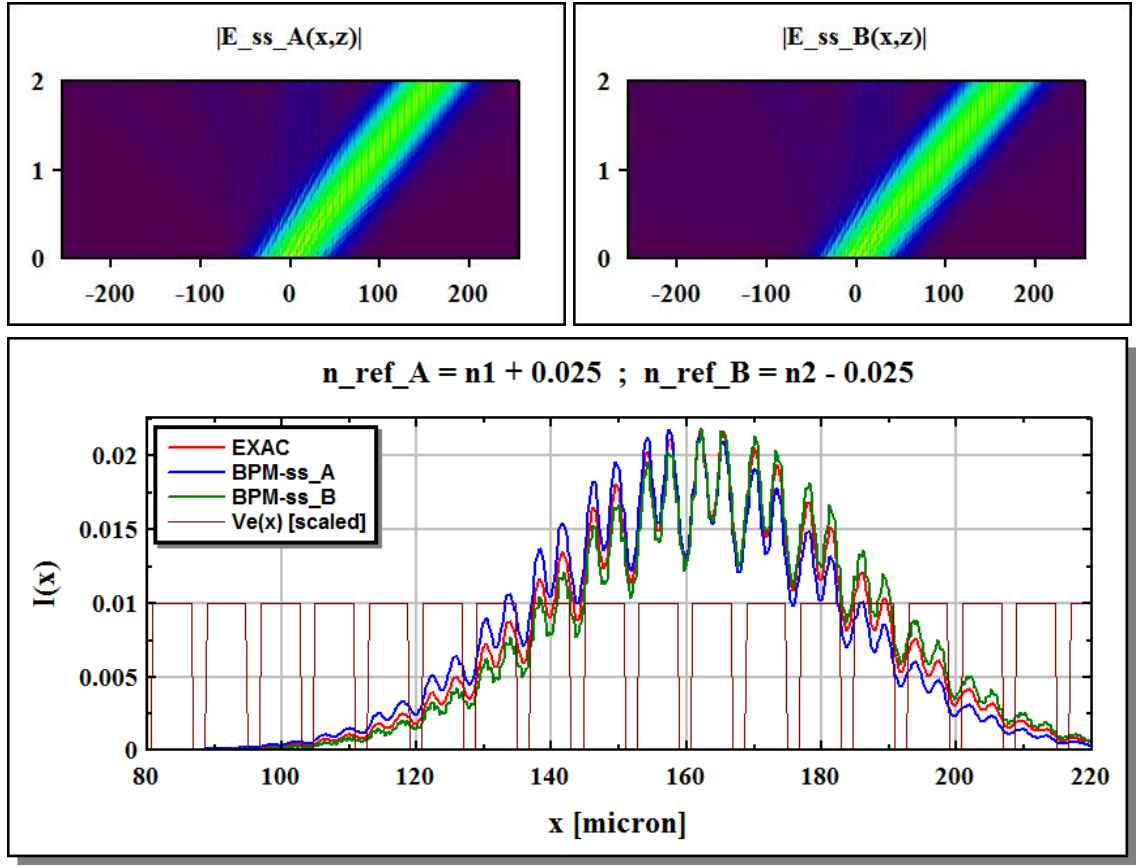

*Figure A9*: Dependence of BPM-ss calculated electrical field distribution on the selection of the reference index, for a tilted wide Gaussian. See text in *Appendix 5* for the respective reference index values. Despite the relatively large swing in reference index values selected, both cross-section A (blue) and cross-section B (green) are fairly close to the EXAC cross-section (red). Looking at both *Figure A8* and *Figure A9*, we see that for the type of exciting-field selected, BPM-ss is more accurate than BPM-fd (see *Figure A10* for calculated cross-sectional distances). Note that for a narrow on-axis Gaussian as the exciting field ("discrete diffraction"), BPM-fd cross-section scores **better** (shorter distance from the EXAC cross-section, i.e BPM-fd map is more accurate) than the BPM-ss cross-section (cf. *Figure 24*).

We find that for the type of exciting field selected, BPM-ss calculated distribution is more accurate than BPM-fd calculated distribution (see *Figure A10*).

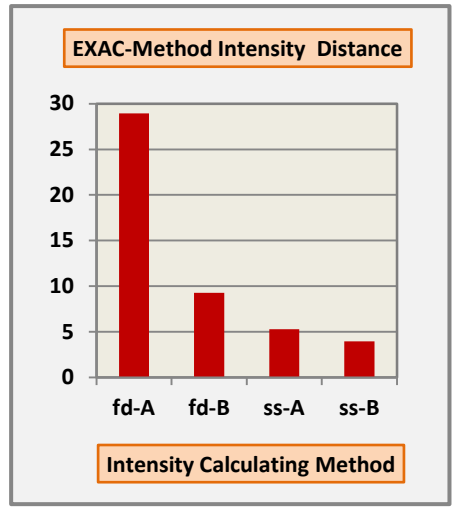

*Figure A10*:  *Cross-sectional distances (to the EXAC cross-section, see equation (56) for the mathematical definition of "distance") for the four cases of Appendix 5 and Appendix 6. Here we see far shorter distances for BPM-ss (vs. BPM-fd distances). In the case of narrow on-axis Gaussian as the input field, shorter distances were calculated for BPM-fd (cf. Figure 24). We conclude than that relative accuracy of the BPM methods depends on excitation conditions.*

Looking at both *Figure 24* in the main body and *Figure A10* of this appendix, we conclude than that relative accuracy of the BPM methods depends on excitation conditions.

## *Appendix 7:* Tips for numerical calculations

### Appendix 7-a: Generation of the waveguide array

Do not use "where" for generation of the WG array [$n(x)$]. "Where" may result in a non-equidistance and/or non equi-width array (by one pixel for each occurrence). Such non-periodic array will "quickly" (very small "degree of randomness") result in erroneous electrical field distributions (Anderson Localization).

Instead – select a fixed pixel width (in the "$x$" direction) and select two integers - one for the WG width and one for the WG-to-WG separation.

### Appendix 7-b: Solving the eigenvalue equation

By "eigenvalue equation" we refer to the implicit equation - $LHS(k_g) = RHS_e(K^2)$ [equation *(68)* in *Appendix 1*]. Here are the steps we suggest in solving the equation (calculate the eigenvalue bands):
- Find the $+1$ and the $-1$ crossings of the $RHS_e(K^2)$ function
- Work out the "book-keeping" for pairing the $+1$ and the $-1$ crossings to individual bands.
- Run the $LHS(k_g)$ function (across the $k_g$ values) and repeatedly find the solution (to the implicit equation) for each of the paired bands [$aroot(y,x)$ in NUMERIT **[27]**, or the equivalent in Matlab].